\documentclass[%
]{revtex4-1}

\usepackage{graphicx}
\usepackage{dcolumn}
\usepackage{bm}
\usepackage{float}
\usepackage{rotating}

\begin{document}
\maxdeadcycles=100000

\begin{center} 
{\bf\Large The 1223 new periodic orbits of planar three-body problem \\ with unequal mass  and zero angular momentum }

\vspace{0.5cm}

Xiaoming Li$^1$   and Yipeng Jing$^{2}$  and  Shijun Liao$^{1, 3, *}$ \\ 
\vspace{0.25cm}
$^1$ School of Naval Architecture, Ocean and Civil Engineering, Shanghai Jiaotong University, China \\  
$^2$ Dept. of Astronomy, School of Physics and Astronomy,  Shanghai Jiaotong University, China\\
$^3$ Ministry-of-Education Key Laboratory in Scientific and Engineering Computing, Shanghai 200240, China\\  
\vspace{0.cm}

* The corresponding author:  sjliao@sjtu.edu.cn
\end{center}
 {\bf Abstract}
{\em
We present 1349  families of Newtonian periodic  planar three-body orbits with unequal mass and zero angular momentum and the initial conditions in case of isosceles collinear configurations.  These 1349 families of the periodic collisionless orbits can be divided into seven classes according to their geometric and algebraic symmetries.  Among these 1349 families, 1223 families are entirely new, to the best of our knowledge.  Furthermore, some periodic orbits have the same free group element for different mass.  We find that the scale-invariant average period linearly increases with the mass of one body when the masses of the other two are constant.  It is suggested that the masses of bodies may play an important role in periodic three-body systems.   The movies of these periodic orbits are given on the website  \url{http://numericaltank.sjtu.edu.cn/three-body/three-body-unequal-mass.htm} . 

}

\vspace{0.5cm}

The study of periodic three-body problem has received lots of attention in recent years.  In 2013,  \v{S}uvakov and Dmitra\v{s}inovi\'{c} \cite{Suvakov2013} found 13 new distinct collisionless periodic orbits of Newtonian planar three-body problem with equal mass and zero angular momentum.  In 2014, Dmitra\v{s}inovi\'{c} et al. \cite{Dmitrasinovic2014} investigated gravitational waves from these periodic three-body systems.  
In the same year, Iasko and Orlov\cite{Iasko2014} found nine new close to periodic orbits of three-body problem with equal mass and \v{S}uvakov \citep{Suvakov2014a} found eleven solutions in the vicinity of the figure-eight orbits with equal mass.  In 2015, Hudomal \cite{Hudomal2015} reported 25 families of periodic orbits Newtonian planar three-body problem with equal mass, including the 11 families found in \cite{Suvakov2013}.   In 2016, Rose \cite{Rose2016} gained 90 periodic planar collisionless orbits with equal mass in case of isosceles collinear configurations.  Recently, Li and Liao \cite{Li2017-SciChina} found more than six hundred new families of planar three-body problem with equal mass.  
These researchers focused on periodic three-body problem with equal mass.  However, the three-body problem with unequal mass is more general. 
In 2002, Gal\'an  et al. \cite{Galan2002} studied stability properties of figure-eight \cite{More1993}\cite{Chenciner2000} as the masses are unequal.  Doedel et al. \cite{Doedel2003} gained periodic orbits from the figure-eight as the mass of one body is varied.     
In 2015, Yan et al. \cite{Yan2015} investigated the spatial isosceles three-body problem with unequal masses in case of one body moves up and down on a vertical line.
However, little attention has been paid to search for the collisionless periodic orbits of the planar three-body problem with unequal mass and zero angular momentum.  In this paper,  we  present results of collisionless periodic orbits in Newtonian planar three-body problem with unequal mass and zero angular momentum.  

The motions of Newtonian planar three-body system can be described by Newtonian second law and gravitational law:
$
  \ddot{\bm{r}}_{i}=\sum_{j=1,j\neq i}^{3} \frac{G m_{j}(\bm{r}_{j}-\bm{r}_{i})}{| \bm{r}_i-\bm{r}_j |^{3}}, 
$
where ${\bm r}_i$ and $m_i$ denote vector position and mass of the $i$th body $(i=1,2,3)$, $G$ denotes the Newtonian gravity coefficient,  and  the dot denotes the derivative with respect to the time $t$, respectively.  We consider a planar three-body system with zero angular momentum and unequal mass ($m_1 = m_2  \neq m_3$) in the case of $G=1$ and the initial conditions in case of the isosceles collinear configurations:
\begin{equation}
\left\{ 
\begin{array} {l}
  \bm{r}_1(0)=(x_1,x_2)=-\bm{r}_2(0), \;\;  \bm{r}_3(0)=(0,0), \\
 \dot{\bm{r}}_1(0)= \dot{\bm{r}}_2(0)=(v_1,v_2), \;\; \dot{\bm{r}}_3(0)=-\frac{m_1+m_2}{m_3}\dot{\bm{r}}_1(0)
 \end{array}
 \right.  \label{initial}
 \end{equation}

With these configurations, if  ${\bm r}_i(t)$ $(i=1,2,3)$ with ${m}_i$ $(i=1,2,3)$ denotes a periodic orbit with the period $T$ of a three-body system, then 
\begin{equation}
{\bm r}'_i(t') = {\bm r}_i(t),\;\;  {\bm v}'_i(t') = \alpha \;{\bm v}_i(t), \;\;  t' = t/\sqrt{\alpha}, \;\; {m}'_i = \alpha^2 \;{m}_i, \label{scaling}
\end{equation}
has a same periodic orbit with the period $T' = T/\sqrt{\alpha}$ for arbitrary $\alpha>0$. 
Therefore, without loss of generality,  we consider $m_1=m_2=1$ and $m_3$ is varied.  Note that 11 families were found by  \v{S}uvakov and Dmitra\v{s}inovi\'{c} \cite{Suvakov2013} and more than six hundred new families of periodic orbits were found by Li and Liao \citep{Li2017-SciChina} in case of $m_3=1$.  In this paper, we search for periodic orbits  in case of $m_3 \neq 1$.

For any given $m_3$, the orbits are determined by the four parameters $(x_1, x_2, v_1, v_2)$.   Write $\bm{y}(t)=(\bm{r}_1(t), \dot{\bm{r}}_1(t))$.
A periodic solution with the period $T_0$ is the root of the equation $\bm{y}(T_0)-\bm{y}(0)=0$, where $T_0$ is unknown.  

Firstly,  we use the grid search method to find approximated initial conditions to satisfy the equation $\bm{y}(T_0)-\bm{y}(0)=0$.  We set the initial positions $x_1=-1$ and $x_2=0$ and we search for initial conditions in a square plane: $v_1\in [0,1]$ and $v_2 \in [0,1]$.  We set 4000 points in each dimension and thus have 16 million  grid points in the square search plane.  With these different 16 million initial conditions, the motion equations are integrated up to the time $T_0 = 200$ by means of the ODE solver dop853 developed by Hairer et al. \citep{Hairer1993},  which is based on an explicit Runge-Kutta method of order 8(5,3) in double precision with adaptive step size control.   The corresponding initial conditions and the period $T_0$  are chosen as the candidates when the return proximity function
\begin{equation}
|\bm{y}(t)-\bm{y}(0)| = \sqrt{\sum_{i=1}^{4}(y_i(t)-y_i(0))^2}   \label{def:return-proximity-function}
\end{equation}
is less than $10^{-1}$.  

Secondly, we improve  these  candidates by  means  of  the Newton-Raphson method \citep{Farantos1995, Lara2002,  Abad2011}.  At this stage, the motion equations are solved numerically by means of the same ODE solver dop853 \cite{Hairer1993}.  The candidates are corrected until the level of the return proximity function (\ref{def:return-proximity-function}) is less than $10^{-3}$.  As mentioned by in Li and Liao \cite{Li2017-SciChina}, many periodic orbits might be lost by means of traditional algorithms in double precision. Thus, we further integrate the motion equations by means of ``Clean Numerical Simulation'' (CNS) \citep{Liao2009, Liao2013-3b,Liao2014-SciChina, Liao2015-IJBC} with negligible numerical noises in a long enough interval of time,  which is based on the arbitrary order of Taylor expansion method \citep{Barton1971, Corliss1982, Chang1994, Barrio2005}  in  arbitrary  precision \citep{Oyanarte1990, Viswanath2004}, plus a convergence check by means of an additional computation with even smaller numerical noises.  A  periodic  orbit  is  found when  the level of the return proximity function (\ref{def:return-proximity-function}) is less than $10^{-6}$.   
Note that the initial positions ${\bm r}_1 =(x_1,x_2)$ will depart from $(-1,0)$ a little.  

It is well-known that, if  ${\bm r}_i(t)$ $(i=1,2,3)$  denotes a periodic orbit with the period $T$ of a three-body system, then 
\begin{equation}
{\bm r}'_i(t') = \alpha \; {\bm r}_i(t),\;\; {\bm v}'_i(t') = {\bm v}_i(t)/\sqrt{\alpha}, \;\;  t' = \alpha^{3/2} \; t, \label{scaling}
\end{equation}
is also a periodic orbit with the period $T' = \alpha^{3/2}\; T$ for arbitrary $\alpha>0$.   Thus, through coordinate transformation and then the scaling of the spatial and temporal coordinates, we can always enforce (-1,0), (1,0) and (0,0) as the initial positions of the body-1, 2 and 3, respectively.  In this way, the periodic orbits are only dependent upon two physical parameters ($v_1$,$v_2$), the initial velocity of Body-1.   

\begin{table*}
\tabcolsep 0pt \caption{The initial velocities and periods $T$ of some newly-found periodic orbits of the three-body system with unequal mass and zero angular momentum  in the case of  $\bm{r}_1(0)=(-1,0)=-\bm{r}_2(0)$,  $\dot{\bm{r}}_1(0)=(v_1,v_2)=\dot{\bm{r}}_2(0)$ and $\bm{r}_3(0)=(0,0)$, $\dot{\bm{r}}_3(0)=(-2v_1/m_3, -2v_2/m_3)$  when $G=1$ and $m_1=m_2=1$, where $T^*=T |E|^{3/2}$ is its scale-invariant period, $L_f$ is the length of the free group word (element). Here, the superscript $i.c.$ indicates the case of the initial conditions with isosceles collinear configuration, due to the fact that there exist periodic orbits in many other cases.  } \label{table1} \vspace*{-12pt}
\begin{center}
\def\temptablewidth{1\textwidth}
{\rule{\temptablewidth}{1pt}}
\begin{tabular*}{\temptablewidth}{@{\extracolsep{\fill}}lccccc}
\hline
Class, number and $m_3$ & $v_1$ & $v_2$  & $T$ & $T^*$ & $L_f$\\
\hline
I.A$_{4}^{i.c.}$(0.5) &     0.2009656237 &     0.2431076328 &    19.0134164290 &     19.086 &  16\\
I.A$_{68}^{i.c.}$(0.5) &     0.2138410831 &     0.0542938396 &    83.8472907647 &    118.112 &  104\\
I.B$_{59}^{i.c.}$(0.75) &     0.4101378717 &     0.1341894173 &   121.0976361440 &    183.067 &  102\\
I.A$_{1}^{i.c.}$(2) &     0.6649107583 &     0.8324167864 &    12.6489061509 &     42.121 &  8\\
II.D$_{2}^{i.c.}$(2) &     0.3057224330 &     0.5215124257 &     8.8237067653 &     64.567 &  12\\
I.A$_{2}^{i.c.}$(4) &     0.9911981217 &     0.7119472124 &    17.6507807837 &    276.852 &  24\\
\hline
\end{tabular*}
{\rule{\temptablewidth}{1pt}}
\end{center}
\end{table*}

\begin{figure*}[ht]
  \centering \includegraphics[scale=0.25]{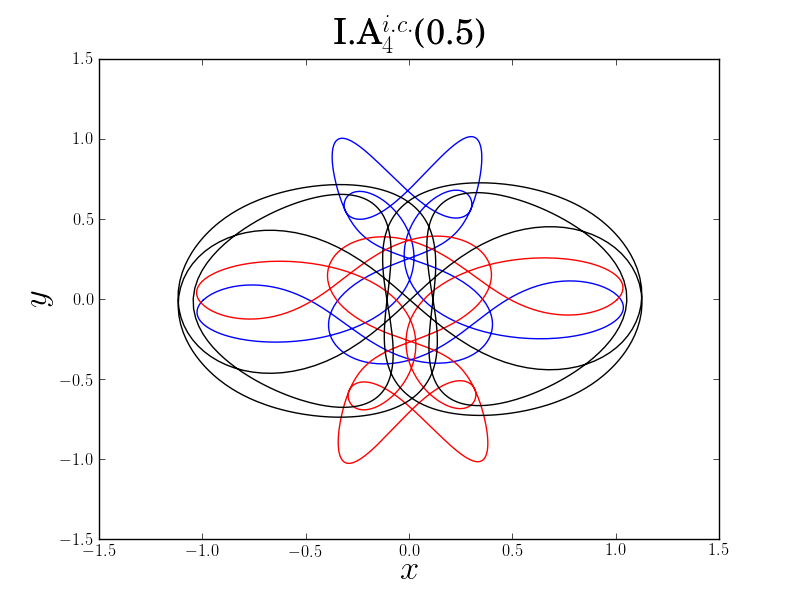}
  \centering \includegraphics[scale=0.25]{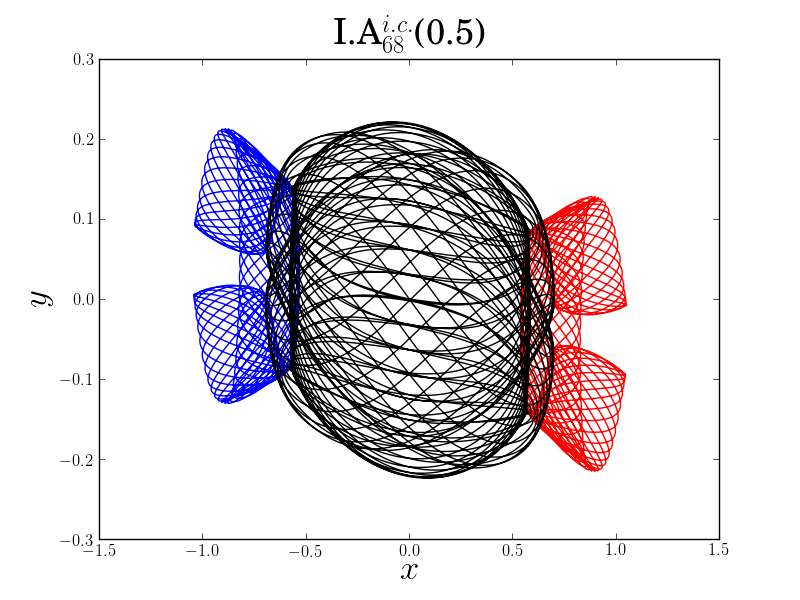}
  \centering \includegraphics[scale=0.25]{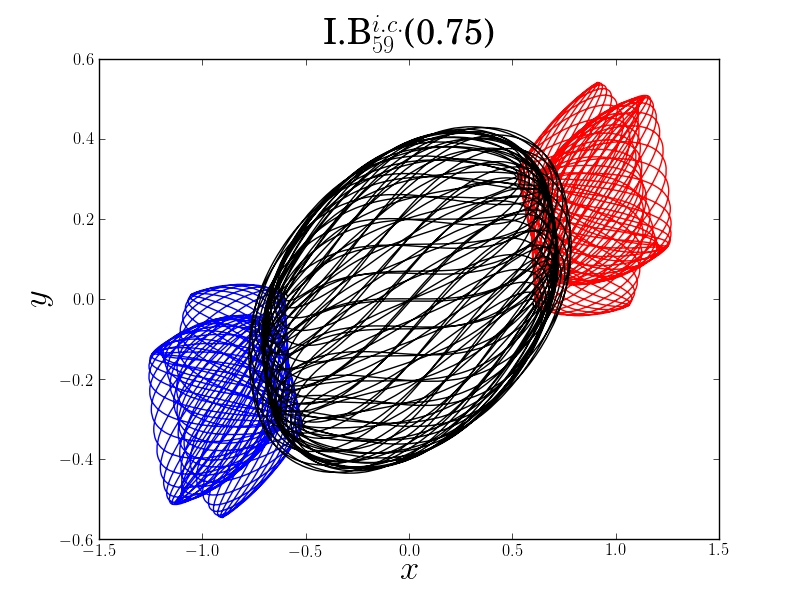}
  \centering \includegraphics[scale=0.25]{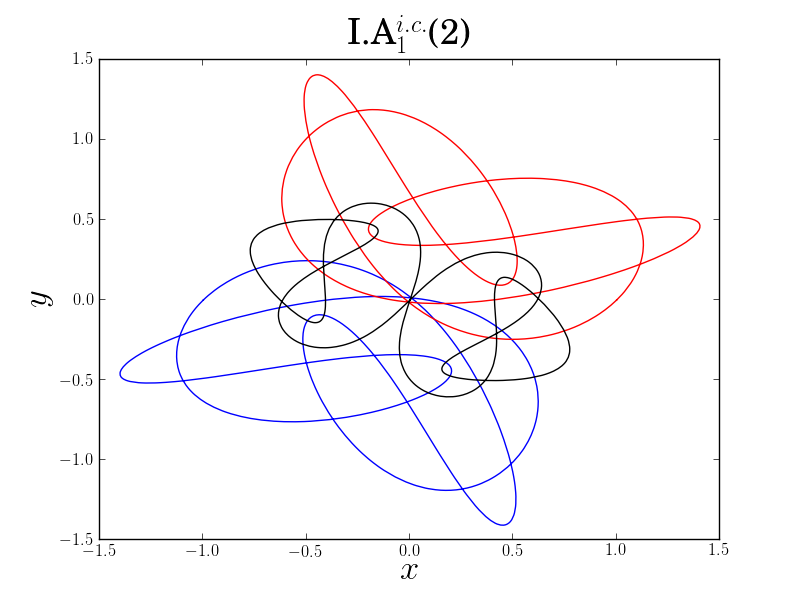}
  \centering \includegraphics[scale=0.25]{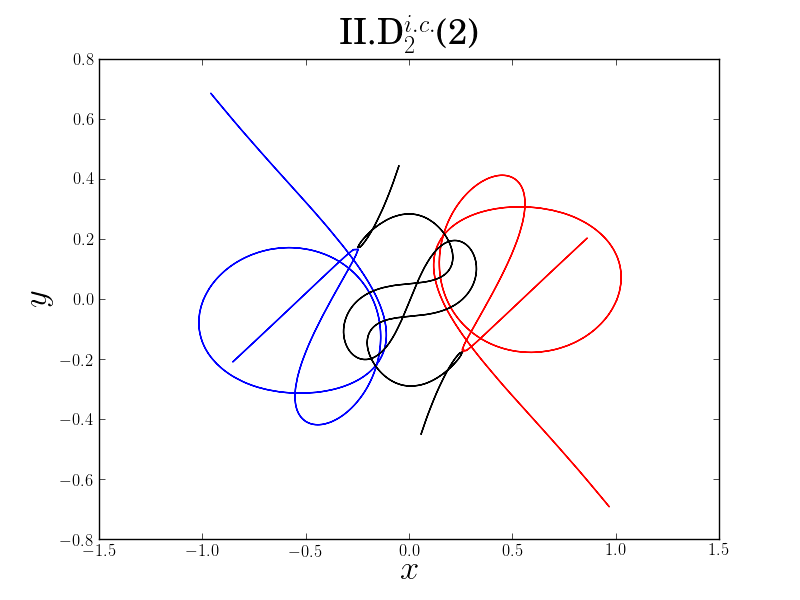}
  \centering \includegraphics[scale=0.25]{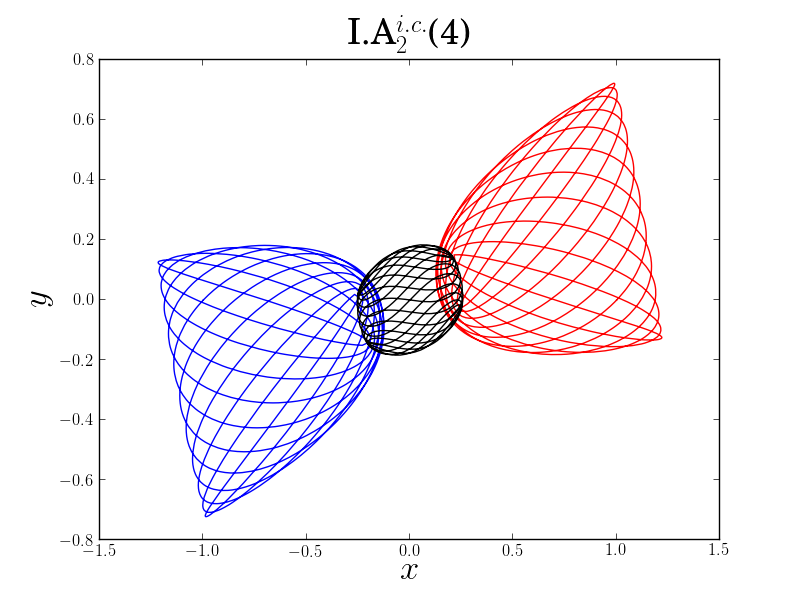}
  \caption{(color online.) Brief overview of the six newly-found families of periodic three-body orbits.}
  \label{fig1}
\end{figure*}

We identify these periodic orbits by means of Montgomery's topological identification and classification method  \citep{Montgomery1998}.  The positions $\bm{r_1}, \bm{r_2}$ and $\bm{r_3}$ of the three-body corresponds to a unit vector $\bm{n}$ in the so-called ``shape sphere'' with the Cartesian components 
\[ n_x=\frac{2\bm{\rho}\cdot\bm{\lambda}}{R^2}, \;\; n_y=\frac{\lambda^2-\rho^2}{R^2}, \;\;  n_z=\frac{2(\bm{\rho}\times\bm{\lambda}) \cdot \bm{e}_z }{R^2},\] 
where 
$   \bm{\rho}=\frac{1}{\sqrt{2}}(\bm{r_1}-\bm{r_2})$, $ \bm{\lambda}=\frac{1}{\sqrt{6}}(\bm{r_1}+\bm{r_2}-2\bm{r_3})$ and the hyper-radius $R=\sqrt{\rho^2+\lambda^2}$.  Then a periodic orbit is associated with a closed curve on the shape sphere, which can be characterized by its topology with three punctures (two-body collision points).  With one of the punctures as the ``north pole", the sphere can be mapped onto a plane by a stereographic projection.  And a closed curve can be mapped onto a plane with two punctures and its topology can be described by the so-called ``free group element'' (word) with letters $a$ (a clockwise around right-hand side puncture),  $b$ (a counter-clockwise around left-hand side puncture) and their inverses $a^{-1}=A$ and $b^{-1}=B$.  For details, please refer  to  \cite{Suvakov2014a}.   

\begin{figure*}[!ht]
  \centering \includegraphics[scale=0.25]{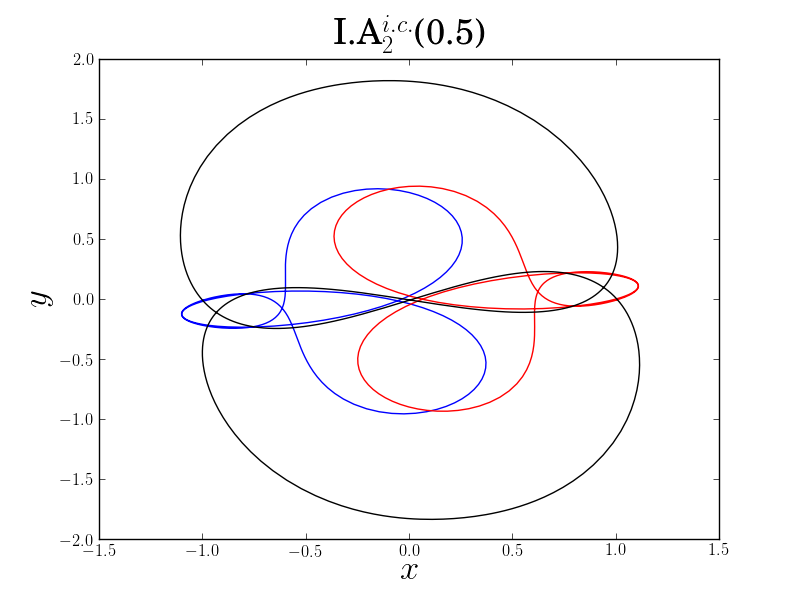}
  \centering \includegraphics[scale=0.25]{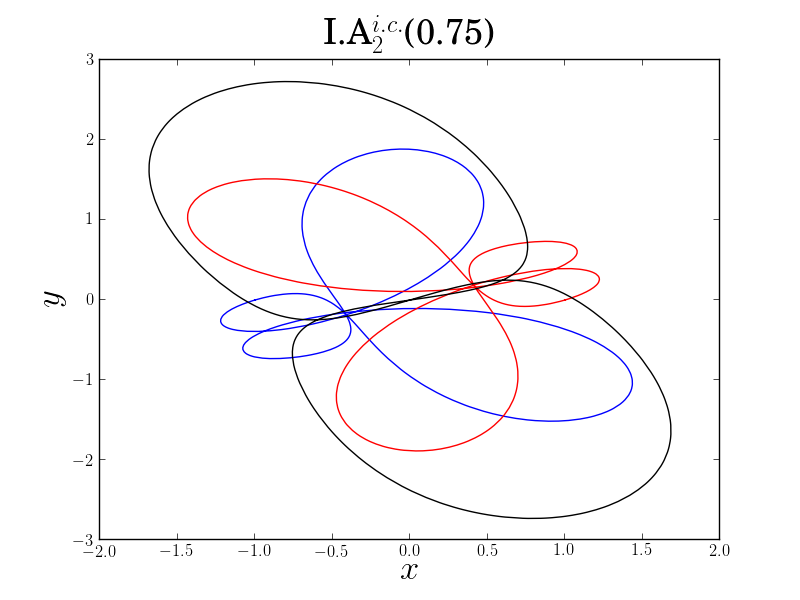}
  \centering \includegraphics[scale=0.25]{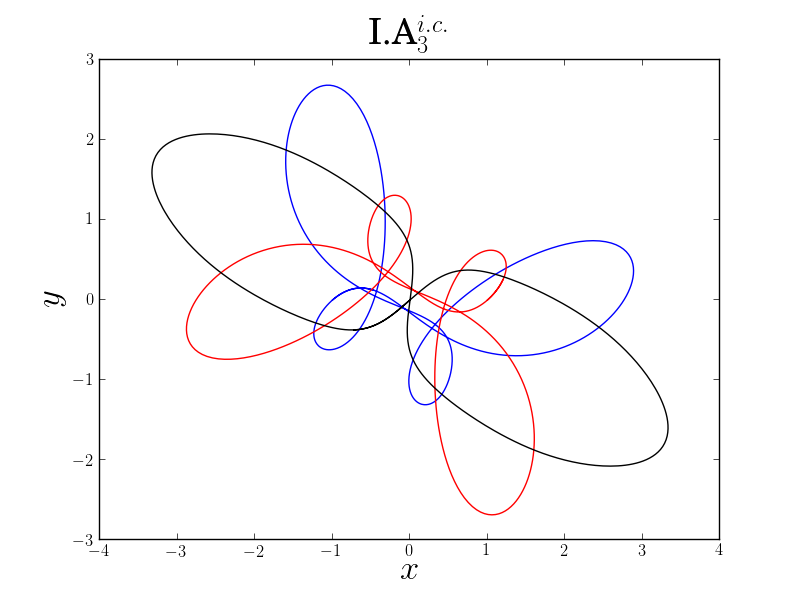}
  \caption{(color online.) The periodic orbits with the same free group element for $m_3$ = 0.5, 0.75 and 1.}
  \label{fig2}
\end{figure*}

The periodic orbits can be divided into different classes on basis of their geometric and algebraic symmetries \cite{Suvakov2013}.   There are two types of geometric symmetries in the shape space: (I) the reflection symmetries of two orthogonal axes --- the equator and the zeroth meridian passing through the ``far'' collision point; and (II) a central reflection symmetry about one point --- the intersection of the equator and the aforementioned zeroth meridian.  Besides, \v{S}uvakov and Dmitra\v{s}inovi\'{c} \cite{Suvakov2013} mentioned three types of algebraic exchange symmetries for the free group elements: (A) the free group elements are symmetric with $a \leftrightarrow A$ and $b \leftrightarrow B$,  (B) free group elements are symmetric with $a \leftrightarrow b$ and $A \leftrightarrow B$, and (C) free group elements are not symmetric under either (A) or (B). 
However, in this paper, we find a new algebraic symmetry class (D) with free group words symmetric under $a \leftrightarrow B$ and $b \leftrightarrow A$. And then (C) can be regarded as free group words are not symmetric under one of (A), (B) and (D).  We observe that the algebraic symmetry class (D) always corresponds to the geometric class (II)  for present orbits.   

\begin{figure}[!ht]
  \centering
  \includegraphics[scale=0.3]{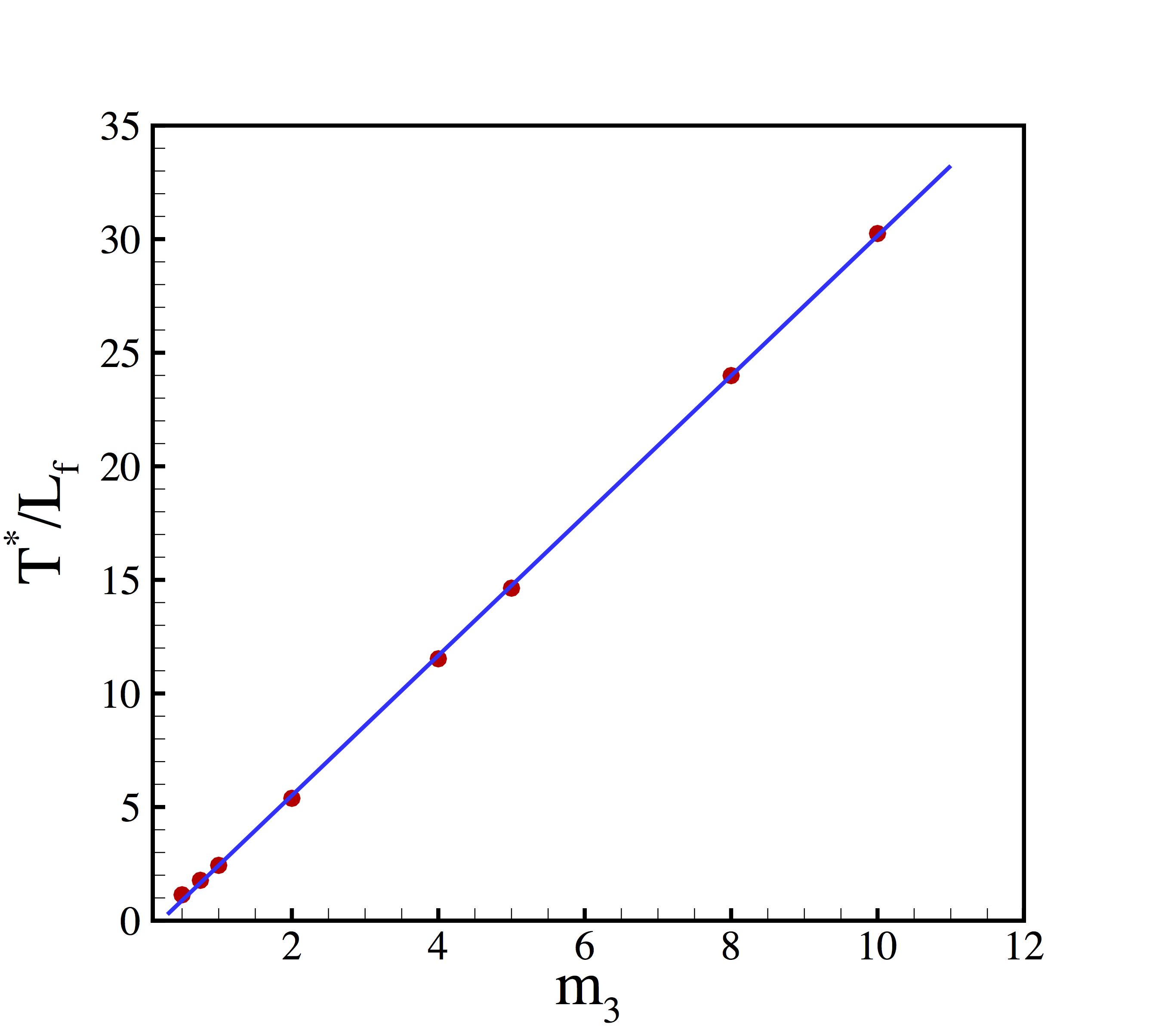}
  \caption{(color online). The scale-invariant  average  period $T^*/L_f = T |E|^{3/2}/L_f$  versus $m_3$.  Symbols: computed results; line: $T^*/L_f=3.074m_3-0.617$. }
  \label{m3}
\end{figure}
 
Within the period  $T\in[0,200]$, we find 565, 401, 237, 85, 35, 17 and 9 families of periodic orbits in case of $m_3$ = 0.5, 0.75, 2, 4, 5, 8 and 10, respectively.  Among these 1349 families of periodic orbits, 1223 families of periodic orbits are entirely new, to the best of our knowledge. These 1349 families of the periodic collisionless orbits can be divided into seven classes:  I.A,  I.B, I.C, II.A, II.B, II.C amd II.D,  as listed in Table~S~I-LXXXI in Supplementary material \cite{Supplement}.   The initial velocities of the periodic orbits are listed in Tables~S~I-XXIX in \cite{Supplement}.  The free group elements of the 1349 families are listed in Table S XXIX-LXXXI in \cite{Supplement}. 
Due to the limited length,  only six new families are listed in Table~\ref{table1},  and their orbits in real space are shown in FIG.~\ref{fig1}. 
Note that Doedel et al. \cite{Doedel2003} found periodic orbits which have the same topology with the figure-eight as the mass of one body is varied.  Here we also find that some periodic orbits have the same free group element for different $m_3$.  For instance, as shown in FIG.~\ref{fig2}, the periodic orbits have the same free group element ($BaBabAbA$) for $m_3$ = 0.5, 0.75 and 1.   The movies of these periodic orbits are given on the website:
 \url{http://numericaltank.sjtu.edu.cn/three-body/three-body-unequal-mass.htm} . 

As mentioned by Li and Liao \cite{Li2017-SciChina}, the scale-invariant  average  period $\bar{T}^* = \bar{T} |E|^{3/2}$ is approximately equal to a {\em universal} constant in case of $m_3=1$, i.e.  $ \bar{T}^* \approx 2.433 \pm 0.075$, with the definition of the average period $\bar{T} = T/L_f$, where $L_f$ is the length of free group element of periodic orbit of a three-body system.  Here the scale-invariant average period is equal to 1.14, 1.77, 2.43, 5.39, 11.53, 14.64, 23.99 and 30.25 for $m_3$ = 0.5, 0.75, 2, 4, 5, 8 and 10, respectively.  It is found that the scale-invariant average period agree well with the formula $\bar{T}^*=3.074m_3-0.617$ as shown in FIG.~\ref{m3}.  The standard deviation is $\sigma$ = 0.135.  It is suggested that the scale-invariant average period linearly increases with $m_3$.       

In this paper, we totally find 1349 families of Newtonian periodic planar three-body orbits with unequal mass and zero angular momentum and the initial conditions in case of isosceles collinear configurations.  These 1349 families of the periodic collisionless orbits can be divided into seven classes according to their geometric and algebraic symmetries.  Among these 1349 families, 1223 families are entirely new, to the best of our knowledge.    Furthermore, some periodic orbits have the same free group element for different $m_3$.  It is found that the scale-invariant average period linearly increases with $m_3$.  It is suggested that the masses of bodies may play an important role in periodic three-body systems.

This work was carried out on TH-2 at National Supercomputer Centre in Guangzhou, China.  It is partly supported by National Natural Science Foundation of China (Approval No. 11432009).


\bibliography{ref}

\newpage

\setcounter{table}{0}
\setcounter{figure}{0}
\renewcommand{\figurename}{FIG. S.}
\renewcommand{\tablename}{Table S.}

\begin{center}

{\bf \Large Supplementary information for\\  ``The 1223 new periodic orbits of planar three-body problem \\with unequal mass  and zero angular momentum "}

\end{center}

\begin{table}[b]
\tabcolsep 0pt \caption{Initial conditions and periods $T$ of the periodic three-body orbits for class I.A in the case of  $\bm{r}_1(0)=(-1,0)=-\bm{r}_2(0)$,  $\dot{\bm{r}}_1(0)=(v_1,v_2)=\dot{\bm{r}}_2(0)$ and $\bm{r}_3(0)=(0,0)$, $\dot{\bm{r}}_3(0)=(-2v_1/m_3, -2v_2/m_3)$ when $G=1$ and $m_1=m_2=1$ and $m_3=0.5$ by means of the search grid $4000\times 4000$ in the interval $T_0\in[0,200]$, where $T^*=T |E|^{3/2}$ is its scale-invariant period, $L_f$ is the length of the free group element.  } \label{table-S5} \vspace*{-12pt}
\begin{center}
\def\temptablewidth{1\textwidth}
{\rule{\temptablewidth}{1pt}}
\begin{tabular*}{\temptablewidth}{@{\extracolsep{\fill}}lccccc}
\hline
Class and number  & $v_1$ & $v_2$  & $T$  & $T^*$ & $L_f$\\
\hline
I.A$_{1}^{i.c.}$(0.5) &     0.2869236336 &     0.0791847624 &     4.1761292190 &      4.538 &  4\\
I.A$_{2}^{i.c.}$(0.5) &     0.3420307307 &     0.1809369236 &    13.9153339459 &      9.063 &  8\\
I.A$_{3}^{i.c.}$(0.5) &     0.3697718457 &     0.1910065395 &    25.9441952945 &     13.095 &  12\\
I.A$_{4}^{i.c.}$(0.5) &     0.2009656237 &     0.2431076328 &    19.0134164290 &     19.086 &  16\\
I.A$_{5}^{i.c.}$(0.5) &     0.2613236072 &     0.2356235235 &    28.4358575383 &     23.513 &  20\\
I.A$_{6}^{i.c.}$(0.5) &     0.1908428490 &     0.1150772110 &    15.9682350284 &     22.361 &  20\\
I.A$_{7}^{i.c.}$(0.5) &     0.1579313682 &     0.0949852732 &    14.5766076405 &     22.363 &  20\\
I.A$_{8}^{i.c.}$(0.5) &     0.0979965852 &     0.0369408875 &    15.6059191780 &     27.112 &  24\\
I.A$_{9}^{i.c.}$(0.5) &     0.3589116510 &     0.0578397225 &    35.2777168591 &     27.120 &  24\\
I.A$_{10}^{i.c.}$(0.5) &     0.2066204352 &     0.1123859298 &    22.8770013381 &     30.956 &  28\\
I.A$_{11}^{i.c.}$(0.5) &     0.3095805649 &     0.1012188182 &    37.8353981553 &     36.122 &  32\\
I.A$_{12}^{i.c.}$(0.5) &     0.2935606362 &     0.2168613674 &    54.5846159117 &     41.571 &  36\\
I.A$_{13}^{i.c.}$(0.5) &     0.2614113685 &     0.1097599351 &    39.2849176561 &     45.204 &  40\\
I.A$_{14}^{i.c.}$(0.5) &     0.3049866810 &     0.0979042378 &    46.1065937257 &     45.210 &  40\\
I.A$_{15}^{i.c.}$(0.5) &     0.1644199050 &     0.0637816144 &    29.0215071279 &     45.244 &  40\\
I.A$_{16}^{i.c.}$(0.5) &     0.2698142826 &     0.0360688014 &    37.8687787781 &     45.458 &  40\\
I.A$_{17}^{i.c.}$(0.5) &     0.1451647294 &     0.0318334148 &    30.5079373557 &     49.973 &  44\\
I.A$_{18}^{i.c.}$(0.5) &     0.3467747647 &     0.0474429378 &    59.7722919460 &     49.973 &  44\\
I.A$_{19}^{i.c.}$(0.5) &     0.3025694869 &     0.0951546278 &    54.5401904272 &     54.294 &  48\\
I.A$_{20}^{i.c.}$(0.5) &     0.2726720005 &     0.0478754379 &    46.2148464304 &     54.543 &  48\\
I.A$_{21}^{i.c.}$(0.5) &     0.2997637007 &     0.0934329270 &    62.7105115603 &     63.376 &  56\\
I.A$_{22}^{i.c.}$(0.5) &     0.2747511246 &     0.0544869553 &    54.5744279508 &     63.626 &  56\\
I.A$_{23}^{i.c.}$(0.5) &     0.2867479329 &     0.0521752523 &    57.0633556930 &     63.626 &  56\\
I.A$_{24}^{i.c.}$(0.5) &     0.2172290935 &     0.0383448898 &    45.1666009592 &     63.631 &  56\\
I.A$_{25}^{i.c.}$(0.5) &     0.3108794721 &     0.1023369865 &    71.4799545005 &     67.698 &  60\\
I.A$_{26}^{i.c.}$(0.5) &     0.2979925625 &     0.0918951185 &    70.9842467059 &     72.456 &  64\\
I.A$_{27}^{i.c.}$(0.5) &     0.2366779591 &     0.0914177522 &    56.6833946453 &     72.485 &  64\\
I.A$_{28}^{i.c.}$(0.5) &     0.1628551705 &     0.0589464762 &    46.2097799724 &     72.484 &  64\\
I.A$_{29}^{i.c.}$(0.5) &     0.2763361520 &     0.0588302447 &    62.9447137981 &     72.706 &  64\\
I.A$_{30}^{i.c.}$(0.5) &     0.1936757357 &     0.0730232621 &    49.7181917085 &     72.488 &  64\\
I.A$_{31}^{i.c.}$(0.5) &     0.3017504100 &     0.1030778699 &    77.7653686390 &     76.789 &  68\\
I.A$_{32}^{i.c.}$(0.5) &     0.1671144104 &     0.0438815944 &    49.1675240282 &     77.185 &  68\\
I.A$_{33}^{i.c.}$(0.5) &     0.3274705985 &     0.0612651208 &    84.0143824473 &     77.185 &  68\\
I.A$_{34}^{i.c.}$(0.5) &     0.2668455153 &     0.0138391891 &    63.2419174415 &     77.282 &  68\\
I.A$_{35}^{i.c.}$(0.5) &     0.3220251063 &     0.0754954232 &    87.6821712800 &     81.574 &  72\\
I.A$_{36}^{i.c.}$(0.5) &     0.2965579937 &     0.0906370328 &    79.2439520137 &     81.536 &  72\\
I.A$_{37}^{i.c.}$(0.5) &     0.2775882955 &     0.0619333069 &    71.3240509215 &     81.786 &  72\\
I.A$_{38}^{i.c.}$(0.5) &     0.3558062278 &     0.0405108521 &   108.4971611336 &     86.349 &  76\\
I.A$_{39}^{i.c.}$(0.5) &     0.3060017590 &     0.0986219478 &    88.0890690057 &     85.876 &  76\\
I.A$_{40}^{i.c.}$(0.5) &     0.2689229383 &     0.0312527426 &    71.5697332821 &     86.372 &  76\\
I.A$_{41}^{i.c.}$(0.5) &     0.1317126561 &     0.0254909293 &    51.5736578935 &     86.350 &  76\\
I.A$_{42}^{i.c.}$(0.5) &     0.1428972736 &     0.0445901978 &    55.4783856796 &     90.717 &  80\\
I.A$_{43}^{i.c.}$(0.5) &     0.3132151994 &     0.1046181562 &    96.6632360131 &     90.178 &  80\\
I.A$_{44}^{i.c.}$(0.5) &     0.2954964679 &     0.0895434067 &    87.5354654697 &     90.615 &  80\\
I.A$_{45}^{i.c.}$(0.5) &     0.2749526022 &     0.0648656500 &    78.6571356819 &     90.866 &  80\\

\hline
\end{tabular*}
{\rule{\temptablewidth}{1pt}}
\end{center}
\end{table}

\begin{table*}
\tabcolsep 0pt \caption{Initial conditions and periods $T$ of the periodic three-body orbits for class I.A in the case of  $\bm{r}_1(0)=(-1,0)=-\bm{r}_2(0)$,  $\dot{\bm{r}}_1(0)=(v_1,v_2)=\dot{\bm{r}}_2(0)$ and $\bm{r}_3(0)=(0,0)$, $\dot{\bm{r}}_3(0)=(-2v_1/m_3, -2v_2/m_3)$ when $G=1$ and $m_1=m_2=1$ and $m_3=0.5$ by means of the search grid $4000\times 4000$ in the interval $T_0\in[0,200]$, where $T^*=T |E|^{3/2}$ is its scale-invariant period, $L_f$ is the length of the free group element.  } \label{table-S5} \vspace*{-12pt}
\begin{center}
\def\temptablewidth{1\textwidth}
{\rule{\temptablewidth}{1pt}}
\begin{tabular*}{\temptablewidth}{@{\extracolsep{\fill}}lccccc}
\hline
Class and number  & $v_1$ & $v_2$  & $T$  & $T^*$ & $L_f$\\
\hline
I.A$_{46}^{i.c.}$(0.5) &     0.2153858894 &     0.0499108529 &    64.5903822323 &     90.874 &  80\\
I.A$_{47}^{i.c.}$(0.5) &     0.2770512888 &     0.1061164594 &    87.0243497655 &     94.961 &  84\\
I.A$_{48}^{i.c.}$(0.5) &     0.3213655238 &     0.0897415420 &   103.6411578855 &     94.961 &  84\\
I.A$_{49}^{i.c.}$(0.5) &     0.2706252379 &     0.0398693716 &    79.9076627886 &     95.459 &  84\\
I.A$_{50}^{i.c.}$(0.5) &     0.2786174957 &     0.2292777524 &   131.7357122012 &    103.058 &  88\\
I.A$_{51}^{i.c.}$(0.5) &     0.2469591661 &     0.0904031712 &    80.4191164038 &     99.719 &  88\\
I.A$_{52}^{i.c.}$(0.5) &     0.3228250487 &     0.0968450484 &   110.3291503458 &     99.273 &  88\\
I.A$_{53}^{i.c.}$(0.5) &     0.2954791318 &     0.1096218279 &    98.7725671166 &     99.273 &  88\\
I.A$_{54}^{i.c.}$(0.5) &     0.2945920946 &     0.0886163420 &    95.8143486677 &     99.694 &  88\\
I.A$_{55}^{i.c.}$(0.5) &     0.1945402732 &     0.0706830008 &    68.4003313978 &     99.727 &  88\\
I.A$_{56}^{i.c.}$(0.5) &     0.1889789414 &     0.0744407920 &    70.7522226916 &    104.113 &  92\\
I.A$_{57}^{i.c.}$(0.5) &     0.3011607127 &     0.0965243843 &   104.0576058514 &    104.046 &  92\\
I.A$_{58}^{i.c.}$(0.5) &     0.2938398485 &     0.0878089007 &   104.0941488385 &    108.772 &  96\\
I.A$_{59}^{i.c.}$(0.5) &     0.3323949017 &     0.0892628147 &   125.2963737946 &    108.361 &  96\\
I.A$_{60}^{i.c.}$(0.5) &     0.1303739201 &     0.0110880478 &    64.8930053804 &    109.158 &  96\\
I.A$_{61}^{i.c.}$(0.5) &     0.3388203286 &     0.0534486578 &   130.4329576227 &    113.548 &  100\\
I.A$_{62}^{i.c.}$(0.5) &     0.1528319560 &     0.0357128745 &    70.2840899540 &    113.548 &  100\\
I.A$_{63}^{i.c.}$(0.5) &     0.2931864677 &     0.0871031920 &   112.3665360659 &    117.850 &  104\\
I.A$_{64}^{i.c.}$(0.5) &     0.2712558612 &     0.0702892333 &   101.3437101345 &    118.101 &  104\\
I.A$_{65}^{i.c.}$(0.5) &     0.3227971415 &     0.0926025930 &   129.6770893175 &    117.455 &  104\\
I.A$_{66}^{i.c.}$(0.5) &     0.2862860474 &     0.1079517198 &   112.0449969617 &    117.455 &  104\\
I.A$_{67}^{i.c.}$(0.5) &     0.3078743981 &     0.0999722100 &   121.7962271737 &    117.455 &  104\\
I.A$_{68}^{i.c.}$(0.5) &     0.2138410831 &     0.0542938396 &    83.8472907647 &    118.112 &  104\\
I.A$_{69}^{i.c.}$(0.5) &     0.2672229781 &     0.0183673134 &    96.9436256746 &    118.196 &  104\\
I.A$_{70}^{i.c.}$(0.5) &     0.2762141023 &     0.1007012568 &   110.7531694029 &    122.210 &  108\\
I.A$_{71}^{i.c.}$(0.5) &     0.3151460645 &     0.0881775189 &   129.0111010994 &    122.210 &  108\\
I.A$_{72}^{i.c.}$(0.5) &     0.2816347029 &     0.1165460881 &   115.5464074541 &    121.753 &  108\\
I.A$_{73}^{i.c.}$(0.5) &     0.1897313629 &     0.0569798644 &    82.3194421170 &    122.548 &  108\\
I.A$_{74}^{i.c.}$(0.5) &     0.2365156031 &     0.0783479856 &    94.4496491279 &    122.548 &  108\\
I.A$_{75}^{i.c.}$(0.5) &     0.1892308321 &     0.0662636060 &    85.7557681263 &    126.964 &  112\\
I.A$_{76}^{i.c.}$(0.5) &     0.3167473355 &     0.0564409230 &   130.6376396627 &    127.209 &  112\\
I.A$_{77}^{i.c.}$(0.5) &     0.3063566174 &     0.0988758502 &   130.0686378697 &    126.543 &  112\\
I.A$_{78}^{i.c.}$(0.5) &     0.2686058325 &     0.0293186861 &   105.2715049896 &    127.286 &  112\\
I.A$_{79}^{i.c.}$(0.5) &     0.3115828668 &     0.1029439863 &   138.7657196429 &    130.849 &  116\\
I.A$_{80}^{i.c.}$(0.5) &     0.1926049152 &     0.0732402826 &    89.8940701833 &    131.356 &  116\\
I.A$_{81}^{i.c.}$(0.5) &     0.2678790398 &     0.0910268984 &   117.9225138007 &    136.004 &  120\\
I.A$_{82}^{i.c.}$(0.5) &     0.2769745628 &     0.1073466261 &   124.4856938122 &    135.627 &  120\\
I.A$_{83}^{i.c.}$(0.5) &     0.2817808026 &     0.0707431739 &   121.7428951384 &    136.256 &  120\\
I.A$_{84}^{i.c.}$(0.5) &     0.2887331369 &     0.1103223170 &   135.3850884074 &    139.941 &  124\\
I.A$_{85}^{i.c.}$(0.5) &     0.2820080025 &     0.0568015273 &   124.4362856628 &    140.873 &  124\\
I.A$_{86}^{i.c.}$(0.5) &     0.1635241804 &     0.0417506200 &    88.9883269271 &    140.761 &  124\\
I.A$_{87}^{i.c.}$(0.5) &     0.2167821354 &     0.0426860123 &   100.0973395584 &    140.885 &  124\\
I.A$_{88}^{i.c.}$(0.5) &     0.3111417615 &     0.0939330679 &   151.0501281052 &    144.714 &  128\\
I.A$_{89}^{i.c.}$(0.5) &     0.2821974598 &     0.0715109410 &   130.1650998679 &    145.334 &  128\\
I.A$_{90}^{i.c.}$(0.5) &     0.2277884543 &     0.0646616649 &   107.9084014726 &    145.349 &  128\\
I.A$_{91}^{i.c.}$(0.5) &     0.2708795826 &     0.0409737605 &   121.9472637204 &    145.460 &  128\\
I.A$_{92}^{i.c.}$(0.5) &     0.2744503104 &     0.0976059036 &   133.9471506576 &    149.452 &  132\\
I.A$_{93}^{i.c.}$(0.5) &     0.3114904704 &     0.0864935535 &   154.6502149192 &    149.452 &  132\\
I.A$_{94}^{i.c.}$(0.5) &     0.1794922768 &     0.0605219047 &   101.6064275160 &    154.197 &  136\\
I.A$_{95}^{i.c.}$(0.5) &     0.3128476865 &     0.1042021457 &   163.9473666361 &    153.329 &  136\\
I.A$_{96}^{i.c.}$(0.5) &     0.2718250586 &     0.0447887187 &   130.2931785836 &    154.545 &  136\\
I.A$_{97}^{i.c.}$(0.5) &     0.2773004224 &     0.0612425070 &   138.4568988662 &    159.033 &  140\\
I.A$_{98}^{i.c.}$(0.5) &     0.1882207909 &     0.0737375490 &   110.5070750985 &    162.981 &  144\\
I.A$_{99}^{i.c.}$(0.5) &     0.2829061443 &     0.0727791246 &   147.0223286431 &    163.488 &  144\\
I.A$_{100}^{i.c.}$(0.5) &     0.2176616157 &     0.0285927952 &   115.8358005700 &    163.643 &  144\\
\hline
\end{tabular*}
{\rule{\temptablewidth}{1pt}}
\end{center}
\end{table*}

\begin{table*}
\tabcolsep 0pt \caption{Initial conditions and periods $T$ of the periodic three-body orbits for class I.A in the case of  $\bm{r}_1(0)=(-1,0)=-\bm{r}_2(0)$,  $\dot{\bm{r}}_1(0)=(v_1,v_2)=\dot{\bm{r}}_2(0)$ and $\bm{r}_3(0)=(0,0)$, $\dot{\bm{r}}_3(0)=(-2v_1/m_3, -2v_2/m_3)$ when $G=1$ and $m_1=m_2=1$ and $m_3=0.5$ by means of the search grid $4000\times 4000$ in the interval $T_0\in[0,200]$, where $T^*=T |E|^{3/2}$ is its scale-invariant period, $L_f$ is the length of the free group element.  } \label{table-S5} \vspace*{-12pt}
\begin{center}
\def\temptablewidth{1\textwidth}
{\rule{\temptablewidth}{1pt}}
\begin{tabular*}{\temptablewidth}{@{\extracolsep{\fill}}lccccc}
\hline
Class and number  & $v_1$ & $v_2$  & $T$  & $T^*$ & $L_f$\\
\hline
I.A$_{101}^{i.c.}$(0.5) &     0.3242506160 &     0.0631949983 &   180.1794835948 &    167.981 &  148\\
I.A$_{102}^{i.c.}$(0.5) &     0.2684432435 &     0.0282692084 &   138.9734917593 &    168.200 &  148\\
I.A$_{103}^{i.c.}$(0.5) &     0.2778616536 &     0.0625774643 &   146.8402992438 &    168.113 &  148\\
I.A$_{104}^{i.c.}$(0.5) &     0.3051413550 &     0.0931938899 &   174.3071401892 &    171.964 &  152\\
I.A$_{105}^{i.c.}$(0.5) &     0.1629885654 &     0.0522640671 &   109.4669718191 &    172.377 &  152\\
I.A$_{106}^{i.c.}$(0.5) &     0.1462535243 &     0.0217299404 &   105.3228173310 &    172.730 &  152\\
I.A$_{107}^{i.c.}$(0.5) &     0.2906493900 &     0.0841512530 &   161.9536048399 &    172.315 &  152\\
I.A$_{108}^{i.c.}$(0.5) &     0.3307845107 &     0.0282510145 &   186.8989238269 &    172.762 &  152\\
I.A$_{109}^{i.c.}$(0.5) &     0.2734339548 &     0.0504407955 &   147.0001395753 &    172.712 &  152\\
I.A$_{110}^{i.c.}$(0.5) &     0.1864663107 &     0.0314460085 &   116.6973116737 &    177.222 &  156\\
I.A$_{111}^{i.c.}$(0.5) &     0.3054857593 &     0.0982549763 &   180.3047709490 &    176.296 &  156\\
I.A$_{112}^{i.c.}$(0.5) &     0.1556984762 &     0.0375406067 &   110.2475386504 &    177.124 &  156\\
I.A$_{113}^{i.c.}$(0.5) &     0.2693793982 &     0.0338185849 &   147.3066947737 &    177.288 &  156\\
I.A$_{114}^{i.c.}$(0.5) &     0.1759573432 &     0.0582869204 &   118.5242718315 &    181.429 &  160\\
I.A$_{115}^{i.c.}$(0.5) &     0.3063956054 &     0.0665241524 &   179.3796751444 &    181.667 &  160\\
I.A$_{116}^{i.c.}$(0.5) &     0.1844792914 &     0.0664186417 &   124.2585706817 &    185.835 &  164\\
I.A$_{117}^{i.c.}$(0.5) &     0.2788305752 &     0.0647737865 &   163.6184241261 &    186.271 &  164\\
I.A$_{118}^{i.c.}$(0.5) &     0.3181202953 &     0.0262108895 &   189.0552591547 &    186.385 &  164\\
I.A$_{119}^{i.c.}$(0.5) &     0.2760366688 &     0.1001106086 &   172.0404786725 &    190.126 &  168\\
I.A$_{120}^{i.c.}$(0.5) &     0.2901045319 &     0.0834748902 &   178.4639514916 &    190.469 &  168\\
I.A$_{121}^{i.c.}$(0.5) &     0.2710039691 &     0.0415005602 &   163.9870417061 &    195.461 &  172\\
I.A$_{122}^{i.c.}$(0.5) &     0.2792522152 &     0.0656898524 &   172.0127626343 &    195.350 &  172\\
I.A$_{123}^{i.c.}$(0.5) &     0.2797022972 &     0.0856462692 &   179.7442254623 &    199.546 &  176\\
I.A$_{124}^{i.c.}$(0.5) &     0.1897491024 &     0.0464972761 &   133.3930483814 &    199.814 &  176\\
I.A$_{125}^{i.c.}$(0.5) &     0.2955617289 &     0.0815450298 &   190.9002964594 &    199.546 &  176\\
I.A$_{126}^{i.c.}$(0.5) &     0.2991146914 &     0.0928880688 &   196.3997389850 &    199.208 &  176\\
I.A$_{127}^{i.c.}$(0.5) &     0.2753246510 &     0.0561181124 &   172.0902983279 &    199.958 &  176\\
I.A$_{128}^{i.c.}$(0.5) &     0.1608928391 &     0.0423566492 &   128.6102391540 &    204.338 &  180\\
I.A$_{129}^{i.c.}$(0.5) &     0.2858473085 &     0.1123019709 &   194.7380952367 &    203.092 &  180\\
I.A$_{130}^{i.c.}$(0.5) &     0.1632254337 &     0.0372831480 &   128.8601185998 &    204.338 &  180\\
I.A$_{131}^{i.c.}$(0.5) &     0.2203920025 &     0.0720677540 &   149.5147366218 &    204.249 &  180\\
I.A$_{132}^{i.c.}$(0.5) &     0.1747313962 &     0.0572750530 &   135.8934495248 &    208.659 &  184\\
I.A$_{133}^{i.c.}$(0.5) &     0.2841883501 &     0.0749607778 &   189.2225836059 &    208.874 &  184\\
I.A$_{134}^{i.c.}$(0.5) &     0.2799959773 &     0.0672515431 &   188.8112079689 &    213.506 &  188\\
I.A$_{135}^{i.c.}$(0.5) &     0.1477960023 &     0.0331559448 &   130.9407612438 &    213.493 &  188\\
I.A$_{136}^{i.c.}$(0.5) &     0.2843861676 &     0.0752851658 &   197.6692863336 &    217.951 &  192\\
I.A$_{137}^{i.c.}$(0.5) &     0.2166248717 &     0.0438265532 &   155.0134625948 &    218.138 &  192\\
I.A$_{138}^{i.c.}$(0.5) &     0.2141903816 &     0.0534864829 &   158.0856536626 &    222.606 &  196\\
I.A$_{139}^{i.c.}$(0.5) &     0.2550856649 &     0.0780342905 &   185.5037127808 &    227.045 &  200\\
I.A$_{140}^{i.c.}$(0.5) &     0.2027586331 &     0.0547976409 &   156.8933680291 &    227.054 &  200\\
I.A$_{141}^{i.c.}$(0.5) &     0.2735238924 &     0.0507318572 &   197.3931765300 &    231.797 &  204\\
I.A$_{142}^{i.c.}$(0.5) &     0.2174237597 &     0.0354281968 &   170.8625404977 &    240.899 &  212\\
I.A$_{143}^{i.c.}$(0.5) &     0.1702863136 &     0.0542463019 &   163.7282370536 &    254.054 &  224\\
I.A$_{144}^{i.c.}$(0.5) &     0.2114296338 &     0.0633519827 &   183.7049132401 &    258.716 &  228\\
I.A$_{145}^{i.c.}$(0.5) &     0.2040531625 &     0.0308354022 &   183.6432714881 &    268.146 &  236\\
I.A$_{146}^{i.c.}$(0.5) &     0.2128981089 &     0.0560440442 &   196.4519555242 &    277.080 &  244\\
I.A$_{147}^{i.c.}$(0.5) &     0.1914804176 &     0.0167542245 &   198.8845878317 &    300.002 &  264\\
\hline
\end{tabular*}
{\rule{\temptablewidth}{1pt}}
\end{center}
\end{table*}

\begin{table*}
\tabcolsep 0pt \caption{Initial conditions and periods $T$ of the periodic three-body orbits for class I.B in the case of  $\bm{r}_1(0)=(-1,0)=-\bm{r}_2(0)$,  $\dot{\bm{r}}_1(0)=(v_1,v_2)=\dot{\bm{r}}_2(0)$ and $\bm{r}_3(0)=(0,0)$, $\dot{\bm{r}}_3(0)=(-2v_1/m_3, -2v_2/m_3)$ when $G=1$ and $m_1=m_2=1$ and $m_3=0.5$ by means of the search grid $4000\times 4000$ in the interval $T_0\in[0,200]$, where $T^*=T |E|^{3/2}$ is its scale-invariant period, $L_f$ is the length of the free group element.  } \label{table-S5} \vspace*{-12pt}
\begin{center}
\def\temptablewidth{1\textwidth}
{\rule{\temptablewidth}{1pt}}
\begin{tabular*}{\temptablewidth}{@{\extracolsep{\fill}}lccccc}
\hline
Class and number  & $v_1$ & $v_2$  & $T$  & $T^*$ & $L_f$\\
\hline
I.B$_{1}^{i.c.}$(0.5) &     0.2374365149 &     0.2536896353 &     8.5581422789 &      7.262 &  6\\
I.B$_{2}^{i.c.}$(0.5) &     0.2707702758 &     0.2974619413 &    19.9858290667 &     11.480 &  10\\
I.B$_{3}^{i.c.}$(0.5) &     0.1804341862 &     0.0774390466 &    10.5764781985 &     15.808 &  14\\
I.B$_{4}^{i.c.}$(0.5) &     0.0548520001 &     0.3291535443 &    20.9927052014 &     19.231 &  14\\
I.B$_{5}^{i.c.}$(0.5) &     0.2817159946 &     0.3093138094 &    31.1291374576 &     15.374 &  14\\
I.B$_{6}^{i.c.}$(0.5) &     0.2679384847 &     0.0246961144 &    16.8511048757 &     20.457 &  18\\
I.B$_{7}^{i.c.}$(0.5) &     0.2674226718 &     0.2139289499 &    23.9372167355 &     20.903 &  18\\
I.B$_{8}^{i.c.}$(0.5) &     0.2878430093 &     0.3151477978 &    42.2778567687 &     19.118 &  18\\
I.B$_{9}^{i.c.}$(0.5) &     0.3030963188 &     0.0966599779 &    25.1036320618 &     24.876 &  22\\
I.B$_{10}^{i.c.}$(0.5) &     0.1099852485 &     0.0308448543 &    14.5241562996 &     24.961 &  22\\
I.B$_{11}^{i.c.}$(0.5) &     0.2291294485 &     0.2119828182 &    24.7437423714 &     25.221 &  22\\
I.B$_{12}^{i.c.}$(0.5) &     0.3692649167 &     0.0417694147 &    34.2771859316 &     24.965 &  22\\
I.B$_{13}^{i.c.}$(0.5) &     0.2737871583 &     0.0515706441 &    25.1965811355 &     29.542 &  26\\
I.B$_{14}^{i.c.}$(0.5) &     0.2988140236 &     0.0926281921 &    33.4227154398 &     33.958 &  30\\
I.B$_{15}^{i.c.}$(0.5) &     0.2788426894 &     0.0601828236 &    33.7944313342 &     38.623 &  34\\
I.B$_{16}^{i.c.}$(0.5) &     0.2790832581 &     0.2276230751 &    50.6192127242 &     39.773 &  34\\
I.B$_{17}^{i.c.}$(0.5) &     0.2163072511 &     0.0457824324 &    27.4550707427 &     38.627 &  34\\
I.B$_{18}^{i.c.}$(0.5) &     0.2631918196 &     0.0971022505 &    36.9742814503 &     43.036 &  38\\
I.B$_{19}^{i.c.}$(0.5) &     0.1940565085 &     0.0716923953 &    29.5220753291 &     43.054 &  38\\
I.B$_{20}^{i.c.}$(0.5) &     0.2790460219 &     0.0652447476 &    41.9537927832 &     47.703 &  42\\
I.B$_{21}^{i.c.}$(0.5) &     0.1166234512 &     0.0062473070 &    27.8908904718 &     47.784 &  42\\
I.B$_{22}^{i.c.}$(0.5) &     0.2941922385 &     0.0882015308 &    49.9756827423 &     52.116 &  46\\
I.B$_{23}^{i.c.}$(0.5) &     0.1482620649 &     0.0469161409 &    32.2082149709 &     52.162 &  46\\
I.B$_{24}^{i.c.}$(0.5) &     0.3383350929 &     0.0633820096 &    60.3349978637 &     52.165 &  46\\
I.B$_{25}^{i.c.}$(0.5) &     0.3086986007 &     0.1005760508 &    58.8260422524 &     56.455 &  50\\
I.B$_{26}^{i.c.}$(0.5) &     0.2804820502 &     0.0682367913 &    50.3545241044 &     56.781 &  50\\
I.B$_{27}^{i.c.}$(0.5) &     0.2928957198 &     0.0867806393 &    58.2512450352 &     61.194 &  54\\
I.B$_{28}^{i.c.}$(0.5) &     0.3121801357 &     0.0954940881 &    68.9135784414 &     65.543 &  58\\
I.B$_{29}^{i.c.}$(0.5) &     0.2815530004 &     0.0703169410 &    58.7667019161 &     65.859 &  58\\
I.B$_{30}^{i.c.}$(0.5) &     0.2702462061 &     0.0720377706 &    56.4003264775 &     65.859 &  58\\
I.B$_{31}^{i.c.}$(0.5) &     0.2131987197 &     0.0555522173 &    46.7195262030 &     65.865 &  58\\
I.B$_{32}^{i.c.}$(0.5) &     0.1832565230 &     0.0627595131 &    46.7521066510 &     70.290 &  62\\
I.B$_{33}^{i.c.}$(0.5) &     0.2919082649 &     0.0856563914 &    66.5195343482 &     70.272 &  62\\
I.B$_{34}^{i.c.}$(0.5) &     0.2703631586 &     0.0386904791 &    58.8880393273 &     70.458 &  62\\
I.B$_{35}^{i.c.}$(0.5) &     0.1859380756 &     0.0731374234 &    50.3601763743 &     74.679 &  66\\
I.B$_{36}^{i.c.}$(0.5) &     0.2489094150 &     0.0992895950 &    61.2653969718 &     74.663 &  66\\
I.B$_{37}^{i.c.}$(0.5) &     0.2823888686 &     0.0718580778 &    67.1889348357 &     74.936 &  66\\
I.B$_{38}^{i.c.}$(0.5) &     0.1615485916 &     0.0277748182 &    46.9823741559 &     74.980 &  66\\
I.B$_{39}^{i.c.}$(0.5) &     0.2911309221 &     0.0847384936 &    74.7822861358 &     79.349 &  70\\
I.B$_{40}^{i.c.}$(0.5) &     0.1771098445 &     0.0784887228 &    52.5734020100 &     79.040 &  70\\
I.B$_{41}^{i.c.}$(0.5) &     0.1838565013 &     0.0752899792 &    53.1629113586 &     79.040 &  70\\
I.B$_{42}^{i.c.}$(0.5) &     0.2143486812 &     0.0918266864 &    58.0738450139 &     79.038 &  70\\
I.B$_{43}^{i.c.}$(0.5) &     0.3247862862 &     0.0695897171 &    85.9588123017 &     79.387 &  70\\
I.B$_{44}^{i.c.}$(0.5) &     0.1610459823 &     0.0516375322 &    50.2179896145 &     79.382 &  70\\
I.B$_{45}^{i.c.}$(0.5) &     0.2722601340 &     0.0464077813 &    67.2341769532 &     79.543 &  70\\
I.B$_{46}^{i.c.}$(0.5) &     0.3192498609 &     0.0879575434 &    90.1594809343 &     83.709 &  74\\
I.B$_{47}^{i.c.}$(0.5) &     0.2830620767 &     0.0730518241 &    75.6195474605 &     84.014 &  74\\
I.B$_{48}^{i.c.}$(0.5) &     0.2905035826 &     0.0839717300 &    83.0409561905 &     88.427 &  78\\
I.B$_{49}^{i.c.}$(0.5) &     0.3002585565 &     0.1059795568 &    88.9553085721 &     88.032 &  78\\
I.B$_{50}^{i.c.}$(0.5) &     0.2174816082 &     0.0342966208 &    62.8453538957 &     88.634 &  78\\
\hline
\end{tabular*}
{\rule{\temptablewidth}{1pt}}
\end{center}
\end{table*}

\begin{table*}
\tabcolsep 0pt \caption{Initial conditions and periods $T$ of the periodic three-body orbits for class I.B in the case of  $\bm{r}_1(0)=(-1,0)=-\bm{r}_2(0)$,  $\dot{\bm{r}}_1(0)=(v_1,v_2)=\dot{\bm{r}}_2(0)$ and $\bm{r}_3(0)=(0,0)$, $\dot{\bm{r}}_3(0)=(-2v_1/m_3, -2v_2/m_3)$ when $G=1$ and $m_1=m_2=1$ and $m_3=0.5$ by means of the search grid $4000\times 4000$ in the interval $T_0\in[0,200]$, where $T^*=T |E|^{3/2}$ is its scale-invariant period, $L_f$ is the length of the free group element.  } \label{table-S5} \vspace*{-12pt}
\begin{center}
\def\temptablewidth{1\textwidth}
{\rule{\temptablewidth}{1pt}}
\begin{tabular*}{\temptablewidth}{@{\extracolsep{\fill}}lccccc}
\hline
Class and number  & $v_1$ & $v_2$  & $T$  & $T^*$ & $L_f$\\
\hline
I.B$_{51}^{i.c.}$(0.5) &     0.2913964797 &     0.0479668990 &    80.7592905815 &     88.627 &  78\\
I.B$_{52}^{i.c.}$(0.5) &     0.2960647431 &     0.0951352334 &    90.5465691204 &     92.793 &  82\\
I.B$_{53}^{i.c.}$(0.5) &     0.2836170353 &     0.0740064147 &    84.0569632087 &     93.091 &  82\\
I.B$_{54}^{i.c.}$(0.5) &     0.2679928311 &     0.0763276805 &    79.4113786013 &     93.090 &  82\\
I.B$_{55}^{i.c.}$(0.5) &     0.2899831050 &     0.0833218568 &    91.2952846897 &     97.504 &  86\\
I.B$_{56}^{i.c.}$(0.5) &     0.3080355719 &     0.1000897653 &   100.8065558913 &     97.122 &  86\\
I.B$_{57}^{i.c.}$(0.5) &     0.1751508803 &     0.0576966004 &    63.5831421300 &     97.522 &  86\\
I.B$_{58}^{i.c.}$(0.5) &     0.2341908442 &     0.0837260648 &    75.0584041949 &     97.525 &  86\\
I.B$_{59}^{i.c.}$(0.5) &     0.2750441739 &     0.0553294312 &    83.9531808683 &     97.709 &  86\\
I.B$_{60}^{i.c.}$(0.5) &     0.3442855217 &     0.0202360484 &   113.5049294912 &     97.768 &  86\\
I.B$_{61}^{i.c.}$(0.5) &     0.3138785714 &     0.1054199562 &   109.2112794427 &    101.417 &  90\\
I.B$_{62}^{i.c.}$(0.5) &     0.2840833689 &     0.0747873466 &    92.4999175257 &    102.168 &  90\\
I.B$_{63}^{i.c.}$(0.5) &     0.1918308588 &     0.0726163450 &    69.5933264452 &    101.922 &  90\\
I.B$_{64}^{i.c.}$(0.5) &     0.2967400478 &     0.0807114885 &   102.3821088719 &    106.581 &  94\\
I.B$_{65}^{i.c.}$(0.5) &     0.1671740666 &     0.0535233977 &    68.2728344880 &    106.607 &  94\\
I.B$_{66}^{i.c.}$(0.5) &     0.3062135700 &     0.0987733794 &   109.0789742536 &    106.210 &  94\\
I.B$_{67}^{i.c.}$(0.5) &     0.2852343987 &     0.1063718482 &   100.6294915837 &    106.209 &  94\\
I.B$_{68}^{i.c.}$(0.5) &     0.3169586307 &     0.0723491117 &   111.3358326383 &    106.613 &  94\\
I.B$_{69}^{i.c.}$(0.5) &     0.2760984711 &     0.0582105683 &    92.3235074759 &    106.790 &  94\\
I.B$_{70}^{i.c.}$(0.5) &     0.2977432358 &     0.0916649424 &   108.5470086473 &    110.955 &  98\\
I.B$_{71}^{i.c.}$(0.5) &     0.2844794178 &     0.0754369919 &   100.9465701326 &    111.245 &  98\\
I.B$_{72}^{i.c.}$(0.5) &     0.3249845265 &     0.0202864165 &   116.6138500488 &    111.383 &  98\\
I.B$_{73}^{i.c.}$(0.5) &     0.2694680508 &     0.0342913622 &    92.5876783108 &    111.373 &  98\\
I.B$_{74}^{i.c.}$(0.5) &     0.2891747738 &     0.0822826126 &   107.7974854583 &    115.658 &  102\\
I.B$_{75}^{i.c.}$(0.5) &     0.2885302862 &     0.0933767620 &   113.1508640370 &    120.034 &  106\\
I.B$_{76}^{i.c.}$(0.5) &     0.2666426613 &     0.0786440441 &   102.3898928785 &    120.321 &  106\\
I.B$_{77}^{i.c.}$(0.5) &     0.2848187359 &     0.0759836961 &   109.3953593575 &    120.322 &  106\\
I.B$_{78}^{i.c.}$(0.5) &     0.1977132984 &     0.0525053406 &    82.0586018240 &    120.336 &  106\\
I.B$_{79}^{i.c.}$(0.5) &     0.2707787504 &     0.0405404036 &   100.9274275269 &    120.459 &  106\\
I.B$_{80}^{i.c.}$(0.5) &     0.2449409545 &     0.0850729560 &    99.3632704793 &    124.756 &  110\\
I.B$_{81}^{i.c.}$(0.5) &     0.1773510110 &     0.0248360734 &    80.6083002912 &    124.987 &  110\\
I.B$_{82}^{i.c.}$(0.5) &     0.2888572586 &     0.0818643534 &   116.0480068229 &    124.734 &  110\\
I.B$_{83}^{i.c.}$(0.5) &     0.2181512383 &     0.0833244460 &    94.9920652210 &    129.161 &  114\\
I.B$_{84}^{i.c.}$(0.5) &     0.2904092593 &     0.1083815356 &   125.0092177942 &    128.699 &  114\\
I.B$_{85}^{i.c.}$(0.5) &     0.2851109699 &     0.0764474074 &   117.8446993799 &    129.398 &  114\\
I.B$_{86}^{i.c.}$(0.5) &     0.2703553256 &     0.0851792459 &   116.2605246289 &    133.811 &  118\\
I.B$_{87}^{i.c.}$(0.5) &     0.2985222840 &     0.0787345637 &   129.2429556733 &    133.811 &  118\\
I.B$_{88}^{i.c.}$(0.5) &     0.3138600006 &     0.0908763661 &   140.5350980635 &    133.464 &  118\\
I.B$_{89}^{i.c.}$(0.5) &     0.3135219744 &     0.1049833885 &   142.8584537333 &    132.992 &  118\\
I.B$_{90}^{i.c.}$(0.5) &     0.3098615763 &     0.0844309289 &   141.5123187727 &    138.190 &  122\\
I.B$_{91}^{i.c.}$(0.5) &     0.2853633746 &     0.0768427234 &   126.2931104761 &    138.475 &  122\\
I.B$_{92}^{i.c.}$(0.5) &     0.2176335170 &     0.0299228620 &    98.1757759947 &    138.640 &  122\\
I.B$_{93}^{i.c.}$(0.5) &     0.2883498729 &     0.0811835015 &   132.5546327228 &    142.888 &  126\\
I.B$_{94}^{i.c.}$(0.5) &     0.1749022486 &     0.0766378437 &    94.0590542640 &    142.271 &  126\\
I.B$_{95}^{i.c.}$(0.5) &     0.1636510704 &     0.0474589703 &    90.6558775234 &    142.963 &  126\\
I.B$_{96}^{i.c.}$(0.5) &     0.2150995281 &     0.0509041474 &   101.7118032345 &    143.121 &  126\\
I.B$_{97}^{i.c.}$(0.5) &     0.2822851529 &     0.0922191464 &   135.1262821798 &    147.271 &  130\\
I.B$_{98}^{i.c.}$(0.5) &     0.2855816329 &     0.0771807489 &   134.7392792763 &    147.552 &  130\\
I.B$_{99}^{i.c.}$(0.5) &     0.2947320378 &     0.0887615187 &   141.6521915390 &    147.271 &  130\\
I.B$_{100}^{i.c.}$(0.5) &     0.3344339309 &     0.0263022352 &   162.8324323639 &    147.763 &  130\\
\hline
\end{tabular*}
{\rule{\temptablewidth}{1pt}}
\end{center}
\end{table*}

\begin{table*}
\tabcolsep 0pt \caption{Initial conditions and periods $T$ of the periodic three-body orbits for class I.B in the case of  $\bm{r}_1(0)=(-1,0)=-\bm{r}_2(0)$,  $\dot{\bm{r}}_1(0)=(v_1,v_2)=\dot{\bm{r}}_2(0)$ and $\bm{r}_3(0)=(0,0)$, $\dot{\bm{r}}_3(0)=(-2v_1/m_3, -2v_2/m_3)$ when $G=1$ and $m_1=m_2=1$ and $m_3=0.5$ by means of the search grid $4000\times 4000$ in the interval $T_0\in[0,200]$, where $T^*=T |E|^{3/2}$ is its scale-invariant period, $L_f$ is the length of the free group element.  } \label{table-S5} \vspace*{-12pt}
\begin{center}
\def\temptablewidth{1\textwidth}
{\rule{\temptablewidth}{1pt}}
\begin{tabular*}{\temptablewidth}{@{\extracolsep{\fill}}lccccc}
\hline
Class and number  & $v_1$ & $v_2$  & $T$  & $T^*$ & $L_f$\\
\hline
I.B$_{101}^{i.c.}$(0.5) &     0.3258205442 &     0.0948987232 &   170.1648469173 &    151.183 &  134\\
I.B$_{102}^{i.c.}$(0.5) &     0.2887071810 &     0.1116133434 &   146.5524883273 &    151.183 &  134\\
I.B$_{103}^{i.c.}$(0.5) &     0.3136039168 &     0.0546949113 &   153.7578002255 &    152.216 &  134\\
I.B$_{104}^{i.c.}$(0.5) &     0.2881475675 &     0.0809077918 &   140.8129334279 &    151.965 &  134\\
I.B$_{105}^{i.c.}$(0.5) &     0.2690556508 &     0.0320231034 &   126.2887308493 &    152.287 &  134\\
I.B$_{106}^{i.c.}$(0.5) &     0.2857703636 &     0.0774702399 &   143.1821020744 &    156.629 &  138\\
I.B$_{107}^{i.c.}$(0.5) &     0.2845169706 &     0.0979582675 &   149.9521586362 &    160.710 &  142\\
I.B$_{108}^{i.c.}$(0.5) &     0.3360387142 &     0.0408766303 &   180.7765714891 &    161.357 &  142\\
I.B$_{109}^{i.c.}$(0.5) &     0.2667815506 &     0.0845599684 &   138.0065476191 &    161.041 &  142\\
I.B$_{110}^{i.c.}$(0.5) &     0.2700536880 &     0.0372395731 &   134.6253858410 &    161.374 &  142\\
I.B$_{111}^{i.c.}$(0.5) &     0.1620074707 &     0.0567511584 &   105.1701997563 &    165.454 &  146\\
I.B$_{112}^{i.c.}$(0.5) &     0.2846001734 &     0.0542198926 &   147.7254345023 &    165.875 &  146\\
I.B$_{113}^{i.c.}$(0.5) &     0.2937225183 &     0.0876854814 &   158.2087324044 &    165.427 &  146\\
I.B$_{114}^{i.c.}$(0.5) &     0.2752695212 &     0.0559644171 &   142.7112029555 &    165.875 &  146\\
I.B$_{115}^{i.c.}$(0.5) &     0.2170568347 &     0.0402768809 &   117.8057908660 &    165.889 &  146\\
I.B$_{116}^{i.c.}$(0.5) &     0.3256765948 &     0.0919189282 &   189.6279855460 &    169.365 &  150\\
I.B$_{117}^{i.c.}$(0.5) &     0.2878232371 &     0.0804604016 &   157.3437190277 &    170.118 &  150\\
I.B$_{118}^{i.c.}$(0.5) &     0.2709508831 &     0.0412767654 &   142.9671399385 &    170.460 &  150\\
I.B$_{119}^{i.c.}$(0.5) &     0.2325404803 &     0.0634100827 &   128.0692487359 &    170.359 &  150\\
I.B$_{120}^{i.c.}$(0.5) &     0.2777928735 &     0.0911581231 &   157.0896063985 &    174.505 &  154\\
I.B$_{121}^{i.c.}$(0.5) &     0.1884011033 &     0.0613519263 &   120.4032477307 &    179.229 &  158\\
I.B$_{122}^{i.c.}$(0.5) &     0.2876940968 &     0.0802805105 &   165.6171419591 &    179.195 &  158\\
I.B$_{123}^{i.c.}$(0.5) &     0.2981478673 &     0.0920364541 &   175.3905472876 &    178.871 &  158\\
I.B$_{124}^{i.c.}$(0.5) &     0.2717607209 &     0.0445435233 &   151.3128521285 &    179.545 &  158\\
I.B$_{125}^{i.c.}$(0.5) &     0.2694901377 &     0.0990271302 &   157.6461325706 &    178.868 &  158\\
I.B$_{126}^{i.c.}$(0.5) &     0.2780311897 &     0.0798240174 &   163.4425288092 &    183.859 &  162\\
I.B$_{127}^{i.c.}$(0.5) &     0.1535774570 &     0.0490036967 &   114.5530130654 &    183.703 &  162\\
I.B$_{128}^{i.c.}$(0.5) &     0.3084553159 &     0.0875379516 &   191.9788446279 &    187.951 &  166\\
I.B$_{129}^{i.c.}$(0.5) &     0.2819693119 &     0.0959873338 &   173.1019417858 &    187.951 &  166\\
I.B$_{130}^{i.c.}$(0.5) &     0.2724981752 &     0.0472634419 &   159.6637122720 &    188.629 &  166\\
I.B$_{131}^{i.c.}$(0.5) &     0.2812272902 &     0.0696965706 &   167.8845519750 &    188.499 &  166\\
I.B$_{132}^{i.c.}$(0.5) &     0.2851403139 &     0.1114556037 &   183.1661242112 &    191.850 &  170\\
I.B$_{133}^{i.c.}$(0.5) &     0.2925384369 &     0.0863782875 &   183.0243366435 &    192.661 &  170\\
I.B$_{134}^{i.c.}$(0.5) &     0.1940041780 &     0.0442273745 &   132.9965797155 &    197.594 &  174\\
I.B$_{135}^{i.c.}$(0.5) &     0.2874879398 &     0.0799909840 &   182.1812064952 &    197.349 &  174\\
I.B$_{136}^{i.c.}$(0.5) &     0.2743698810 &     0.0902336260 &   179.0226701835 &    201.738 &  178\\
I.B$_{137}^{i.c.}$(0.5) &     0.2691006601 &     0.0969556708 &   180.8849024635 &    206.107 &  182\\
I.B$_{138}^{i.c.}$(0.5) &     0.1872185341 &     0.0408400425 &   136.8385909602 &    206.678 &  182\\
I.B$_{139}^{i.c.}$(0.5) &     0.2708877517 &     0.0813214333 &   182.9605968884 &    211.089 &  186\\
I.B$_{140}^{i.c.}$(0.5) &     0.2632104741 &     0.0837121394 &   182.1621606443 &    215.501 &  190\\
I.B$_{141}^{i.c.}$(0.5) &     0.2821312739 &     0.0713901022 &   193.1416258963 &    215.731 &  190\\
I.B$_{142}^{i.c.}$(0.5) &     0.2394605930 &     0.0614826122 &   168.8258165112 &    220.374 &  194\\
I.B$_{143}^{i.c.}$(0.5) &     0.1924591719 &     0.0733547400 &   150.2890500922 &    219.658 &  194\\
I.B$_{144}^{i.c.}$(0.5) &     0.2124129905 &     0.0567064578 &   159.2830391611 &    224.832 &  198\\
I.B$_{145}^{i.c.}$(0.5) &     0.2716732997 &     0.0442108364 &   193.3516822884 &    229.546 &  202\\
I.B$_{146}^{i.c.}$(0.5) &     0.1784558095 &     0.0599818427 &   156.5079579799 &    238.104 &  210\\
I.B$_{147}^{i.c.}$(0.5) &     0.2168487625 &     0.0421574292 &   172.7342330299 &    243.143 &  214\\
I.B$_{148}^{i.c.}$(0.5) &     0.2116222333 &     0.0574328794 &   178.3098811885 &    252.068 &  222\\
I.B$_{149}^{i.c.}$(0.5) &     0.1734501113 &     0.0584514203 &   172.5037487239 &    265.336 &  234\\
I.B$_{150}^{i.c.}$(0.5) &     0.1773937613 &     0.0586759020 &   173.8818026019 &    265.336 &  234\\
I.B$_{151}^{i.c.}$(0.5) &     0.1589938384 &     0.0147834329 &   168.2376674277 &    270.519 &  238\\
I.B$_{152}^{i.c.}$(0.5) &     0.1732220167 &     0.0304907626 &   193.6476848963 &    302.170 &  266\\
\hline
\end{tabular*}
{\rule{\temptablewidth}{1pt}}
\end{center}
\end{table*}

\begin{table*}
\tabcolsep 0pt \caption{Initial conditions and periods $T$ of the periodic three-body orbits for class II.A and II.B in the case of  $\bm{r}_1(0)=(-1,0)=-\bm{r}_2(0)$,  $\dot{\bm{r}}_1(0)=(v_1,v_2)=\dot{\bm{r}}_2(0)$ and $\bm{r}_3(0)=(0,0)$, $\dot{\bm{r}}_3(0)=(-2v_1/m_3, -2v_2/m_3)$ when $G=1$ and $m_1=m_2=1$ and $m_3=0.5$ by means of the search grid $4000\times 4000$ in the interval $T_0\in[0,200]$, where $T^*=T |E|^{3/2}$ is its scale-invariant period, $L_f$ is the length of the free group element.  } \label{table-S5} \vspace*{-12pt}
\begin{center}
\def\temptablewidth{1\textwidth}
{\rule{\temptablewidth}{1pt}}
\begin{tabular*}{\temptablewidth}{@{\extracolsep{\fill}}lccccc}
\hline
Class and number  & $v_1$ & $v_2$  & $T$  & $T^*$ & $L_f$\\
\hline
II.A$_{1}^{i.c.}$(0.5) &     0.3225427028 &     0.1048724883 &    30.3858430513 &     27.025 &  24\\
II.A$_{2}^{i.c.}$(0.5) &     0.2131394477 &     0.2536523446 &    41.0020774214 &     38.035 &  32\\
II.A$_{3}^{i.c.}$(0.5) &     0.2831992418 &     0.2242675073 &    71.8227958466 &     56.038 &  48\\
II.A$_{4}^{i.c.}$(0.5) &     0.2890737487 &     0.1189680270 &    57.6132727750 &     58.603 &  52\\
II.A$_{5}^{i.c.}$(0.5) &     0.1160519606 &     0.0374080642 &    37.3113944214 &     63.514 &  56\\
II.A$_{6}^{i.c.}$(0.5) &     0.2814144757 &     0.0894379180 &   115.6439696541 &    126.928 &  112\\
II.A$_{7}^{i.c.}$(0.5) &     0.2647972881 &     0.0999834687 &   180.6781521304 &    208.284 &  184\\
II.B$_{1}^{i.c.}$(0.5) &     0.2522118365 &     0.2382777564 &    63.7808362694 &     54.281 &  46\\
II.B$_{2}^{i.c.}$(0.5) &     0.2921124451 &     0.1099360087 &   117.3258510268 &    119.607 &  106\\
II.B$_{3}^{i.c.}$(0.5) &     0.1843047650 &     0.0735504184 &   111.0333400751 &    165.168 &  146\\
II.B$_{4}^{i.c.}$(0.5) &     0.3051194885 &     0.0996833215 &   191.8933811547 &    187.542 &  166\\
\hline
\end{tabular*}
{\rule{\temptablewidth}{1pt}}
\end{center}
\end{table*}

\begin{table*}
\tabcolsep 0pt \caption{Initial conditions and periods $T$ of the periodic three-body orbits for class II.C in the case of  $\bm{r}_1(0)=(-1,0)=-\bm{r}_2(0)$,  $\dot{\bm{r}}_1(0)=(v_1,v_2)=\dot{\bm{r}}_2(0)$ and $\bm{r}_3(0)=(0,0)$, $\dot{\bm{r}}_3(0)=(-2v_1/m_3, -2v_2/m_3)$ when $G=1$ and $m_1=m_2=1$ and $m_3=0.5$ by means of the search grid $4000\times 4000$ in the interval $T_0\in[0,200]$, where $T^*=T |E|^{3/2}$ is its scale-invariant period, $L_f$ is the length of the free group element.  } \label{table-S5} \vspace*{-12pt}
\begin{center}
\def\temptablewidth{1\textwidth}
{\rule{\temptablewidth}{1pt}}
\begin{tabular*}{\temptablewidth}{@{\extracolsep{\fill}}lccccc}
\hline
Class and number  & $v_1$ & $v_2$  & $T$  & $T^*$ & $L_f$\\
\hline
II.C$_{1}^{i.c.}$(0.5) &     0.2057599772 &     0.2910772545 &    16.4482452694 &     13.225 &  10\\
II.C$_{2}^{i.c.}$(0.5) &     0.0621756721 &     0.0261903906 &     6.2854133740 &     11.285 &  10\\
II.C$_{3}^{i.c.}$(0.5) &     0.0657658390 &     0.1124034346 &     8.0617096205 &     13.572 &  12\\
II.C$_{4}^{i.c.}$(0.5) &     0.0169747300 &     0.0752136850 &     8.8877784855 &     15.845 &  14\\
II.C$_{5}^{i.c.}$(0.5) &     0.2251711660 &     0.3443496457 &    32.8998590701 &     17.385 &  14\\
II.C$_{6}^{i.c.}$(0.5) &     0.1194732446 &     0.0207612630 &    10.6764792286 &     18.190 &  16\\
II.C$_{7}^{i.c.}$(0.5) &     0.0551050358 &     0.3697557117 &    28.5843693576 &     20.500 &  16\\
II.C$_{8}^{i.c.}$(0.5) &     0.0108740090 &     0.3192205525 &    25.7212160836 &     25.333 &  18\\
II.C$_{9}^{i.c.}$(0.5) &     0.3448435595 &     0.0697188543 &    24.6216211722 &     20.364 &  18\\
II.C$_{10}^{i.c.}$(0.5) &     0.0117243540 &     0.3525284661 &    28.5310959155 &     23.470 &  18\\
II.C$_{11}^{i.c.}$(0.5) &     0.1780039242 &     0.2004919509 &    16.5294086907 &     20.135 &  18\\
II.C$_{12}^{i.c.}$(0.5) &     0.1282291061 &     0.3649372977 &    40.6015240363 &     26.471 &  20\\
II.C$_{13}^{i.c.}$(0.5) &     0.3234173131 &     0.2471156877 &    44.4546188773 &     24.471 &  22\\
II.C$_{14}^{i.c.}$(0.5) &     0.2104677107 &     0.1057158539 &    18.2820987595 &     24.716 &  22\\
II.C$_{15}^{i.c.}$(0.5) &     0.1983865989 &     0.1226004003 &    19.7881493182 &     26.930 &  24\\
II.C$_{16}^{i.c.}$(0.5) &     0.2287935035 &     0.0923297799 &    22.5126057181 &     29.433 &  26\\
II.C$_{17}^{i.c.}$(0.5) &     0.0913811473 &     0.0093493963 &    16.8086808175 &     29.586 &  26\\
II.C$_{18}^{i.c.}$(0.5) &     0.2383747181 &     0.2447155547 &    35.0733137252 &     30.771 &  26\\
II.C$_{19}^{i.c.}$(0.5) &     0.0992256924 &     0.0269686482 &    18.2257172127 &     31.728 &  28\\
II.C$_{20}^{i.c.}$(0.5) &     0.3345991412 &     0.0565118924 &    35.7753676955 &     31.788 &  28\\
II.C$_{21}^{i.c.}$(0.5) &     0.1490452956 &     0.0531097638 &    21.0786048660 &     33.979 &  30\\
II.C$_{22}^{i.c.}$(0.5) &     0.2263132666 &     0.0992959988 &    25.8750904330 &     33.785 &  30\\
II.C$_{23}^{i.c.}$(0.5) &     0.1822531374 &     0.1005779979 &    22.9574138909 &     33.376 &  30\\
II.C$_{24}^{i.c.}$(0.5) &     0.3130939094 &     0.0872917851 &    35.4487192804 &     33.958 &  30\\
II.C$_{25}^{i.c.}$(0.5) &     0.2379407604 &     0.2522685220 &    44.6369551897 &     38.031 &  32\\
II.C$_{26}^{i.c.}$(0.5) &     0.1230951336 &     0.0496740170 &    21.5462119008 &     36.147 &  32\\
II.C$_{27}^{i.c.}$(0.5) &     0.1326230917 &     0.0945397322 &    24.0011090101 &     38.376 &  34\\
II.C$_{28}^{i.c.}$(0.5) &     0.1440416968 &     0.1072211552 &    24.7755961375 &     38.378 &  34\\
II.C$_{29}^{i.c.}$(0.5) &     0.1444470268 &     0.0693298831 &    25.2083739245 &     40.495 &  36\\
II.C$_{30}^{i.c.}$(0.5) &     0.3196418099 &     0.1040424691 &    47.3539072837 &     42.814 &  38\\
\hline
\end{tabular*}
{\rule{\temptablewidth}{1pt}}
\end{center}
\end{table*}

\begin{table*}
\tabcolsep 0pt \caption{Initial conditions and periods $T$ of the periodic three-body orbits for class II.C in the case of  $\bm{r}_1(0)=(-1,0)=-\bm{r}_2(0)$,  $\dot{\bm{r}}_1(0)=(v_1,v_2)=\dot{\bm{r}}_2(0)$ and $\bm{r}_3(0)=(0,0)$, $\dot{\bm{r}}_3(0)=(-2v_1/m_3, -2v_2/m_3)$ when $G=1$ and $m_1=m_2=1$ and $m_3=0.5$ by means of the search grid $4000\times 4000$ in the interval $T_0\in[0,200]$, where $T^*=T |E|^{3/2}$ is its scale-invariant period, $L_f$ is the length of the free group element.  } \label{table-S5} \vspace*{-12pt}
\begin{center}
\def\temptablewidth{1\textwidth}
{\rule{\temptablewidth}{1pt}}
\begin{tabular*}{\temptablewidth}{@{\extracolsep{\fill}}lccccc}
\hline
Class and number  & $v_1$ & $v_2$  & $T$  & $T^*$ & $L_f$\\
\hline
II.C$_{31}^{i.c.}$(0.5) &     0.0987975586 &     0.0968340577 &    25.8114857482 &     42.954 &  38\\
II.C$_{32}^{i.c.}$(0.5) &     0.2432511280 &     0.0916195182 &    34.3617337272 &     43.051 &  38\\
II.C$_{33}^{i.c.}$(0.5) &     0.3382801410 &     0.0351009613 &    48.8084195841 &     43.188 &  38\\
II.C$_{34}^{i.c.}$(0.5) &     0.0852972761 &     0.0255811336 &    24.1174290003 &     42.561 &  38\\
II.C$_{35}^{i.c.}$(0.5) &     0.3220538773 &     0.0646621470 &    48.2270219401 &     45.398 &  40\\
II.C$_{36}^{i.c.}$(0.5) &     0.2718709077 &     0.0946174473 &    42.0581440899 &     47.577 &  42\\
II.C$_{37}^{i.c.}$(0.5) &     0.3183080806 &     0.0764243838 &    50.2696384479 &     47.594 &  42\\
II.C$_{38}^{i.c.}$(0.5) &     0.1710053276 &     0.0753437523 &    31.0798039439 &     47.424 &  42\\
II.C$_{39}^{i.c.}$(0.5) &     0.2934877947 &     0.2212457298 &    65.1670851943 &     48.795 &  42\\
II.C$_{40}^{i.c.}$(0.5) &     0.1428288108 &     0.0553224707 &    30.6121545358 &     49.772 &  44\\
II.C$_{41}^{i.c.}$(0.5) &     0.2635213411 &     0.1069239583 &    43.3809654410 &     49.749 &  44\\
II.C$_{42}^{i.c.}$(0.5) &     0.1937551146 &     0.0588647172 &    36.9605508290 &     54.466 &  48\\
II.C$_{43}^{i.c.}$(0.5) &     0.2501765655 &     0.0899074157 &    46.1520234471 &     56.667 &  50\\
II.C$_{44}^{i.c.}$(0.5) &     0.0819979788 &     0.3405135958 &    79.1849364880 &     66.109 &  50\\
II.C$_{45}^{i.c.}$(0.5) &     0.1421031951 &     0.1000963218 &    36.0751418819 &     56.519 &  50\\
II.C$_{46}^{i.c.}$(0.5) &     0.1769229292 &     0.0317210521 &    36.6796720789 &     56.798 &  50\\
II.C$_{47}^{i.c.}$(0.5) &     0.2231965871 &     0.0917189066 &    44.2893524922 &     58.865 &  52\\
II.C$_{48}^{i.c.}$(0.5) &     0.3149842289 &     0.0687327481 &    60.7978969392 &     59.011 &  52\\
II.C$_{49}^{i.c.}$(0.5) &     0.1615942955 &     0.1179674433 &    40.9841942386 &     60.738 &  54\\
II.C$_{50}^{i.c.}$(0.5) &     0.2810932036 &     0.2296505498 &    82.0206638443 &     63.285 &  54\\
II.C$_{51}^{i.c.}$(0.5) &     0.1735416113 &     0.0688826379 &    39.9994858929 &     61.055 &  54\\
II.C$_{52}^{i.c.}$(0.5) &     0.1035547024 &     0.0422685017 &    35.0762402950 &     60.451 &  54\\
II.C$_{53}^{i.c.}$(0.5) &     0.2690033537 &     0.0919983273 &    53.3422504976 &     61.194 &  54\\
II.C$_{54}^{i.c.}$(0.5) &     0.1173397763 &     0.0109376998 &    38.5590034737 &     65.976 &  58\\
II.C$_{55}^{i.c.}$(0.5) &     0.1315056462 &     0.0526807760 &    39.5934146422 &     65.562 &  58\\
II.C$_{56}^{i.c.}$(0.5) &     0.2613378486 &     0.1103086095 &    56.9874074323 &     65.537 &  58\\
II.C$_{57}^{i.c.}$(0.5) &     0.3281658348 &     0.0363891724 &    73.0735492094 &     68.188 &  60\\
II.C$_{58}^{i.c.}$(0.5) &     0.1395436149 &     0.0265973488 &    41.1870639981 &     68.161 &  60\\
II.C$_{59}^{i.c.}$(0.5) &     0.1878900821 &     0.0561739494 &    45.5265171501 &     68.082 &  60\\
II.C$_{60}^{i.c.}$(0.5) &     0.2635734120 &     0.1081725286 &    61.2354053056 &     70.083 &  62\\
II.C$_{61}^{i.c.}$(0.5) &     0.2540232703 &     0.0883181524 &    57.8669161621 &     70.283 &  62\\
II.C$_{62}^{i.c.}$(0.5) &     0.1821497093 &     0.0381106799 &    46.0695631799 &     70.412 &  62\\
II.C$_{63}^{i.c.}$(0.5) &     0.2811013873 &     0.1118997053 &    67.8894384900 &     72.243 &  64\\
II.C$_{64}^{i.c.}$(0.5) &     0.2669232153 &     0.0902025793 &    64.5730278183 &     74.810 &  66\\
II.C$_{65}^{i.c.}$(0.5) &     0.1792102190 &     0.1389127769 &    53.8787781562 &     74.660 &  66\\
II.C$_{66}^{i.c.}$(0.5) &     0.3085501674 &     0.0768386796 &    75.4497521560 &     74.824 &  66\\
II.C$_{67}^{i.c.}$(0.5) &     0.1564466743 &     0.0570450740 &    48.4839903391 &     77.011 &  68\\
II.C$_{68}^{i.c.}$(0.5) &     0.2374442052 &     0.0698194855 &    60.9419117383 &     79.483 &  70\\
II.C$_{69}^{i.c.}$(0.5) &     0.1495364869 &     0.0339284734 &    50.3005990140 &     81.760 &  72\\
II.C$_{70}^{i.c.}$(0.5) &     0.3151482492 &     0.0494781713 &    82.8987639982 &     81.803 &  72\\
II.C$_{71}^{i.c.}$(0.5) &     0.3414149024 &     0.0554960419 &    95.4752935994 &     81.760 &  72\\
II.C$_{72}^{i.c.}$(0.5) &     0.1863427441 &     0.0555295217 &    54.4276379358 &     81.697 &  72\\
II.C$_{73}^{i.c.}$(0.5) &     0.3285842514 &     0.0936843712 &    95.1841177804 &     83.484 &  74\\
II.C$_{74}^{i.c.}$(0.5) &     0.2562981114 &     0.0869856262 &    69.5011518434 &     83.898 &  74\\
II.C$_{75}^{i.c.}$(0.5) &     0.1322093913 &     0.0211571451 &    51.5542382142 &     86.350 &  76\\
II.C$_{76}^{i.c.}$(0.5) &     0.3185759283 &     0.0928737708 &    92.8294620014 &     85.876 &  76\\
II.C$_{77}^{i.c.}$(0.5) &     0.3226279066 &     0.1036036069 &    96.1122549182 &     85.628 &  76\\
II.C$_{78}^{i.c.}$(0.5) &     0.1206929092 &     0.0421948480 &    50.9428983671 &     86.044 &  76\\
II.C$_{79}^{i.c.}$(0.5) &     0.2763819973 &     0.1013126179 &    80.1001367274 &     88.252 &  78\\
II.C$_{80}^{i.c.}$(0.5) &     0.2654060732 &     0.0888677277 &    75.7855698562 &     88.426 &  78\\
II.C$_{81}^{i.c.}$(0.5) &     0.3057951517 &     0.0766709356 &    88.0351418174 &     88.439 &  78\\
II.C$_{82}^{i.c.}$(0.5) &     0.2823465042 &     0.1166804112 &    85.8497303391 &     90.177 &  80\\
II.C$_{83}^{i.c.}$(0.5) &     0.2638802619 &     0.0961802884 &    77.9476788146 &     90.613 &  80\\
II.C$_{84}^{i.c.}$(0.5) &     0.1581811787 &     0.1114689424 &    61.7013854081 &     92.806 &  82\\
II.C$_{85}^{i.c.}$(0.5) &     0.2475778287 &     0.0746419321 &    73.9745774361 &     93.099 &  82\\
\hline
\end{tabular*}
{\rule{\temptablewidth}{1pt}}
\end{center}
\end{table*}

\begin{table*}
\tabcolsep 0pt \caption{Initial conditions and periods $T$ of the periodic three-body orbits for class II.C in the case of  $\bm{r}_1(0)=(-1,0)=-\bm{r}_2(0)$,  $\dot{\bm{r}}_1(0)=(v_1,v_2)=\dot{\bm{r}}_2(0)$ and $\bm{r}_3(0)=(0,0)$, $\dot{\bm{r}}_3(0)=(-2v_1/m_3, -2v_2/m_3)$ when $G=1$ and $m_1=m_2=1$ and $m_3=0.5$ by means of the search grid $4000\times 4000$ in the interval $T_0\in[0,200]$, where $T^*=T |E|^{3/2}$ is its scale-invariant period, $L_f$ is the length of the free group element.  } \label{table-S5} \vspace*{-12pt}
\begin{center}
\def\temptablewidth{1\textwidth}
{\rule{\temptablewidth}{1pt}}
\begin{tabular*}{\temptablewidth}{@{\extracolsep{\fill}}lccccc}
\hline
Class and number  & $v_1$ & $v_2$  & $T$  & $T^*$ & $L_f$\\
\hline
II.C$_{86}^{i.c.}$(0.5) &     0.1516949119 &     0.0244769729 &    57.3481255424 &     93.166 &  82\\
II.C$_{87}^{i.c.}$(0.5) &     0.3220495093 &     0.0374666542 &    96.8850696521 &     93.189 &  82\\
II.C$_{88}^{i.c.}$(0.5) &     0.1858022899 &     0.0553034082 &    63.4160528904 &     95.312 &  84\\
II.C$_{89}^{i.c.}$(0.5) &     0.2655776786 &     0.1010849772 &    84.7984660370 &     97.331 &  86\\
II.C$_{90}^{i.c.}$(0.5) &     0.3207800454 &     0.0938599983 &   106.3208891400 &     97.122 &  86\\
II.C$_{91}^{i.c.}$(0.5) &     0.1834880615 &     0.0537582335 &    66.0505896692 &     99.853 &  88\\
II.C$_{92}^{i.c.}$(0.5) &     0.3018070276 &     0.0795462027 &   100.0943865475 &    102.041 &  90\\
II.C$_{93}^{i.c.}$(0.5) &     0.3038401134 &     0.0764563615 &   100.6645786318 &    102.054 &  90\\
II.C$_{94}^{i.c.}$(0.5) &     0.3052322511 &     0.1102837421 &   105.6095849909 &    101.417 &  90\\
II.C$_{95}^{i.c.}$(0.5) &     0.3566421417 &     0.0401336571 &   132.0028545187 &    104.539 &  92\\
II.C$_{96}^{i.c.}$(0.5) &     0.2657655740 &     0.1056700425 &    91.2995867266 &    104.043 &  92\\
II.C$_{97}^{i.c.}$(0.5) &     0.1446403514 &     0.0463024934 &    64.0268715528 &    104.324 &  92\\
II.C$_{98}^{i.c.}$(0.5) &     0.1257825213 &     0.0270171780 &    61.9602861500 &    104.540 &  92\\
II.C$_{99}^{i.c.}$(0.5) &     0.2636263542 &     0.0943925785 &    89.3740530822 &    104.231 &  92\\
II.C$_{100}^{i.c.}$(0.5) &     0.3189489999 &     0.0684963314 &   109.6451976549 &    104.408 &  92\\
II.C$_{101}^{i.c.}$(0.5) &     0.2425203938 &     0.0657640364 &    81.0766276092 &    104.493 &  92\\
II.C$_{102}^{i.c.}$(0.5) &     0.2865282765 &     0.0260626161 &    94.3794078874 &    106.829 &  94\\
II.C$_{103}^{i.c.}$(0.5) &     0.3294573946 &     0.0567093697 &   119.3411186852 &    108.973 &  96\\
II.C$_{104}^{i.c.}$(0.5) &     0.3300293300 &     0.0595323594 &   120.0149738801 &    108.973 &  96\\
II.C$_{105}^{i.c.}$(0.5) &     0.1887576280 &     0.0377729761 &    72.3298978206 &    109.035 &  96\\
II.C$_{106}^{i.c.}$(0.5) &     0.1665864480 &     0.0397866710 &    69.2172422426 &    108.973 &  96\\
II.C$_{107}^{i.c.}$(0.5) &     0.3289575551 &     0.0668810663 &   122.6577661872 &    111.174 &  98\\
II.C$_{108}^{i.c.}$(0.5) &     0.2749671985 &     0.1116521323 &   103.5125111065 &    112.909 &  100\\
II.C$_{109}^{i.c.}$(0.5) &     0.2252171434 &     0.0831314124 &    84.9641254583 &    113.344 &  100\\
II.C$_{110}^{i.c.}$(0.5) &     0.3035132687 &     0.0742550960 &   111.4832387743 &    113.467 &  100\\
II.C$_{111}^{i.c.}$(0.5) &     0.2995384163 &     0.0855174656 &   110.8383172246 &    113.311 &  100\\
II.C$_{112}^{i.c.}$(0.5) &     0.2634205915 &     0.0870091576 &    98.2072237371 &    115.656 &  102\\
II.C$_{113}^{i.c.}$(0.5) &     0.3024538187 &     0.0762577398 &   113.3596459196 &    115.668 &  102\\
II.C$_{114}^{i.c.}$(0.5) &     0.3110478575 &     0.0862561185 &   119.2212947649 &    115.494 &  102\\
II.C$_{115}^{i.c.}$(0.5) &     0.3526389241 &     0.0428144462 &   145.8125822163 &    118.133 &  104\\
II.C$_{116}^{i.c.}$(0.5) &     0.1918597167 &     0.0736315430 &    80.4647582647 &    117.735 &  104\\
II.C$_{117}^{i.c.}$(0.5) &     0.2698687648 &     0.1023315358 &   104.3314555538 &    117.667 &  104\\
II.C$_{118}^{i.c.}$(0.5) &     0.1414881922 &     0.0354161821 &    71.8047330418 &    118.134 &  104\\
II.C$_{119}^{i.c.}$(0.5) &     0.1619102285 &     0.0244352124 &    74.0209337785 &    118.169 &  104\\
II.C$_{120}^{i.c.}$(0.5) &     0.2502762644 &     0.0696607780 &    94.2261347512 &    118.110 &  104\\
II.C$_{121}^{i.c.}$(0.5) &     0.1870717535 &     0.0147811565 &    77.5861543862 &    118.190 &  104\\
II.C$_{122}^{i.c.}$(0.5) &     0.1905368730 &     0.0303149606 &    79.9609883712 &    120.424 &  106\\
II.C$_{123}^{i.c.}$(0.5) &     0.1675018862 &     0.0354981141 &    76.4546465382 &    120.379 &  106\\
II.C$_{124}^{i.c.}$(0.5) &     0.2382138646 &     0.0760779564 &    94.7217192960 &    122.548 &  108\\
II.C$_{125}^{i.c.}$(0.5) &     0.3251008379 &     0.0627976669 &   132.0053020644 &    122.583 &  108\\
II.C$_{126}^{i.c.}$(0.5) &     0.3268797519 &     0.0654741271 &   133.5845577449 &    122.583 &  108\\
II.C$_{127}^{i.c.}$(0.5) &     0.2824206151 &     0.1120611754 &   117.3283797375 &    124.152 &  110\\
II.C$_{128}^{i.c.}$(0.5) &     0.3355862027 &     0.0688344773 &   142.8830057530 &    124.693 &  110\\
II.C$_{129}^{i.c.}$(0.5) &     0.2650997364 &     0.0972324666 &   107.7796238869 &    124.572 &  110\\
II.C$_{130}^{i.c.}$(0.5) &     0.3155083362 &     0.0780524043 &   130.0655193752 &    124.605 &  110\\
II.C$_{131}^{i.c.}$(0.5) &     0.2911251331 &     0.0747106544 &   116.3271632708 &    124.860 &  110\\
II.C$_{132}^{i.c.}$(0.5) &     0.3023181843 &     0.0747198638 &   124.2563024241 &    127.082 &  112\\
II.C$_{133}^{i.c.}$(0.5) &     0.2877451269 &     0.0973212810 &   119.7416237253 &    126.751 &  112\\
II.C$_{134}^{i.c.}$(0.5) &     0.2730233157 &     0.0958101195 &   114.8135152620 &    129.113 &  114\\
II.C$_{135}^{i.c.}$(0.5) &     0.2522438389 &     0.2397521752 &   158.1926187312 &    133.822 &  114\\
II.C$_{136}^{i.c.}$(0.5) &     0.1972498965 &     0.0367328629 &    87.4975613393 &    129.499 &  114\\
II.C$_{137}^{i.c.}$(0.5) &     0.2551006961 &     0.0719840482 &   106.9936965308 &    131.727 &  116\\
II.C$_{138}^{i.c.}$(0.5) &     0.1466907115 &     0.0325820647 &    80.6365220274 &    131.733 &  116\\
II.C$_{139}^{i.c.}$(0.5) &     0.3114989143 &     0.0680246052 &   133.2896024822 &    131.635 &  116\\
II.C$_{140}^{i.c.}$(0.5) &     0.3295340331 &     0.0484649621 &   143.3806235446 &    131.777 &  116\\
II.C$_{141}^{i.c.}$(0.5) &     0.1909835958 &     0.0219691369 &    87.3877632060 &    131.809 &  116\\
II.C$_{142}^{i.c.}$(0.5) &     0.2413919468 &     0.0306502812 &    99.7934250857 &    131.831 &  116\\
II.C$_{143}^{i.c.}$(0.5) &     0.1733103989 &     0.0446120548 &    87.8145595559 &    136.194 &  120\\
\hline
\end{tabular*}
{\rule{\temptablewidth}{1pt}}
\end{center}
\end{table*}

\begin{table*}
\tabcolsep 0pt \caption{Initial conditions and periods $T$ of the periodic three-body orbits for class II.C in the case of  $\bm{r}_1(0)=(-1,0)=-\bm{r}_2(0)$,  $\dot{\bm{r}}_1(0)=(v_1,v_2)=\dot{\bm{r}}_2(0)$ and $\bm{r}_3(0)=(0,0)$, $\dot{\bm{r}}_3(0)=(-2v_1/m_3, -2v_2/m_3)$ when $G=1$ and $m_1=m_2=1$ and $m_3=0.5$ by means of the search grid $4000\times 4000$ in the interval $T_0\in[0,200]$, where $T^*=T |E|^{3/2}$ is its scale-invariant period, $L_f$ is the length of the free group element.  } \label{table-S5} \vspace*{-12pt}
\begin{center}
\def\temptablewidth{1\textwidth}
{\rule{\temptablewidth}{1pt}}
\begin{tabular*}{\temptablewidth}{@{\extracolsep{\fill}}lccccc}
\hline
Class and number  & $v_1$ & $v_2$  & $T$  & $T^*$ & $L_f$\\
\hline

II.C$_{144}^{i.c.}$(0.5) &     0.2930149393 &     0.0749995913 &   130.0780110428 &    138.475 &  122\\
II.C$_{145}^{i.c.}$(0.5) &     0.3010872261 &     0.0827870509 &   137.9566276722 &    140.544 &  124\\
II.C$_{146}^{i.c.}$(0.5) &     0.2837900515 &     0.1048066565 &   131.7075390111 &    140.171 &  124\\
II.C$_{147}^{i.c.}$(0.5) &     0.1654894951 &     0.0372403932 &    89.1316363022 &    140.762 &  124\\
II.C$_{148}^{i.c.}$(0.5) &     0.2005754999 &     0.0307687587 &    95.6979960354 &    140.884 &  124\\
II.C$_{149}^{i.c.}$(0.5) &     0.1950195608 &     0.0382427268 &    96.2751827640 &    143.118 &  126\\
II.C$_{150}^{i.c.}$(0.5) &     0.3069226729 &     0.1114459071 &   151.7018490077 &    144.231 &  128\\
II.C$_{151}^{i.c.}$(0.5) &     0.2583666188 &     0.0736044941 &   119.5432346546 &    145.343 &  128\\
II.C$_{152}^{i.c.}$(0.5) &     0.2136148238 &     0.0824671746 &   105.2690821399 &    144.976 &  128\\
II.C$_{153}^{i.c.}$(0.5) &     0.3026649056 &     0.0527959420 &   139.5203274022 &    145.428 &  128\\
II.C$_{154}^{i.c.}$(0.5) &     0.1869669478 &     0.0727253609 &    99.4296116997 &    147.169 &  130\\
II.C$_{155}^{i.c.}$(0.5) &     0.3253795688 &     0.0922726657 &   166.6893561658 &    149.032 &  132\\
II.C$_{156}^{i.c.}$(0.5) &     0.3510221862 &     0.0469440817 &   183.8293814238 &    149.920 &  132\\
II.C$_{157}^{i.c.}$(0.5) &     0.2976026689 &     0.0679239870 &   142.5318450479 &    149.883 &  132\\
II.C$_{158}^{i.c.}$(0.5) &     0.2476930572 &     0.0788426863 &   119.5507824703 &    149.779 &  132\\
II.C$_{159}^{i.c.}$(0.5) &     0.1744790844 &     0.0483169316 &    96.9984259826 &    149.806 &  132\\
II.C$_{160}^{i.c.}$(0.5) &     0.2329477926 &     0.0962157533 &   116.2632004744 &    149.480 &  132\\
II.C$_{161}^{i.c.}$(0.5) &     0.2049799809 &     0.0695040711 &   105.1805254163 &    149.783 &  132\\
II.C$_{162}^{i.c.}$(0.5) &     0.2000398001 &     0.0570752651 &   102.9916394933 &    149.783 &  132\\
II.C$_{163}^{i.c.}$(0.5) &     0.3208285254 &     0.0707218166 &   161.5242818190 &    152.011 &  134\\
II.C$_{164}^{i.c.}$(0.5) &     0.2323718985 &     0.0470731914 &   113.1431229338 &    152.265 &  134\\
II.C$_{165}^{i.c.}$(0.5) &     0.2732770465 &     0.1112359067 &   139.7843087122 &    153.574 &  136\\
II.C$_{166}^{i.c.}$(0.5) &     0.1973164730 &     0.0325297824 &   104.2330202025 &    154.504 &  136\\
II.C$_{167}^{i.c.}$(0.5) &     0.3555507477 &     0.0410620694 &   193.9086747432 &    154.510 &  136\\
II.C$_{168}^{i.c.}$(0.5) &     0.3011590042 &     0.0819088183 &   151.2083294223 &    154.160 &  136\\
II.C$_{169}^{i.c.}$(0.5) &     0.3069577168 &     0.0877702567 &   156.2447738532 &    153.992 &  136\\
II.C$_{170}^{i.c.}$(0.5) &     0.3243977377 &     0.0286054600 &   161.8647903378 &    154.573 &  136\\
II.C$_{171}^{i.c.}$(0.5) &     0.2708273208 &     0.0936225609 &   137.4958739955 &    156.348 &  138\\
II.C$_{172}^{i.c.}$(0.5) &     0.3032598179 &     0.0211927097 &   148.3161503321 &    156.832 &  138\\
II.C$_{173}^{i.c.}$(0.5) &     0.2499690317 &     0.0880524793 &   128.9029384970 &    158.722 &  140\\
II.C$_{174}^{i.c.}$(0.5) &     0.2439808630 &     0.0927858470 &   128.9093834720 &    160.913 &  142\\
II.C$_{175}^{i.c.}$(0.5) &     0.2921786881 &     0.1158238005 &   158.5701925931 &    160.019 &  142\\
II.C$_{176}^{i.c.}$(0.5) &     0.3187109289 &     0.0697453074 &   171.6434882025 &    163.419 &  144\\
II.C$_{177}^{i.c.}$(0.5) &     0.2345393298 &     0.0733775891 &   124.6080587586 &    163.399 &  144\\
II.C$_{178}^{i.c.}$(0.5) &     0.2817443085 &     0.0820058547 &   149.7102508372 &    165.580 &  146\\
II.C$_{179}^{i.c.}$(0.5) &     0.2683324131 &     0.0800302736 &   142.0849053423 &    165.705 &  146\\
II.C$_{180}^{i.c.}$(0.5) &     0.2911513179 &     0.1023124996 &   159.3225451639 &    165.048 &  146\\
II.C$_{181}^{i.c.}$(0.5) &     0.2450691759 &     0.0560195533 &   128.7590915568 &    165.887 &  146\\
II.C$_{182}^{i.c.}$(0.5) &     0.1637533270 &     0.0120392511 &   104.0378908319 &    165.953 &  146\\
II.C$_{183}^{i.c.}$(0.5) &     0.3189840056 &     0.0986343953 &   182.4517653543 &    166.971 &  148\\
II.C$_{184}^{i.c.}$(0.5) &     0.3277428450 &     0.0422660249 &   180.4850020783 &    168.174 &  148\\
II.C$_{185}^{i.c.}$(0.5) &     0.2822423292 &     0.0943666677 &   154.2006195402 &    167.612 &  148\\
II.C$_{186}^{i.c.}$(0.5) &     0.1529038054 &     0.0482036149 &   105.9017717709 &    170.094 &  150\\
II.C$_{187}^{i.c.}$(0.5) &     0.2502629663 &     0.0420236439 &   133.3212700219 &    170.461 &  150\\
II.C$_{188}^{i.c.}$(0.5) &     0.2595662993 &     0.0702980306 &   140.2477320721 &    170.352 &  150\\
II.C$_{189}^{i.c.}$(0.5) &     0.2719916833 &     0.1024099935 &   153.7109398488 &    171.961 &  152\\
II.C$_{190}^{i.c.}$(0.5) &     0.3318880756 &     0.0584426073 &   191.7202695166 &    172.549 &  152\\
II.C$_{191}^{i.c.}$(0.5) &     0.2642631613 &     0.0846766978 &   148.3494808332 &    174.656 &  154\\
II.C$_{192}^{i.c.}$(0.5) &     0.2950993654 &     0.1054674522 &   171.5975520159 &    173.910 &  154\\
II.C$_{193}^{i.c.}$(0.5) &     0.2952577971 &     0.0652451441 &   164.2296591527 &    174.893 &  154\\
II.C$_{194}^{i.c.}$(0.5) &     0.1919407079 &     0.0544880286 &   119.2943656794 &    177.015 &  156\\
II.C$_{195}^{i.c.}$(0.5) &     0.2341210185 &     0.0775543587 &   135.3588931652 &    177.015 &  156\\
II.C$_{196}^{i.c.}$(0.5) &     0.2740046014 &     0.0751646038 &   154.2765378448 &    177.104 &  156\\
II.C$_{197}^{i.c.}$(0.5) &     0.2741492762 &     0.1053054597 &   161.7072926859 &    178.670 &  158\\
II.C$_{198}^{i.c.}$(0.5) &     0.3114471237 &     0.0925603589 &   186.4022437357 &    178.674 &  158\\
II.C$_{199}^{i.c.}$(0.5) &     0.2517718983 &     0.0599951023 &   142.7458386836 &    179.508 &  158\\
II.C$_{200}^{i.c.}$(0.5) &     0.2780070267 &     0.0823491926 &   159.7348584169 &    179.195 &  158\\
\hline
\end{tabular*}
{\rule{\temptablewidth}{1pt}}
\end{center}
\end{table*}

\begin{table*}
\tabcolsep 0pt \caption{Initial conditions and periods $T$ of the periodic three-body orbits for class II.C in the case of  $\bm{r}_1(0)=(-1,0)=-\bm{r}_2(0)$,  $\dot{\bm{r}}_1(0)=(v_1,v_2)=\dot{\bm{r}}_2(0)$ and $\bm{r}_3(0)=(0,0)$, $\dot{\bm{r}}_3(0)=(-2v_1/m_3, -2v_2/m_3)$ when $G=1$ and $m_1=m_2=1$ and $m_3=0.5$ by means of the search grid $4000\times 4000$ in the interval $T_0\in[0,200]$, where $T^*=T |E|^{3/2}$ is its scale-invariant period, $L_f$ is the length of the free group element.  } \label{table-S5} \vspace*{-12pt}
\begin{center}
\def\temptablewidth{1\textwidth}
{\rule{\temptablewidth}{1pt}}
\begin{tabular*}{\temptablewidth}{@{\extracolsep{\fill}}lccccc}
\hline
Class and number  & $v_1$ & $v_2$  & $T$  & $T^*$ & $L_f$\\
\hline

II.C$_{201}^{i.c.}$(0.5) &     0.3208251493 &     0.0312083220 &   184.9613703373 &    179.574 &  158\\
II.C$_{202}^{i.c.}$(0.5) &     0.2823197687 &     0.1061568833 &   169.2512195359 &    180.837 &  160\\
II.C$_{203}^{i.c.}$(0.5) &     0.3058525824 &     0.0859666113 &   182.5097652558 &    181.230 &  160\\
II.C$_{204}^{i.c.}$(0.5) &     0.2711017290 &     0.0880228048 &   158.5702051413 &    181.391 &  160\\
II.C$_{205}^{i.c.}$(0.5) &     0.2832710026 &     0.0993650437 &   168.4082593848 &    181.046 &  160\\
II.C$_{206}^{i.c.}$(0.5) &     0.2580794098 &     0.0662802067 &   148.3342287109 &    181.742 &  160\\
II.C$_{207}^{i.c.}$(0.5) &     0.2168915835 &     0.0791081052 &   134.0444642533 &    183.636 &  162\\
II.C$_{208}^{i.c.}$(0.5) &     0.3182346683 &     0.0550212092 &   190.1060999565 &    184.007 &  162\\
II.C$_{209}^{i.c.}$(0.5) &     0.2602548035 &     0.0468178359 &   149.1982024781 &    184.087 &  162\\
II.C$_{210}^{i.c.}$(0.5) &     0.2957290452 &     0.1115815297 &   184.7739565799 &    184.904 &  164\\
II.C$_{211}^{i.c.}$(0.5) &     0.2713556472 &     0.1013159584 &   165.2338308800 &    185.584 &  164\\
II.C$_{212}^{i.c.}$(0.5) &     0.1400364517 &     0.0295516193 &   112.7541853074 &    186.294 &  164\\
II.C$_{213}^{i.c.}$(0.5) &     0.2538303347 &     0.0857213734 &   152.5491452121 &    185.952 &  164\\
II.C$_{214}^{i.c.}$(0.5) &     0.1661354659 &     0.0434379136 &   118.3442264866 &    186.158 &  164\\
II.C$_{215}^{i.c.}$(0.5) &     0.2947852597 &     0.0614773180 &   173.9781290999 &    186.283 &  164\\
II.C$_{216}^{i.c.}$(0.5) &     0.2959154222 &     0.1062666806 &   183.5801829920 &    185.154 &  164\\
II.C$_{217}^{i.c.}$(0.5) &     0.1872594057 &     0.0518107976 &   125.4822943491 &    188.412 &  166\\
II.C$_{218}^{i.c.}$(0.5) &     0.1946620874 &     0.0643435760 &   130.1369103426 &    190.634 &  168\\
II.C$_{219}^{i.c.}$(0.5) &     0.1960305611 &     0.0313147606 &   129.8525440051 &    193.128 &  170\\
II.C$_{220}^{i.c.}$(0.5) &     0.2756620575 &     0.0823970177 &   170.3420494464 &    192.810 &  170\\
II.C$_{221}^{i.c.}$(0.5) &     0.3006677095 &     0.0801050248 &   190.4434520843 &    195.007 &  172\\
II.C$_{222}^{i.c.}$(0.5) &     0.2138125453 &     0.0765286284 &   140.8953101446 &    195.052 &  172\\
II.C$_{223}^{i.c.}$(0.5) &     0.2817233384 &     0.1045112371 &   181.1173423050 &    194.465 &  172\\
II.C$_{224}^{i.c.}$(0.5) &     0.2826464621 &     0.0982734891 &   180.3452656908 &    194.667 &  172\\
II.C$_{225}^{i.c.}$(0.5) &     0.2632094008 &     0.0725802686 &   165.0883588774 &    197.586 &  174\\
II.C$_{226}^{i.c.}$(0.5) &     0.2731816009 &     0.0806117521 &   172.4907007388 &    197.474 &  174\\
II.C$_{227}^{i.c.}$(0.5) &     0.2960515524 &     0.1072263897 &   195.1428066451 &    196.397 &  174\\
II.C$_{228}^{i.c.}$(0.5) &     0.2951231894 &     0.0577880334 &   184.2949392209 &    197.670 &  174\\
II.C$_{229}^{i.c.}$(0.5) &     0.2700922029 &     0.0499192216 &   166.1828634798 &    197.713 &  174\\
II.C$_{230}^{i.c.}$(0.5) &     0.2677573881 &     0.0945684253 &   173.5769232952 &    199.385 &  176\\
II.C$_{231}^{i.c.}$(0.5) &     0.2434451873 &     0.0747898054 &   156.7395420237 &    199.814 &  176\\
II.C$_{232}^{i.c.}$(0.5) &     0.2936385507 &     0.0678957262 &   188.9849393321 &    202.125 &  178\\
II.C$_{233}^{i.c.}$(0.5) &     0.2624985469 &     0.0847688537 &   170.4292589966 &    201.886 &  178\\
II.C$_{234}^{i.c.}$(0.5) &     0.1968157177 &     0.0738072529 &   139.3969959798 &    201.651 &  178\\
II.C$_{235}^{i.c.}$(0.5) &     0.2823384174 &     0.1064870414 &   188.3906435625 &    201.170 &  178\\
II.C$_{236}^{i.c.}$(0.5) &     0.2136874355 &     0.0627089113 &   145.7957129717 &    204.251 &  180\\
II.C$_{237}^{i.c.}$(0.5) &     0.2861638798 &     0.0874577519 &   191.0899162207 &    206.277 &  182\\
II.C$_{238}^{i.c.}$(0.5) &     0.2926689949 &     0.0405343684 &   190.6416032233 &    209.088 &  184\\
II.C$_{239}^{i.c.}$(0.5) &     0.1910600294 &     0.0654138331 &   142.9115531752 &    210.871 &  186\\
II.C$_{240}^{i.c.}$(0.5) &     0.1794185340 &     0.0647107348 &   142.2515470251 &    215.270 &  190\\
II.C$_{241}^{i.c.}$(0.5) &     0.2749406093 &     0.0726374464 &   188.1053317808 &    215.731 &  190\\
II.C$_{242}^{i.c.}$(0.5) &     0.2498197638 &     0.0791829559 &   175.1478065921 &    217.861 &  192\\
II.C$_{243}^{i.c.}$(0.5) &     0.2465045273 &     0.0440978667 &   170.5920826149 &    220.463 &  194\\
II.C$_{244}^{i.c.}$(0.5) &     0.2632419381 &     0.0497931425 &   182.7043137437 &    222.714 &  196\\
II.C$_{245}^{i.c.}$(0.5) &     0.2549695079 &     0.0579934165 &   184.1495051302 &    229.515 &  202\\
II.C$_{246}^{i.c.}$(0.5) &     0.2632833046 &     0.0670452882 &   194.5495654808 &    233.984 &  206\\
II.C$_{247}^{i.c.}$(0.5) &     0.2515183994 &     0.0784843668 &   198.0785646371 &    245.200 &  216\\
II.C$_{248}^{i.c.}$(0.5) &     0.2462955611 &     0.0455335432 &   189.9687042853 &    245.464 &  216\\
II.C$_{249}^{i.c.}$(0.5) &     0.2069465170 &     0.0690747434 &   172.8812870460 &    245.099 &  216\\
II.C$_{250}^{i.c.}$(0.5) &     0.2182720723 &     0.0536172219 &   177.7299779781 &    247.614 &  218\\
II.C$_{251}^{i.c.}$(0.5) &     0.1421095213 &     0.0311726007 &   154.5748576016 &    254.455 &  224\\
II.C$_{252}^{i.c.}$(0.5) &     0.1540640475 &     0.0494956780 &   164.2313972673 &    263.085 &  232\\
II.C$_{253}^{i.c.}$(0.5) &     0.1575278637 &     0.0383819193 &   170.1977938681 &    272.489 &  240\\
II.C$_{254}^{i.c.}$(0.5) &     0.1495219644 &     0.0367502784 &   170.6390914389 &    277.069 &  244\\
II.C$_{255}^{i.c.}$(0.5) &     0.1964997354 &     0.0214843951 &   196.7183026566 &    293.141 &  258\\
\hline
\end{tabular*}
{\rule{\temptablewidth}{1pt}}
\end{center}
\end{table*}

\begin{table*}
\tabcolsep 0pt \caption{Initial conditions and periods $T$ of the periodic three-body orbits for class I.A in the case of  $\bm{r}_1(0)=(-1,0)=-\bm{r}_2(0)$,  $\dot{\bm{r}}_1(0)=(v_1,v_2)=\dot{\bm{r}}_2(0)$ and $\bm{r}_3(0)=(0,0)$, $\dot{\bm{r}}_3(0)=(-2v_1/m_3, -2v_2/m_3)$ when $G=1$ and $m_1=m_2=1$ and $m_3=0.75$ by means of the search grid $4000\times 4000$ in the interval $T_0\in[0,200]$, where $T^*=T |E|^{3/2}$ is its scale-invariant period, $L_f$ is the length of the free group element.  } \label{table-S5} \vspace*{-12pt}
\begin{center}
\def\temptablewidth{1\textwidth}
{\rule{\temptablewidth}{1pt}}
\begin{tabular*}{\temptablewidth}{@{\extracolsep{\fill}}lccccc}
\hline
Class and number  & $v_1$ & $v_2$  & $T$  & $T^*$ & $L_f$\\
\hline
I.A$_{1}^{i.c.}$(0.75) &     0.4227625247 &     0.2533646387 &     5.9858187252 &      6.993 &  4\\
I.A$_{2}^{i.c.}$(0.75) &     0.5337490177 &     0.3041674607 &    25.0460382386 &     12.114 &  8\\
I.A$_{3}^{i.c.}$(0.75) &     0.3662988976 &     0.1741689247 &    17.3598689063 &     28.658 &  16\\
I.A$_{4}^{i.c.}$(0.75) &     0.4329138577 &     0.3286773639 &    30.2157279993 &     26.520 &  16\\
I.A$_{5}^{i.c.}$(0.75) &     0.2658200014 &     0.3876978560 &    20.4071584640 &     26.484 &  16\\
I.A$_{6}^{i.c.}$(0.75) &     0.0848079590 &     0.0632351933 &    10.5323245447 &     28.878 &  16\\
I.A$_{7}^{i.c.}$(0.75) &     0.1632908635 &     0.4709992142 &    36.5167219251 &     41.488 &  24\\
I.A$_{8}^{i.c.}$(0.75) &     0.3986630485 &     0.2098556495 &    30.3699479925 &     42.738 &  24\\
I.A$_{9}^{i.c.}$(0.75) &     0.3972439494 &     0.1439208014 &    32.1901761672 &     50.237 &  28\\
I.A$_{10}^{i.c.}$(0.75) &     0.1902461690 &     0.1147396347 &    20.4870964105 &     50.262 &  28\\
I.A$_{11}^{i.c.}$(0.75) &     0.4062910529 &     0.2262182980 &    42.8004488374 &     56.762 &  32\\
I.A$_{12}^{i.c.}$(0.75) &     0.3444998604 &     0.0249623855 &    29.4488166879 &     57.520 &  32\\
I.A$_{13}^{i.c.}$(0.75) &     0.3588572795 &     0.1923161209 &    39.1877154517 &     64.373 &  36\\
I.A$_{14}^{i.c.}$(0.75) &     0.3585145568 &     0.1395242890 &    40.8139997342 &     71.804 &  40\\
I.A$_{15}^{i.c.}$(0.75) &     0.4042841150 &     0.2375692455 &    54.2573733044 &     70.767 &  40\\
I.A$_{16}^{i.c.}$(0.75) &     0.4356001994 &     0.3176227502 &    82.1366960817 &     74.183 &  44\\
I.A$_{17}^{i.c.}$(0.75) &     0.3992933028 &     0.2021813618 &    55.1053036718 &     78.451 &  44\\
I.A$_{18}^{i.c.}$(0.75) &     0.3465301653 &     0.0611178943 &    41.1425735253 &     79.087 &  44\\
I.A$_{19}^{i.c.}$(0.75) &     0.4747944843 &     0.0277185939 &    62.4983813817 &     79.156 &  44\\
I.A$_{20}^{i.c.}$(0.75) &     0.1788553784 &     0.1250024592 &    34.8754522497 &     86.012 &  48\\
I.A$_{21}^{i.c.}$(0.75) &     0.4123273935 &     0.2418549526 &    67.6579019767 &     84.763 &  48\\
I.A$_{22}^{i.c.}$(0.75) &     0.4002811362 &     0.2152589233 &    66.7743434315 &     92.493 &  52\\
I.A$_{23}^{i.c.}$(0.75) &     0.3570732994 &     0.1330553625 &    52.5165797298 &     93.368 &  52\\
I.A$_{24}^{i.c.}$(0.75) &     0.3204248773 &     0.1988777895 &    55.6667301422 &    100.077 &  56\\
I.A$_{25}^{i.c.}$(0.75) &     0.4200665164 &     0.2432598967 &    81.5599720155 &     98.754 &  56\\
I.A$_{26}^{i.c.}$(0.75) &     0.3477173243 &     0.0739384079 &    52.8403128515 &    100.651 &  56\\
I.A$_{27}^{i.c.}$(0.75) &     0.4043929787 &     0.2232119633 &    79.2711602148 &    106.517 &  60\\
I.A$_{28}^{i.c.}$(0.75) &     0.3642842051 &     0.1636761967 &    63.8886666871 &    107.559 &  60\\
I.A$_{29}^{i.c.}$(0.75) &     0.4181714913 &     0.2478074981 &    93.4047379916 &    112.743 &  64\\
I.A$_{30}^{i.c.}$(0.75) &     0.4353986367 &     0.1115724837 &    81.3432979489 &    114.946 &  64\\
I.A$_{31}^{i.c.}$(0.75) &     0.3561647035 &     0.1287849051 &    64.2195094868 &    114.931 &  64\\
I.A$_{32}^{i.c.}$(0.75) &     0.4012766820 &     0.1981267054 &    80.1768711298 &    114.162 &  64\\
I.A$_{33}^{i.c.}$(0.75) &     0.1901620868 &     0.0896947011 &    46.1543229949 &    115.000 &  64\\
I.A$_{34}^{i.c.}$(0.75) &     0.4064136540 &     0.2296155572 &    91.5695991198 &    120.529 &  68\\
I.A$_{35}^{i.c.}$(0.75) &     0.3662485364 &     0.1831531016 &    74.6480334841 &    121.694 &  68\\
I.A$_{36}^{i.c.}$(0.75) &     0.3484936438 &     0.0810521814 &    64.5396545591 &    122.214 &  68\\
I.A$_{37}^{i.c.}$(0.75) &     0.3625081838 &     0.1561981586 &    75.6201745758 &    129.136 &  72\\
I.A$_{38}^{i.c.}$(0.75) &     0.3555412053 &     0.1257508069 &    75.9229791803 &    136.494 &  76\\
I.A$_{39}^{i.c.}$(0.75) &     0.3490401568 &     0.0856323568 &    76.2397282094 &    143.776 &  80\\
I.A$_{40}^{i.c.}$(0.75) &     0.4023237219 &     0.1953477132 &   105.1410526930 &    149.871 &  84\\
I.A$_{41}^{i.c.}$(0.75) &     0.3368899556 &     0.1858462998 &    89.4831471991 &    157.407 &  88\\
I.A$_{42}^{i.c.}$(0.75) &     0.4037216808 &     0.2221234572 &   115.7611144325 &    156.271 &  88\\
I.A$_{43}^{i.c.}$(0.75) &     0.3637279697 &     0.1884044215 &    96.5977490387 &    157.408 &  88\\
I.A$_{44}^{i.c.}$(0.75) &     0.3550873014 &     0.1234828055 &    87.6269195614 &    158.055 &  88\\
I.A$_{45}^{i.c.}$(0.75) &     0.3651069701 &     0.1674068653 &    98.6376062555 &    164.877 &  92\\
I.A$_{46}^{i.c.}$(0.75) &     0.3984203610 &     0.2064633948 &   115.6749862554 &    163.928 &  92\\
I.A$_{47}^{i.c.}$(0.75) &     0.4262489553 &     0.0672054586 &   109.3699499056 &    165.348 &  92\\
I.A$_{48}^{i.c.}$(0.75) &     0.3494454129 &     0.0888406710 &    87.9401405108 &    165.338 &  92\\
I.A$_{49}^{i.c.}$(0.75) &     0.4164619789 &     0.2419173608 &   143.2372283584 &    176.522 &  100\\
I.A$_{50}^{i.c.}$(0.75) &     0.3547423337 &     0.1217229642 &    99.3312692044 &    179.617 &  100\\

\hline
\end{tabular*}
{\rule{\temptablewidth}{1pt}}
\end{center}
\end{table*}

\begin{table*}
\tabcolsep 0pt \caption{Initial conditions and periods $T$ of the periodic three-body orbits for class I.A in the case of  $\bm{r}_1(0)=(-1,0)=-\bm{r}_2(0)$,  $\dot{\bm{r}}_1(0)=(v_1,v_2)=\dot{\bm{r}}_2(0)$ and $\bm{r}_3(0)=(0,0)$, $\dot{\bm{r}}_3(0)=(-2v_1/m_3, -2v_2/m_3)$ when $G=1$ and $m_1=m_2=1$ and $m_3=0.75$ by means of the search grid $4000\times 4000$ in the interval $T_0\in[0,200]$, where $T^*=T |E|^{3/2}$ is its scale-invariant period, $L_f$ is the length of the free group element.  } \label{table-S5} \vspace*{-12pt}
\begin{center}
\def\temptablewidth{1\textwidth}
{\rule{\temptablewidth}{1pt}}
\begin{tabular*}{\temptablewidth}{@{\extracolsep{\fill}}lccccc}
\hline
Class and number  & $v_1$ & $v_2$  & $T$  & $T^*$ & $L_f$\\
\hline
I.A$_{51}^{i.c.}$(0.75) &     0.3666229117 &     0.1802557053 &   109.4686390383 &    179.012 &  100\\
I.A$_{52}^{i.c.}$(0.75) &     0.4034422709 &     0.2320847945 &   139.2068145953 &    184.296 &  104\\
I.A$_{53}^{i.c.}$(0.75) &     0.4017850173 &     0.1937496080 &   129.5985275936 &    185.581 &  104\\
I.A$_{54}^{i.c.}$(0.75) &     0.1878826213 &     0.1228511711 &    76.2757360731 &    186.545 &  104\\
I.A$_{55}^{i.c.}$(0.75) &     0.3638942950 &     0.1619913559 &   110.3963190576 &    186.458 &  104\\
I.A$_{56}^{i.c.}$(0.75) &     0.3497577254 &     0.0912169635 &    99.6406881029 &    186.899 &  104\\
I.A$_{57}^{i.c.}$(0.75) &     0.3477310897 &     0.1903660735 &   113.6756165257 &    193.118 &  108\\
I.A$_{58}^{i.c.}$(0.75) &     0.4012296139 &     0.2174859716 &   139.6773663749 &    192.002 &  108\\
I.A$_{59}^{i.c.}$(0.75) &     0.3592003644 &     0.1424958332 &   110.7390745623 &    193.846 &  108\\
I.A$_{60}^{i.c.}$(0.75) &     0.3453188013 &     0.0436341225 &   100.0379934281 &    194.129 &  108\\
I.A$_{61}^{i.c.}$(0.75) &     0.3987295714 &     0.2040518543 &   140.4203692852 &    199.641 &  112\\
I.A$_{62}^{i.c.}$(0.75) &     0.3544714361 &     0.1203177331 &   111.0359806099 &    201.178 &  112\\
I.A$_{63}^{i.c.}$(0.75) &     0.4027598089 &     0.1099065756 &   126.7174189742 &    201.180 &  112\\
I.A$_{64}^{i.c.}$(0.75) &     0.3085492973 &     0.1235638992 &    99.8775281408 &    201.180 &  112\\
I.A$_{65}^{i.c.}$(0.75) &     0.3500056587 &     0.0930489404 &   111.3412553208 &    208.461 &  116\\
I.A$_{66}^{i.c.}$(0.75) &     0.3659182597 &     0.1843854011 &   131.8265841519 &    214.729 &  120\\
I.A$_{67}^{i.c.}$(0.75) &     0.3459823155 &     0.0539982359 &   111.7320070042 &    215.696 &  120\\
I.A$_{68}^{i.c.}$(0.75) &     0.1906186812 &     0.1029059018 &    87.2101531089 &    215.531 &  120\\
I.A$_{69}^{i.c.}$(0.75) &     0.3894810083 &     0.0456452590 &   125.3277220883 &    215.697 &  120\\
I.A$_{70}^{i.c.}$(0.75) &     0.3654656585 &     0.1691775243 &   133.3741145779 &    222.194 &  124\\
I.A$_{71}^{i.c.}$(0.75) &     0.3714351327 &     0.1973698603 &   145.6730252269 &    228.826 &  128\\
I.A$_{72}^{i.c.}$(0.75) &     0.2497702662 &     0.0946546767 &   100.3713839877 &    230.057 &  128\\
I.A$_{73}^{i.c.}$(0.75) &     0.3619204754 &     0.1537575193 &   133.8460648608 &    229.609 &  128\\
I.A$_{74}^{i.c.}$(0.75) &     0.3504692196 &     0.1957352300 &   136.7945571599 &    228.826 &  128\\
I.A$_{75}^{i.c.}$(0.75) &     0.3502071722 &     0.0945047358 &   123.0417715598 &    230.022 &  128\\
I.A$_{76}^{i.c.}$(0.75) &     0.3579495351 &     0.1370259459 &   134.1447275594 &    236.977 &  132\\
I.A$_{77}^{i.c.}$(0.75) &     0.3540736229 &     0.1182148529 &   134.4463621214 &    244.301 &  136\\
I.A$_{78}^{i.c.}$(0.75) &     0.3644051317 &     0.1875040842 &   153.7854226046 &    250.442 &  140\\
I.A$_{79}^{i.c.}$(0.75) &     0.3503741196 &     0.0956894365 &   134.7421912222 &    251.583 &  140\\
I.A$_{80}^{i.c.}$(0.75) &     0.2850469461 &     0.1237066414 &   125.9073062746 &    265.878 &  148\\
I.A$_{81}^{i.c.}$(0.75) &     0.3604411206 &     0.1477540307 &   157.2539569259 &    272.751 &  152\\
I.A$_{82}^{i.c.}$(0.75) &     0.2623247173 &     0.0630697393 &   122.9004467283 &    280.406 &  156\\
I.A$_{83}^{i.c.}$(0.75) &     0.3537959008 &     0.1167175132 &   157.8578914671 &    287.422 &  160\\
I.A$_{84}^{i.c.}$(0.75) &     0.3404668705 &     0.1788234152 &   166.8446546886 &    293.645 &  164\\
I.A$_{85}^{i.c.}$(0.75) &     0.3506344970 &     0.0975004375 &   158.1426267292 &    294.705 &  164\\
I.A$_{86}^{i.c.}$(0.75) &     0.3649683941 &     0.1667479259 &   179.9043199632 &    301.096 &  168\\
I.A$_{87}^{i.c.}$(0.75) &     0.1936102478 &     0.0742911296 &   123.6422146647 &    309.182 &  172\\
I.A$_{88}^{i.c.}$(0.75) &     0.2440200098 &     0.1100378530 &   135.0238328296 &    309.180 &  172\\
I.A$_{89}^{i.c.}$(0.75) &     0.3450126559 &     0.0378080014 &   158.9350378593 &    309.170 &  172\\
I.A$_{90}^{i.c.}$(0.75) &     0.1906314309 &     0.1067262314 &   128.1965967535 &    316.058 &  176\\
I.A$_{91}^{i.c.}$(0.75) &     0.3507379026 &     0.0982077679 &   169.8426052773 &    316.265 &  176\\
I.A$_{92}^{i.c.}$(0.75) &     0.3564275786 &     0.1300379619 &   180.9554087549 &    323.231 &  180\\
I.A$_{93}^{i.c.}$(0.75) &     0.3480099279 &     0.0767141118 &   170.2199214581 &    323.516 &  180\\
I.A$_{94}^{i.c.}$(0.75) &     0.2806075517 &     0.1252726560 &   155.4447267514 &    330.581 &  184\\
I.A$_{95}^{i.c.}$(0.75) &     0.3616931000 &     0.1527867827 &   192.0738064899 &    330.082 &  184\\
I.A$_{96}^{i.c.}$(0.75) &     0.3335530219 &     0.1439508032 &   180.7729773204 &    337.454 &  188\\
I.A$_{97}^{i.c.}$(0.75) &     0.3589093395 &     0.1412419876 &   192.3672014822 &    337.454 &  188\\
I.A$_{98}^{i.c.}$(0.75) &     0.3482666391 &     0.0790526219 &   181.9192922993 &    345.079 &  192\\
I.A$_{99}^{i.c.}$(0.75) &     0.3048177204 &     0.0814603730 &   175.3772678451 &    366.643 &  204\\
I.A$_{100}^{i.c.}$(0.75) &     0.1916766819 &     0.0708884572 &   148.9747140897 &    373.907 &  208\\
I.A$_{101}^{i.c.}$(0.75) &     0.1906790534 &     0.1012235584 &   153.9336180218 &    380.799 &  212\\
I.A$_{102}^{i.c.}$(0.75) &     0.1752778915 &     0.0747834674 &   186.0099259874 &    474.459 &  264\\
\hline
\end{tabular*}
{\rule{\temptablewidth}{1pt}}
\end{center}
\end{table*}

\begin{table*}
\tabcolsep 0pt \caption{Initial conditions and periods $T$ of the periodic three-body orbits for class I.B in the case of  $\bm{r}_1(0)=(-1,0)=-\bm{r}_2(0)$,  $\dot{\bm{r}}_1(0)=(v_1,v_2)=\dot{\bm{r}}_2(0)$ and $\bm{r}_3(0)=(0,0)$, $\dot{\bm{r}}_3(0)=(-2v_1/m_3, -2v_2/m_3)$ when $G=1$ and $m_1=m_2=1$ and $m_3=0.75$ by means of the search grid $4000\times 4000$ in the interval $T_0\in[0,200]$, where $T^*=T |E|^{3/2}$ is its scale-invariant period, $L_f$ is the length of the free group element.  } \label{table-S5} \vspace*{-12pt}
\begin{center}
\def\temptablewidth{1\textwidth}
{\rule{\temptablewidth}{1pt}}
\begin{tabular*}{\temptablewidth}{@{\extracolsep{\fill}}lccccc}
\hline
Class and number  & $v_1$ & $v_2$  & $T$  & $T^*$ & $L_f$\\
\hline
I.B$_{1}^{i.c.}$(0.75) &     0.3955115480 &     0.1946666519 &    12.2220545916 &     17.855 &  10\\
I.B$_{2}^{i.c.}$(0.75) &     0.1652925912 &     0.4505132187 &    19.7241358099 &     24.503 &  14\\
I.B$_{3}^{i.c.}$(0.75) &     0.1233328251 &     0.2289900725 &    13.7490351096 &     31.883 &  18\\
I.B$_{4}^{i.c.}$(0.75) &     0.2402669908 &     0.2588885318 &    16.6376620580 &     31.876 &  18\\
I.B$_{5}^{i.c.}$(0.75) &     0.4765383456 &     0.2017941573 &    31.0357663010 &     31.879 &  18\\
I.B$_{6}^{i.c.}$(0.75) &     0.1040025375 &     0.3982558829 &    17.9396240903 &     29.044 &  18\\
I.B$_{7}^{i.c.}$(0.75) &     0.4371771293 &     0.2657669634 &    35.6549923918 &     37.828 &  22\\
I.B$_{8}^{i.c.}$(0.75) &     0.3633412780 &     0.1596559094 &    23.2527357070 &     39.449 &  22\\
I.B$_{9}^{i.c.}$(0.75) &     0.3162906618 &     0.2714988226 &    28.8310339842 &     45.873 &  26\\
I.B$_{10}^{i.c.}$(0.75) &     0.4933323280 &     0.2064499251 &    49.4361887730 &     45.872 &  26\\
I.B$_{11}^{i.c.}$(0.75) &     0.4158908017 &     0.2496571698 &    49.3642077739 &     59.869 &  34\\
I.B$_{12}^{i.c.}$(0.75) &     0.4970901818 &     0.2121794524 &    66.8568048178 &     59.855 &  34\\
I.B$_{13}^{i.c.}$(0.75) &     0.3596003749 &     0.1442041184 &    34.9624207563 &     61.021 &  34\\
I.B$_{14}^{i.c.}$(0.75) &     0.3996784534 &     0.2136076584 &    48.5583118422 &     67.616 &  38\\
I.B$_{15}^{i.c.}$(0.75) &     0.3456671473 &     0.0493751273 &    35.2949160577 &     68.304 &  38\\
I.B$_{16}^{i.c.}$(0.75) &     0.4782489370 &     0.2264695367 &    76.8933412890 &     73.834 &  42\\
I.B$_{17}^{i.c.}$(0.75) &     0.3665299184 &     0.1814212787 &    46.0361237055 &     75.177 &  42\\
I.B$_{18}^{i.c.}$(0.75) &     0.1904438490 &     0.0986924753 &    33.3439218167 &     82.633 &  46\\
I.B$_{19}^{i.c.}$(0.75) &     0.3577021441 &     0.1359199549 &    46.6652691512 &     82.586 &  46\\
I.B$_{20}^{i.c.}$(0.75) &     0.4050030711 &     0.2242764227 &    61.0244495835 &     81.640 &  46\\
I.B$_{21}^{i.c.}$(0.75) &     0.3471929158 &     0.0686233513 &    46.9911527301 &     89.869 &  50\\
I.B$_{22}^{i.c.}$(0.75) &     0.3653163022 &     0.1684258905 &    58.0034052587 &     96.768 &  54\\
I.B$_{23}^{i.c.}$(0.75) &     0.3565724828 &     0.1307224452 &    58.3679719678 &    104.150 &  58\\
I.B$_{24}^{i.c.}$(0.75) &     0.3984106274 &     0.2071944713 &    73.0101745147 &    103.333 &  58\\
I.B$_{25}^{i.c.}$(0.75) &     0.3627539276 &     0.1894607450 &    67.9677626251 &    110.890 &  62\\
I.B$_{26}^{i.c.}$(0.75) &     0.4903485127 &     0.1991992446 &   114.2530489267 &    109.638 &  62\\
I.B$_{27}^{i.c.}$(0.75) &     0.3179795403 &     0.2629379702 &    67.9432993586 &    109.638 &  62\\
I.B$_{28}^{i.c.}$(0.75) &     0.4061419103 &     0.2382710400 &    84.7867751755 &    109.650 &  62\\
I.B$_{29}^{i.c.}$(0.75) &     0.3481423472 &     0.0779310702 &    58.6898548060 &    111.433 &  62\\
I.B$_{30}^{i.c.}$(0.75) &     0.1856050918 &     0.1204393057 &    48.2038106270 &    118.404 &  66\\
I.B$_{31}^{i.c.}$(0.75) &     0.4006546101 &     0.2161876360 &    84.9972461438 &    117.371 &  66\\
I.B$_{32}^{i.c.}$(0.75) &     0.4112452436 &     0.2412703817 &    98.1488111526 &    123.646 &  70\\
I.B$_{33}^{i.c.}$(0.75) &     0.3063648281 &     0.2705939680 &    75.6551445192 &    123.631 &  70\\
I.B$_{34}^{i.c.}$(0.75) &     0.3558263642 &     0.1271496463 &    70.0711814814 &    125.713 &  70\\
I.B$_{35}^{i.c.}$(0.75) &     0.4039901561 &     0.2225640639 &    97.5141753629 &    131.394 &  74\\
I.B$_{36}^{i.c.}$(0.75) &     0.3666436348 &     0.1782820654 &    80.8001762843 &    132.494 &  74\\
I.B$_{37}^{i.c.}$(0.75) &     0.3487887762 &     0.0835642922 &    70.3896308512 &    132.995 &  74\\
I.B$_{38}^{i.c.}$(0.75) &     0.4037169648 &     0.0717293035 &    81.7272612361 &    132.997 &  74\\
I.B$_{39}^{i.c.}$(0.75) &     0.3988539052 &     0.2035444203 &    97.7542523288 &    139.046 &  78\\
I.B$_{40}^{i.c.}$(0.75) &     0.3617854786 &     0.1531862240 &    81.4797594757 &    139.923 &  78\\
I.B$_{41}^{i.c.}$(0.75) &     0.1897966977 &     0.0838158904 &    58.9391568318 &    147.365 &  82\\
I.B$_{42}^{i.c.}$(0.75) &     0.4068544802 &     0.2273814344 &   110.1366017255 &    145.408 &  82\\
I.B$_{43}^{i.c.}$(0.75) &     0.4408681230 &     0.1057452927 &   105.8649119093 &    147.299 &  82\\
I.B$_{44}^{i.c.}$(0.75) &     0.3530121154 &     0.1949096882 &    88.1427535972 &    146.599 &  82\\
I.B$_{45}^{i.c.}$(0.75) &     0.3552976675 &     0.1245404438 &    81.7748946257 &    147.275 &  82\\
I.B$_{46}^{i.c.}$(0.75) &     0.3990215187 &     0.2115568486 &   109.2687466676 &    153.092 &  86\\
I.B$_{47}^{i.c.}$(0.75) &     0.1863329193 &     0.1272516263 &    63.1308826408 &    154.153 &  86\\
I.B$_{48}^{i.c.}$(0.75) &     0.3527830404 &     0.1717732793 &    89.5915066227 &    154.085 &  86\\
I.B$_{49}^{i.c.}$(0.75) &     0.3265683649 &     0.2517165071 &    98.6763921324 &    159.383 &  90\\
I.B$_{50}^{i.c.}$(0.75) &     0.3605452650 &     0.1481927024 &    93.1815353179 &    161.494 &  90\\
I.B$_{51}^{i.c.}$(0.75) &     0.3660526470 &     0.1839372430 &   103.2412757962 &    168.211 &  94\\
I.B$_{52}^{i.c.}$(0.75) &     0.4010926159 &     0.2171880829 &   121.4505154029 &    167.125 &  94\\
I.B$_{53}^{i.c.}$(0.75) &     0.3549037937 &     0.1225506762 &    93.4790465414 &    168.836 &  94\\
I.B$_{54}^{i.c.}$(0.75) &     0.3997914828 &     0.2009996337 &   122.7109281615 &    174.757 &  98\\
I.B$_{55}^{i.c.}$(0.75) &     0.3644847369 &     0.1645612391 &   104.5220019814 &    175.668 &  98\\
I.B$_{56}^{i.c.}$(0.75) &     0.3496110783 &     0.0901110425 &    93.7904060164 &    176.119 &  98\\
I.B$_{57}^{i.c.}$(0.75) &     0.3297350406 &     0.1911985943 &   102.5698998873 &    182.305 &  102\\
I.B$_{58}^{i.c.}$(0.75) &     0.4035305311 &     0.2218058040 &   134.0109575462 &    181.148 &  102\\
I.B$_{59}^{i.c.}$(0.75) &     0.4101378717 &     0.1341894173 &   121.0976361440 &    183.067 &  102\\
I.B$_{60}^{i.c.}$(0.75) &     0.1896138824 &     0.1072260658 &    74.2291889491 &    183.160 &  102\\

\hline
\end{tabular*}
{\rule{\temptablewidth}{1pt}}
\end{center}
\end{table*}

\begin{table*}
\tabcolsep 0pt \caption{Initial conditions and periods $T$ of the periodic three-body orbits for class I.B in the case of  $\bm{r}_1(0)=(-1,0)=-\bm{r}_2(0)$,  $\dot{\bm{r}}_1(0)=(v_1,v_2)=\dot{\bm{r}}_2(0)$ and $\bm{r}_3(0)=(0,0)$, $\dot{\bm{r}}_3(0)=(-2v_1/m_3, -2v_2/m_3)$ when $G=1$ and $m_1=m_2=1$ and $m_3=0.75$ by means of the search grid $4000\times 4000$ in the interval $T_0\in[0,200]$, where $T^*=T |E|^{3/2}$ is its scale-invariant period, $L_f$ is the length of the free group element.  } \label{table-S5} \vspace*{-12pt}
\begin{center}
\def\temptablewidth{1\textwidth}
{\rule{\temptablewidth}{1pt}}
\begin{tabular*}{\temptablewidth}{@{\extracolsep{\fill}}lccccc}
\hline
Class and number  & $v_1$ & $v_2$  & $T$  & $T^*$ & $L_f$\\
\hline
I.B$_{61}^{i.c.}$(0.75) &     0.3654676797 &     0.2013570084 &   114.7433059587 &    182.305 &  102\\
I.B$_{62}^{i.c.}$(0.75) &     0.3449318936 &     0.0361030611 &    94.1918485647 &    183.345 &  102\\
I.B$_{63}^{i.c.}$(0.75) &     0.3415958742 &     0.1781262476 &   108.0595164221 &    189.810 &  106\\
I.B$_{64}^{i.c.}$(0.75) &     0.4877782970 &     0.2017718810 &   193.6578545086 &    187.395 &  106\\
I.B$_{65}^{i.c.}$(0.75) &     0.3984752813 &     0.2084246515 &   133.7314062042 &    188.808 &  106\\
I.B$_{66}^{i.c.}$(0.75) &     0.3665861978 &     0.1770429252 &   115.5336619786 &    189.810 &  106\\
I.B$_{67}^{i.c.}$(0.75) &     0.3545991926 &     0.1209830667 &   105.1835820780 &    190.398 &  106\\
I.B$_{68}^{i.c.}$(0.75) &     0.2542763689 &     0.0906969814 &    86.6765732076 &    197.702 &  110\\
I.B$_{69}^{i.c.}$(0.75) &     0.4385158820 &     0.0640412131 &   136.5399121956 &    197.702 &  110\\
I.B$_{70}^{i.c.}$(0.75) &     0.3498884353 &     0.0921885496 &   105.4909744219 &    197.680 &  110\\
I.B$_{71}^{i.c.}$(0.75) &     0.3195165115 &     0.2668419229 &   126.3377271874 &    201.388 &  114\\
I.B$_{72}^{i.c.}$(0.75) &     0.3588402297 &     0.1409427418 &   116.5906134693 &    204.629 &  114\\
I.B$_{73}^{i.c.}$(0.75) &     0.1882107801 &     0.0759993295 &    84.3610994604 &    212.092 &  118\\
I.B$_{74}^{i.c.}$(0.75) &     0.3543567239 &     0.1197162571 &   116.8884609087 &    211.959 &  118\\
I.B$_{75}^{i.c.}$(0.75) &     0.3789094561 &     0.2014477498 &   142.9781422625 &    218.012 &  122\\
I.B$_{76}^{i.c.}$(0.75) &     0.3263199440 &     0.2689451474 &   138.1775267185 &    215.378 &  122\\
I.B$_{77}^{i.c.}$(0.75) &     0.3861811965 &     0.1515735137 &   136.6224402902 &    218.823 &  122\\
I.B$_{78}^{i.c.}$(0.75) &     0.3501113687 &     0.0938162205 &   117.1915230963 &    219.241 &  122\\
I.B$_{79}^{i.c.}$(0.75) &     0.3368216321 &     0.1814605343 &   127.4112970385 &    225.529 &  126\\
I.B$_{80}^{i.c.}$(0.75) &     0.3582187899 &     0.1382234521 &   128.2933049194 &    226.194 &  126\\
I.B$_{81}^{i.c.}$(0.75) &     0.3462687542 &     0.0578449911 &   117.5796866317 &    226.478 &  126\\
I.B$_{82}^{i.c.}$(0.75) &     0.2922389186 &     0.1233632111 &   112.1079121569 &    233.528 &  130\\
I.B$_{83}^{i.c.}$(0.75) &     0.3650176895 &     0.1669814530 &   139.2710431601 &    232.987 &  130\\
I.B$_{84}^{i.c.}$(0.75) &     0.1892866683 &     0.0221911859 &    94.4736496681 &    240.971 &  134\\
I.B$_{85}^{i.c.}$(0.75) &     0.3467696508 &     0.0639486684 &   129.2760551437 &    248.043 &  138\\
I.B$_{86}^{i.c.}$(0.75) &     0.3665370880 &     0.1763780053 &   150.2600860454 &    247.126 &  138\\
I.B$_{87}^{i.c.}$(0.75) &     0.4083790203 &     0.0486715665 &   153.0421639861 &    248.051 &  138\\
I.B$_{88}^{i.c.}$(0.75) &     0.3690686913 &     0.1623965482 &   153.0450189807 &    254.568 &  142\\
I.B$_{89}^{i.c.}$(0.75) &     0.3539953094 &     0.1177951302 &   140.2991413854 &    255.081 &  142\\
I.B$_{90}^{i.c.}$(0.75) &     0.3504472774 &     0.0962027008 &   140.5923548014 &    262.363 &  146\\
I.B$_{91}^{i.c.}$(0.75) &     0.3607597004 &     0.1491004927 &   151.3989526999 &    261.966 &  146\\
I.B$_{92}^{i.c.}$(0.75) &     0.3238354084 &     0.1371451748 &   140.0376600674 &    269.324 &  150\\
I.B$_{93}^{i.c.}$(0.75) &     0.3572663963 &     0.1339427411 &   151.6983734688 &    269.323 &  150\\
I.B$_{94}^{i.c.}$(0.75) &     0.3538571400 &     0.1170498280 &   152.0049074323 &    276.642 &  154\\
I.B$_{95}^{i.c.}$(0.75) &     0.1558167527 &     0.0551420353 &   105.7114373172 &    276.818 &  154\\
I.B$_{96}^{i.c.}$(0.75) &     0.3116737151 &     0.2686682600 &   172.3506257247 &    279.145 &  158\\
I.B$_{97}^{i.c.}$(0.75) &     0.3601419243 &     0.1464946578 &   163.1073152563 &    283.535 &  158\\
I.B$_{98}^{i.c.}$(0.75) &     0.3666366005 &     0.1799469189 &   172.8962301998 &    282.847 &  158\\
I.B$_{99}^{i.c.}$(0.75) &     0.1903231208 &     0.1095157688 &   115.2343315677 &    283.684 &  158\\
I.B$_{100}^{i.c.}$(0.75) &     0.3954025458 &     0.0884270772 &   171.7557975683 &    283.926 &  158\\
I.B$_{101}^{i.c.}$(0.75) &     0.2741018535 &     0.2554039554 &   158.4790653853 &    286.886 &  162\\
I.B$_{102}^{i.c.}$(0.75) &     0.3475551472 &     0.0723444026 &   152.6716275805 &    291.172 &  162\\
I.B$_{103}^{i.c.}$(0.75) &     0.2818186464 &     0.1243991912 &   140.4484779108 &    298.229 &  166\\
I.B$_{104}^{i.c.}$(0.75) &     0.3626205078 &     0.1566624854 &   174.4937355907 &    297.722 &  166\\
I.B$_{105}^{i.c.}$(0.75) &     0.3027698354 &     0.0259686608 &   139.3368817906 &    298.387 &  166\\
I.B$_{106}^{i.c.}$(0.75) &     0.1894542670 &     0.0321508176 &   120.0636229967 &    305.697 &  170\\
I.B$_{107}^{i.c.}$(0.75) &     0.2539168728 &     0.0672679036 &   135.4481207778 &    312.762 &  174\\
I.B$_{108}^{i.c.}$(0.75) &     0.1902702982 &     0.0922368820 &   125.6625896697 &    312.635 &  174\\
I.B$_{109}^{i.c.}$(0.75) &     0.3619799733 &     0.1540068560 &   186.2125683041 &    319.296 &  178\\
I.B$_{110}^{i.c.}$(0.75) &     0.3663329795 &     0.1827419446 &   195.3412841176 &    318.565 &  178\\
I.B$_{111}^{i.c.}$(0.75) &     0.3660548095 &     0.1724827017 &   196.9056013811 &    326.034 &  182\\
I.B$_{112}^{i.c.}$(0.75) &     0.3507844479 &     0.0985240185 &   175.6925290041 &    327.046 &  182\\
I.B$_{113}^{i.c.}$(0.75) &     0.3044228733 &     0.1358113832 &   166.1997019758 &    334.023 &  186\\
I.B$_{114}^{i.c.}$(0.75) &     0.2542619684 &     0.1152551288 &   152.1697588812 &    341.537 &  190\\
I.B$_{115}^{i.c.}$(0.75) &     0.3560451335 &     0.1282099353 &   198.5101580823 &    355.575 &  198\\
I.B$_{116}^{i.c.}$(0.75) &     0.2804847900 &     0.1262187708 &   170.7534255757 &    362.934 &  202\\
I.B$_{117}^{i.c.}$(0.75) &     0.3534701962 &     0.1149289573 &   198.8305779481 &    362.885 &  202\\
I.B$_{118}^{i.c.}$(0.75) &     0.2555646230 &     0.0547119924 &   184.8709866273 &    427.814 &  238\\
I.B$_{119}^{i.c.}$(0.75) &     0.2195606596 &     0.0417898885 &   192.3469350978 &    471.042 &  262\\
\hline
\end{tabular*}
{\rule{\temptablewidth}{1pt}}
\end{center}
\end{table*}

\begin{table*}
\tabcolsep 0pt \caption{Initial conditions and periods $T$ of the periodic three-body orbits for class I.C, II.A, II.B and II.C in the case of  $\bm{r}_1(0)=(-1,0)=-\bm{r}_2(0)$,  $\dot{\bm{r}}_1(0)=(v_1,v_2)=\dot{\bm{r}}_2(0)$ and $\bm{r}_3(0)=(0,0)$, $\dot{\bm{r}}_3(0)=(-2v_1/m_3, -2v_2/m_3)$ when $G=1$ and $m_1=m_2=1$ and $m_3=0.75$ by means of the search grid $4000\times 4000$ in the interval $T_0\in[0,200]$, where $T^*=T |E|^{3/2}$ is its scale-invariant period, $L_f$ is the length of the free group element.  } \label{table-S5} \vspace*{-12pt}
\begin{center}
\def\temptablewidth{1\textwidth}
{\rule{\temptablewidth}{1pt}}
\begin{tabular*}{\temptablewidth}{@{\extracolsep{\fill}}lccccc}
\hline
Class and number  & $v_1$ & $v_2$  & $T$  & $T^*$ & $L_f$\\
\hline
I.C$_{1}^{i.c.}$(0.75) &     0.3492568012 &     0.0873658756 &   164.1798151269 &    309.114 &  172\\
II.A$_{1}^{i.c.}$(0.75) &     0.0457217669 &     0.4155107550 &    35.7782521925 &     56.700 &  32\\
II.A$_{2}^{i.c.}$(0.75) &     0.3662039254 &     0.1902792904 &   177.3768529401 &    286.154 &  160\\
II.B$_{1}^{i.c.}$(0.75) &     0.4381285967 &     0.3220362889 &    49.4213483013 &     43.320 &  26\\
II.B$_{2}^{i.c.}$(0.75) &     0.0101896437 &     0.4481073728 &    38.7358240153 &     55.005 &  30\\
II.B$_{3}^{i.c.}$(0.75) &     0.4307203115 &     0.3088925608 &    60.0883362355 &     57.396 &  34\\
II.B$_{4}^{i.c.}$(0.75) &     0.4948733005 &     0.2916775784 &    89.3424667418 &     62.744 &  38\\
II.B$_{5}^{i.c.}$(0.75) &     0.2400776886 &     0.2653522221 &    84.1773067006 &    159.382 &  90\\
II.B$_{6}^{i.c.}$(0.75) &     0.4923254515 &     0.2163677765 &   181.8030296771 &    165.584 &  94\\
II.B$_{7}^{i.c.}$(0.75) &     0.3303505354 &     0.2593875979 &   119.0524138670 &    187.395 &  106\\
II.B$_{8}^{i.c.}$(0.75) &     0.2917672411 &     0.2642502660 &   130.2350131232 &    223.133 &  126\\
II.C$_{1}^{i.c.}$(0.75) &     0.4035774909 &     0.4506146940 &    40.5609005922 &     21.662 &  14\\
II.C$_{2}^{i.c.}$(0.75) &     0.1324227283 &     0.5620538107 &    36.1934395342 &     24.808 &  14\\
II.C$_{3}^{i.c.}$(0.75) &     0.1341111080 &     0.4453950464 &    24.2244479627 &     32.110 &  18\\
II.C$_{4}^{i.c.}$(0.75) &     0.4517904481 &     0.3984638333 &    48.3454418351 &     26.479 &  18\\
II.C$_{5}^{i.c.}$(0.75) &     0.2393043211 &     0.5033839293 &    42.5998505567 &     34.028 &  20\\
II.C$_{6}^{i.c.}$(0.75) &     0.1116511450 &     0.2453018959 &    15.2915537330 &     34.906 &  20\\
II.C$_{7}^{i.c.}$(0.75) &     0.3480362654 &     0.3387435791 &    28.8665011010 &     34.911 &  20\\
II.C$_{8}^{i.c.}$(0.75) &     0.0786239428 &     0.2174331235 &    14.4096938296 &     34.914 &  20\\
II.C$_{9}^{i.c.}$(0.75) &     0.0949078347 &     0.0861318506 &    14.6316164321 &     39.529 &  22\\
II.C$_{10}^{i.c.}$(0.75) &     0.5663074233 &     0.2564527847 &    81.3382389484 &     36.201 &  22\\
II.C$_{11}^{i.c.}$(0.75) &     0.3133201544 &     0.2667139185 &    24.0027591665 &     38.878 &  22\\
II.C$_{12}^{i.c.}$(0.75) &     0.2776333523 &     0.3754936101 &    27.4777277295 &     36.138 &  22\\
II.C$_{13}^{i.c.}$(0.75) &     0.3979273741 &     0.4420643033 &    52.2403095602 &     30.782 &  22\\
II.C$_{14}^{i.c.}$(0.75) &     0.2337646460 &     0.5184766402 &    56.5821385868 &     41.552 &  24\\
II.C$_{15}^{i.c.}$(0.75) &     0.3880716012 &     0.2990924298 &    41.2462573723 &     48.876 &  28\\
II.C$_{16}^{i.c.}$(0.75) &     0.4865292170 &     0.3089091578 &    66.3906894676 &     45.926 &  28\\
II.C$_{17}^{i.c.}$(0.75) &     0.1104500350 &     0.0419264385 &    19.8549898630 &     54.017 &  30\\
II.C$_{18}^{i.c.}$(0.75) &     0.4882350663 &     0.2165892200 &    56.7379512652 &     52.865 &  30\\
II.C$_{19}^{i.c.}$(0.75) &     0.4740333642 &     0.2953568360 &    63.5896788482 &     50.380 &  30\\
II.C$_{20}^{i.c.}$(0.75) &     0.1657977156 &     0.2455083068 &    26.0997008776 &     56.742 &  32\\
II.C$_{21}^{i.c.}$(0.75) &     0.1029343451 &     0.0959035704 &    21.4416761259 &     57.375 &  32\\
II.C$_{22}^{i.c.}$(0.75) &     0.3642933566 &     0.2731827113 &    43.3607662662 &     59.855 &  34\\
II.C$_{23}^{i.c.}$(0.75) &     0.4263629526 &     0.2972052504 &    56.5834451288 &     57.398 &  34\\
II.C$_{24}^{i.c.}$(0.75) &     0.2599355931 &     0.2563418408 &    34.3131840064 &     63.752 &  36\\
II.C$_{25}^{i.c.}$(0.75) &     0.1081605506 &     0.2250738084 &    27.0471407539 &     63.765 &  36\\
II.C$_{26}^{i.c.}$(0.75) &     0.3345621166 &     0.2805925385 &    45.0497509698 &     66.843 &  38\\
II.C$_{27}^{i.c.}$(0.75) &     0.4559070521 &     0.2389218004 &    64.0813709053 &     66.847 &  38\\
II.C$_{28}^{i.c.}$(0.75) &     0.4200713008 &     0.2503863362 &    56.1760930747 &     66.862 &  38\\
II.C$_{29}^{i.c.}$(0.75) &     0.1657156665 &     0.1155208590 &    27.0714448548 &     68.140 &  38\\
II.C$_{30}^{i.c.}$(0.75) &     0.2108168025 &     0.4566831086 &    61.2686739131 &     68.034 &  40\\
II.C$_{31}^{i.c.}$(0.75) &     0.3297304506 &     0.2585199450 &    44.7971548880 &     70.759 &  40\\
II.C$_{32}^{i.c.}$(0.75) &     0.4597689840 &     0.1105992605 &    56.0808786052 &     71.891 &  40\\
II.C$_{33}^{i.c.}$(0.75) &     0.4455245438 &     0.2894791201 &    70.8609796509 &     67.167 &  40\\
II.C$_{34}^{i.c.}$(0.75) &     0.2177623204 &     0.4365092717 &    54.8016637683 &     65.608 &  40\\
II.C$_{35}^{i.c.}$(0.75) &     0.3610750806 &     0.4162413089 &    81.4888700221 &     68.038 &  40\\
II.C$_{36}^{i.c.}$(0.75) &     0.3211861550 &     0.1095009097 &    38.0767038800 &     75.462 &  42\\
II.C$_{37}^{i.c.}$(0.75) &     0.4799903735 &     0.2090171894 &    78.3383579606 &     77.757 &  44\\
II.C$_{38}^{i.c.}$(0.75) &     0.3984873246 &     0.3859118015 &    95.4378576535 &     77.672 &  46\\
II.C$_{39}^{i.c.}$(0.75) &     0.3419100775 &     0.1964415407 &    48.0914105413 &     82.226 &  46\\
II.C$_{40}^{i.c.}$(0.75) &     0.1831280973 &     0.2527284011 &    38.7565386929 &     81.609 &  46\\
II.C$_{41}^{i.c.}$(0.75) &     0.4668339454 &     0.2196009821 &    81.7805347369 &     84.753 &  48\\
II.C$_{42}^{i.c.}$(0.75) &     0.2695654045 &     0.1653865705 &    41.2047988880 &     86.007 &  48\\
II.C$_{43}^{i.c.}$(0.75) &     0.2471716099 &     0.4591801620 &    84.6911160699 &     85.058 &  50\\
II.C$_{44}^{i.c.}$(0.75) &     0.2348993023 &     0.0496701443 &    37.5752092509 &     89.884 &  50\\
II.C$_{45}^{i.c.}$(0.75) &     0.4607930537 &     0.2258237872 &    87.1992395234 &     91.746 &  52\\
II.C$_{46}^{i.c.}$(0.75) &     0.4935658319 &     0.2099521465 &    99.8457220227 &     91.744 &  52\\
II.C$_{47}^{i.c.}$(0.75) &     0.2312703709 &     0.2651474356 &    49.7439541861 &     95.631 &  54\\
II.C$_{48}^{i.c.}$(0.75) &     0.4311991946 &     0.3298943041 &   101.2913661831 &     89.267 &  54\\
II.C$_{49}^{i.c.}$(0.75) &     0.2786464975 &     0.4912442516 &   122.3430901926 &     92.589 &  54\\
II.C$_{50}^{i.c.}$(0.75) &     0.2591764961 &     0.1464220096 &    46.3658598456 &    100.521 &  56\\

\hline
\end{tabular*}
{\rule{\temptablewidth}{1pt}}
\end{center}
\end{table*}

\begin{table*}
\tabcolsep 0pt \caption{Initial conditions and periods $T$ of the periodic three-body orbits for class II.C in the case of  $\bm{r}_1(0)=(-1,0)=-\bm{r}_2(0)$,  $\dot{\bm{r}}_1(0)=(v_1,v_2)=\dot{\bm{r}}_2(0)$ and $\bm{r}_3(0)=(0,0)$, $\dot{\bm{r}}_3(0)=(-2v_1/m_3, -2v_2/m_3)$ when $G=1$ and $m_1=m_2=1$ and $m_3=0.75$ by means of the search grid $4000\times 4000$ in the interval $T_0\in[0,200]$, where $T^*=T |E|^{3/2}$ is its scale-invariant period, $L_f$ is the length of the free group element.  } \label{table-S5} \vspace*{-12pt}
\begin{center}
\def\temptablewidth{1\textwidth}
{\rule{\temptablewidth}{1pt}}
\begin{tabular*}{\temptablewidth}{@{\extracolsep{\fill}}lccccc}
\hline
Class and number  & $v_1$ & $v_2$  & $T$  & $T^*$ & $L_f$\\
\hline
II.C$_{51}^{i.c.}$(0.75) &     0.3292245081 &     0.2700636997 &    64.0015597083 &     98.738 &  56\\
II.C$_{52}^{i.c.}$(0.75) &     0.1761148939 &     0.2525176515 &    46.8228354487 &     99.462 &  56\\
II.C$_{53}^{i.c.}$(0.75) &     0.2832288095 &     0.2029755275 &    51.6398129899 &    100.115 &  56\\
II.C$_{54}^{i.c.}$(0.75) &     0.3734550631 &     0.2016974843 &    64.5586481759 &    100.079 &  56\\
II.C$_{55}^{i.c.}$(0.75) &     0.1079453802 &     0.1017385147 &    37.6522550231 &    100.118 &  56\\
II.C$_{56}^{i.c.}$(0.75) &     0.2893615743 &     0.1732851961 &    52.1564228215 &    103.868 &  58\\
II.C$_{57}^{i.c.}$(0.75) &     0.2739490611 &     0.1112476104 &    49.5347587742 &    107.810 &  60\\
II.C$_{58}^{i.c.}$(0.75) &     0.3143025441 &     0.4537167635 &   123.0075052602 &    102.059 &  60\\
II.C$_{59}^{i.c.}$(0.75) &     0.4442192619 &     0.2378705824 &    95.6646185864 &    105.733 &  60\\
II.C$_{60}^{i.c.}$(0.75) &     0.1121048529 &     0.0933447023 &    40.3882383150 &    107.616 &  60\\
II.C$_{61}^{i.c.}$(0.75) &     0.3654261757 &     0.4156535144 &   124.2929450326 &    102.061 &  60\\
II.C$_{62}^{i.c.}$(0.75) &     0.2834284459 &     0.2714433145 &    63.7610535892 &    109.638 &  62\\
II.C$_{63}^{i.c.}$(0.75) &     0.4748987889 &     0.2113333814 &   108.1264470442 &    109.638 &  62\\
II.C$_{64}^{i.c.}$(0.75) &     0.2368845535 &     0.1168010406 &    49.9219991102 &    114.999 &  64\\
II.C$_{65}^{i.c.}$(0.75) &     0.4039679707 &     0.2577569632 &    90.4576937114 &    112.726 &  64\\
II.C$_{66}^{i.c.}$(0.75) &     0.3284638906 &     0.2771383238 &    74.0903119133 &    112.719 &  64\\
II.C$_{67}^{i.c.}$(0.75) &     0.4954331403 &     0.2139970413 &   125.2599481230 &    112.719 &  64\\
II.C$_{68}^{i.c.}$(0.75) &     0.3256664816 &     0.1877076128 &    65.3726386438 &    117.933 &  66\\
II.C$_{69}^{i.c.}$(0.75) &     0.2177444543 &     0.4489972741 &    98.5606081526 &    111.692 &  66\\
II.C$_{70}^{i.c.}$(0.75) &     0.1116166364 &     0.0787235985 &    45.4896538661 &    122.120 &  68\\
II.C$_{71}^{i.c.}$(0.75) &     0.2166794086 &     0.2627405139 &    60.9820654333 &    120.507 &  68\\
II.C$_{72}^{i.c.}$(0.75) &     0.4505347794 &     0.0407260752 &    87.5074099143 &    122.244 &  68\\
II.C$_{73}^{i.c.}$(0.75) &     0.3236624960 &     0.4321844272 &   132.5563445697 &    119.080 &  70\\
II.C$_{74}^{i.c.}$(0.75) &     0.3100482475 &     0.4328437530 &   126.4841220008 &    119.076 &  70\\
II.C$_{75}^{i.c.}$(0.75) &     0.4520015173 &     0.2260302455 &   112.7189121601 &    123.633 &  70\\
II.C$_{76}^{i.c.}$(0.75) &     0.3784298834 &     0.2495376710 &    88.8298126856 &    123.634 &  70\\
II.C$_{77}^{i.c.}$(0.75) &     0.3002068606 &     0.2595660581 &    75.1530170603 &    127.505 &  72\\
II.C$_{78}^{i.c.}$(0.75) &     0.5034392071 &     0.1881587075 &   139.7265179198 &    127.518 &  72\\
II.C$_{79}^{i.c.}$(0.75) &     0.2792793082 &     0.1452471143 &    63.4672809336 &    132.888 &  74\\
II.C$_{80}^{i.c.}$(0.75) &     0.1704946320 &     0.4652537474 &   109.7290454265 &    126.546 &  74\\
II.C$_{81}^{i.c.}$(0.75) &     0.3886930945 &     0.1996476392 &    91.6231246943 &    135.788 &  76\\
II.C$_{82}^{i.c.}$(0.75) &     0.2980200648 &     0.2673702020 &    80.1527850060 &    134.516 &  76\\
II.C$_{83}^{i.c.}$(0.75) &     0.2092067208 &     0.4812270767 &   130.5165320395 &    128.641 &  76\\
II.C$_{84}^{i.c.}$(0.75) &     0.1206988814 &     0.0968566597 &    51.5317621978 &    136.261 &  76\\
II.C$_{85}^{i.c.}$(0.75) &     0.2017659812 &     0.1315138428 &    57.0335074764 &    136.280 &  76\\
II.C$_{86}^{i.c.}$(0.75) &     0.4493814446 &     0.0747851880 &   101.6265270381 &    140.162 &  78\\
II.C$_{87}^{i.c.}$(0.75) &     0.4800092637 &     0.2196511819 &   142.2301391511 &    137.618 &  78\\
II.C$_{88}^{i.c.}$(0.75) &     0.4984728341 &     0.2029987070 &   151.5260848913 &    137.617 &  78\\
II.C$_{89}^{i.c.}$(0.75) &     0.1132373526 &     0.0848401539 &    52.3492264971 &    139.989 &  78\\
II.C$_{90}^{i.c.}$(0.75) &     0.2119580098 &     0.2619382262 &    72.9469032636 &    145.383 &  82\\
II.C$_{91}^{i.c.}$(0.75) &     0.3555624393 &     0.2519592295 &    99.7762660659 &    148.518 &  84\\
II.C$_{92}^{i.c.}$(0.75) &     0.3320163863 &     0.2697721730 &    98.9557389721 &    151.603 &  86\\
II.C$_{93}^{i.c.}$(0.75) &     0.3390466919 &     0.2772682633 &   102.6797843211 &    151.602 &  86\\
II.C$_{94}^{i.c.}$(0.75) &     0.3923869301 &     0.1982688683 &   104.7040889511 &    153.642 &  86\\
II.C$_{95}^{i.c.}$(0.75) &     0.1133073970 &     0.0721736877 &    57.4443319839 &    154.480 &  86\\
II.C$_{96}^{i.c.}$(0.75) &     0.3087791621 &     0.1727360704 &    80.5689065010 &    154.124 &  86\\
II.C$_{97}^{i.c.}$(0.75) &     0.1935268473 &     0.2576335629 &    75.8206239259 &    156.235 &  88\\
II.C$_{98}^{i.c.}$(0.75) &     0.2937269662 &     0.1310053585 &    76.6069327690 &    158.064 &  88\\
II.C$_{99}^{i.c.}$(0.75) &     0.3119641579 &     0.4353298592 &   165.7961110787 &    153.101 &  90\\
II.C$_{100}^{i.c.}$(0.75) &     0.2560099766 &     0.4488404136 &   148.3975005786 &    153.097 &  90\\
II.C$_{101}^{i.c.}$(0.75) &     0.3282610510 &     0.4300507768 &   171.6008711131 &    153.101 &  90\\
II.C$_{102}^{i.c.}$(0.75) &     0.1212986423 &     0.0683081873 &    61.6823006908 &    165.249 &  92\\
II.C$_{103}^{i.c.}$(0.75) &     0.3201104564 &     0.2612000474 &   103.3021060602 &    166.396 &  94\\
II.C$_{104}^{i.c.}$(0.75) &     0.3279000291 &     0.2763797657 &   108.4818233700 &    165.584 &  94\\
II.C$_{105}^{i.c.}$(0.75) &     0.3084839268 &     0.1629915673 &    87.0805961214 &    168.638 &  94\\
II.C$_{106}^{i.c.}$(0.75) &     0.2688296146 &     0.1181832994 &    78.9575107005 &    172.516 &  96\\
II.C$_{107}^{i.c.}$(0.75) &     0.4152991663 &     0.1197329112 &   116.6820260011 &    175.958 &  98\\
II.C$_{108}^{i.c.}$(0.75) &     0.2864563062 &     0.2740950034 &   103.9050178198 &    176.499 &  100\\
II.C$_{109}^{i.c.}$(0.75) &     0.3120705685 &     0.2720598023 &   109.8857400002 &    176.499 &  100\\
II.C$_{110}^{i.c.}$(0.75) &     0.3393296129 &     0.2554121073 &   116.4797850100 &    180.397 &  102\\
\hline
\end{tabular*}
{\rule{\temptablewidth}{1pt}}
\end{center}
\end{table*}

\begin{table*}
\tabcolsep 0pt \caption{Initial conditions and periods $T$ of the periodic three-body orbits for class II.C in the case of  $\bm{r}_1(0)=(-1,0)=-\bm{r}_2(0)$,  $\dot{\bm{r}}_1(0)=(v_1,v_2)=\dot{\bm{r}}_2(0)$ and $\bm{r}_3(0)=(0,0)$, $\dot{\bm{r}}_3(0)=(-2v_1/m_3, -2v_2/m_3)$ when $G=1$ and $m_1=m_2=1$ and $m_3=0.75$ by means of the search grid $4000\times 4000$ in the interval $T_0\in[0,200]$, where $T^*=T |E|^{3/2}$ is its scale-invariant period, $L_f$ is the length of the free group element.  } \label{table-S5} \vspace*{-12pt}
\begin{center}
\def\temptablewidth{1\textwidth}
{\rule{\temptablewidth}{1pt}}
\begin{tabular*}{\temptablewidth}{@{\extracolsep{\fill}}lccccc}
\hline
Class and number  & $v_1$ & $v_2$  & $T$  & $T^*$ & $L_f$\\
\hline
II.C$_{111}^{i.c.}$(0.75) &     0.1181895504 &     0.0758977642 &    68.4251332328 &    183.131 &  102\\
II.C$_{112}^{i.c.}$(0.75) &     0.3762237073 &     0.2525106949 &   131.7379537609 &    183.494 &  104\\
II.C$_{113}^{i.c.}$(0.75) &     0.3960217517 &     0.1959815422 &   130.1263669788 &    189.352 &  106\\
II.C$_{114}^{i.c.}$(0.75) &     0.2794461019 &     0.0678830585 &    86.2018484103 &    190.525 &  106\\
II.C$_{115}^{i.c.}$(0.75) &     0.2193806640 &     0.2606198586 &    96.8102097868 &    191.261 &  108\\
II.C$_{116}^{i.c.}$(0.75) &     0.3584363302 &     0.2496690110 &   131.0937464396 &    194.391 &  110\\
II.C$_{117}^{i.c.}$(0.75) &     0.2555380955 &     0.2505369048 &   104.7207733533 &    198.265 &  112\\
II.C$_{118}^{i.c.}$(0.75) &     0.2717717477 &     0.1211826752 &    94.4766709178 &    204.869 &  114\\
II.C$_{119}^{i.c.}$(0.75) &     0.4557116917 &     0.2243651046 &   186.1251013074 &    201.389 &  114\\
II.C$_{120}^{i.c.}$(0.75) &     0.3368046655 &     0.2601961372 &   130.4398908126 &    201.388 &  114\\
II.C$_{121}^{i.c.}$(0.75) &     0.4716148265 &     0.2169359149 &   197.8478047433 &    201.390 &  114\\
II.C$_{122}^{i.c.}$(0.75) &     0.1983164515 &     0.1289228288 &    85.0121650660 &    204.420 &  114\\
II.C$_{123}^{i.c.}$(0.75) &     0.4314296428 &     0.1085814064 &   144.8128259248 &    208.309 &  116\\
II.C$_{124}^{i.c.}$(0.75) &     0.3984768264 &     0.0833001471 &   126.8834769024 &    208.462 &  116\\
II.C$_{125}^{i.c.}$(0.75) &     0.3508369213 &     0.1499116351 &   119.2495101416 &    211.730 &  118\\
II.C$_{126}^{i.c.}$(0.75) &     0.2382029257 &     0.4600002384 &   199.3101782590 &    204.138 &  120\\
II.C$_{127}^{i.c.}$(0.75) &     0.3539713336 &     0.1844517695 &   127.4592957316 &    214.729 &  120\\
II.C$_{128}^{i.c.}$(0.75) &     0.4679065922 &     0.0492370802 &   169.3015724927 &    219.318 &  122\\
II.C$_{129}^{i.c.}$(0.75) &     0.2897587650 &     0.1322568131 &   107.2426867012 &    222.768 &  124\\
II.C$_{130}^{i.c.}$(0.75) &     0.3194350927 &     0.1437653927 &   117.2090308828 &    226.198 &  126\\
II.C$_{131}^{i.c.}$(0.75) &     0.2856067895 &     0.2664601284 &   129.0520293629 &    223.134 &  126\\
II.C$_{132}^{i.c.}$(0.75) &     0.3172180539 &     0.2641989427 &   140.3245705513 &    226.273 &  128\\
II.C$_{133}^{i.c.}$(0.75) &     0.3500093713 &     0.1856549777 &   136.8114267533 &    232.586 &  130\\
II.C$_{134}^{i.c.}$(0.75) &     0.2402496725 &     0.0392897276 &    99.6934119722 &    237.292 &  132\\
II.C$_{135}^{i.c.}$(0.75) &     0.3685221643 &     0.0304592777 &   131.2899349360 &    240.865 &  134\\
II.C$_{136}^{i.c.}$(0.75) &     0.2976587430 &     0.2715721913 &   144.3157440128 &    240.266 &  136\\
II.C$_{137}^{i.c.}$(0.75) &     0.3818757096 &     0.1993653265 &   162.7392471190 &    246.680 &  138\\
II.C$_{138}^{i.c.}$(0.75) &     0.3749650745 &     0.1872646291 &   158.6249794643 &    250.442 &  140\\
II.C$_{139}^{i.c.}$(0.75) &     0.2533117212 &     0.0765118733 &   110.8909227112 &    255.234 &  142\\
II.C$_{140}^{i.c.}$(0.75) &     0.2938548621 &     0.1099787648 &   123.3013725819 &    258.731 &  144\\
II.C$_{141}^{i.c.}$(0.75) &     0.3418869995 &     0.1513320020 &   144.4057666321 &    261.966 &  146\\
II.C$_{142}^{i.c.}$(0.75) &     0.3568986704 &     0.1789162975 &   157.3345205939 &    264.987 &  148\\
II.C$_{143}^{i.c.}$(0.75) &     0.3507048994 &     0.1879172283 &   158.6337077719 &    268.298 &  150\\
II.C$_{144}^{i.c.}$(0.75) &     0.4289912118 &     0.0740929532 &   183.2158749766 &    273.162 &  152\\
II.C$_{145}^{i.c.}$(0.75) &     0.2052527461 &     0.0562709352 &   110.0586247060 &    273.334 &  152\\
II.C$_{146}^{i.c.}$(0.75) &     0.4153025220 &     0.1261299710 &   184.5272181346 &    276.437 &  154\\
II.C$_{147}^{i.c.}$(0.75) &     0.3985285632 &     0.1078946305 &   171.5880123205 &    276.644 &  154\\
II.C$_{148}^{i.c.}$(0.75) &     0.2587721852 &     0.2603689252 &   155.2429995575 &    286.883 &  162\\
II.C$_{149}^{i.c.}$(0.75) &     0.3015928568 &     0.0717923667 &   137.6823840835 &    291.173 &  162\\
II.C$_{150}^{i.c.}$(0.75) &     0.4240317938 &     0.0881010520 &   193.6334447174 &    291.079 &  162\\
II.C$_{151}^{i.c.}$(0.75) &     0.3829944898 &     0.1198045039 &   176.0142373504 &    294.549 &  164\\
II.C$_{152}^{i.c.}$(0.75) &     0.1313939229 &     0.0902579269 &   114.5276827160 &    301.563 &  168\\
II.C$_{153}^{i.c.}$(0.75) &     0.4089860758 &     0.0422384137 &   188.4670965240 &    305.576 &  170\\
II.C$_{154}^{i.c.}$(0.75) &     0.3176605344 &     0.0885482935 &   152.5558686604 &    309.114 &  172\\
II.C$_{155}^{i.c.}$(0.75) &     0.3712251533 &     0.1994085280 &   198.5463547977 &    311.052 &  174\\
II.C$_{156}^{i.c.}$(0.75) &     0.3918996873 &     0.1432248649 &   196.4917875525 &    312.203 &  174\\
II.C$_{157}^{i.c.}$(0.75) &     0.2050985427 &     0.1005778254 &   128.5282060869 &    312.635 &  174\\
II.C$_{158}^{i.c.}$(0.75) &     0.3419116730 &     0.1826902734 &   182.6440041642 &    318.564 &  178\\
II.C$_{159}^{i.c.}$(0.75) &     0.2830434789 &     0.0744751508 &   147.7926007568 &    323.523 &  180\\
II.C$_{160}^{i.c.}$(0.75) &     0.2776335848 &     0.1338443741 &   154.0847561202 &    327.078 &  182\\
II.C$_{161}^{i.c.}$(0.75) &     0.2865358925 &     0.1200209400 &   161.6167188878 &    341.339 &  190\\
II.C$_{162}^{i.c.}$(0.75) &     0.2700170285 &     0.1122748834 &   162.4518900554 &    355.781 &  198\\
II.C$_{163}^{i.c.}$(0.75) &     0.3047927194 &     0.0555637745 &   176.7118459185 &    373.871 &  208\\
II.C$_{164}^{i.c.}$(0.75) &     0.1380664200 &     0.0670078477 &   142.5493314030 &    377.360 &  210\\
II.C$_{165}^{i.c.}$(0.75) &     0.2454841019 &     0.1186795750 &   166.2145766431 &    377.360 &  210\\
II.C$_{166}^{i.c.}$(0.75) &     0.3174952187 &     0.0732714608 &   191.6755720430 &    391.823 &  218\\
II.C$_{167}^{i.c.}$(0.75) &     0.1696261275 &     0.1004200178 &   171.6162456989 &    434.470 &  242\\
II.C$_{168}^{i.c.}$(0.75) &     0.2044935780 &     0.0801337875 &   178.1873542624 &    438.633 &  244\\
II.C$_{169}^{i.c.}$(0.75) &     0.2196248784 &     0.0651566054 &   196.7913913300 &    478.244 &  266\\
\hline
\end{tabular*}
{\rule{\temptablewidth}{1pt}}
\end{center}
\end{table*}

\begin{table*}
\tabcolsep 0pt \caption{Initial conditions and periods $T$ of the periodic three-body orbits for class I.A in the case of  $\bm{r}_1(0)=(-1,0)=-\bm{r}_2(0)$,  $\dot{\bm{r}}_1(0)=(v_1,v_2)=\dot{\bm{r}}_2(0)$ and $\bm{r}_3(0)=(0,0)$, $\dot{\bm{r}}_3(0)=(-2v_1/m_3, -2v_2/m_3)$ when $G=1$ and $m_1=m_2=1$ and $m_3=2$ by means of the search grid $4000\times 4000$ in the interval $T_0\in[0,200]$, where $T^*=T |E|^{3/2}$ is its scale-invariant period, $L_f$ is the length of the free group element.  } \label{table-S5} \vspace*{-12pt}
\begin{center}
\def\temptablewidth{1\textwidth}
{\rule{\temptablewidth}{1pt}}
\begin{tabular*}{\temptablewidth}{@{\extracolsep{\fill}}lccccc}
\hline
Class and number  & $v_1$ & $v_2$  & $T$  & $T^*$ & $L_f$\\
\hline
I.A$_{1}^{i.c.}$(2) &     0.6649107583 &     0.8324167864 &    12.6489061509 &     42.121 &  8\\
I.A$_{2}^{i.c.}$(2) &     0.6656250225 &     0.4965289014 &    11.7760601476 &     64.923 &  12\\
I.A$_{3}^{i.c.}$(2) &     0.6825287273 &     0.6174444064 &    36.6762382587 &    172.376 &  32\\
I.A$_{4}^{i.c.}$(2) &     0.6994683581 &     0.3498807082 &    36.5607892892 &    216.851 &  40\\
I.A$_{5}^{i.c.}$(2) &     0.4085452441 &     0.0399781234 &    28.1061456733 &    238.731 &  44\\
I.A$_{6}^{i.c.}$(2) &     0.6925444835 &     0.7038306689 &    63.1111348894 &    256.991 &  48\\
I.A$_{7}^{i.c.}$(2) &     0.6852621866 &     0.6403262957 &    61.6549217313 &    279.758 &  52\\
I.A$_{8}^{i.c.}$(2) &     0.6786880593 &     0.5712663380 &    60.3945847774 &    302.292 &  56\\
I.A$_{9}^{i.c.}$(2) &     0.7520356911 &     0.5826701574 &    68.5259154653 &    302.310 &  56\\
I.A$_{10}^{i.c.}$(2) &     0.7731160189 &     0.4217868038 &    68.4901771349 &    346.809 &  64\\
I.A$_{11}^{i.c.}$(2) &     0.6258530965 &     0.6849876905 &    78.6876130536 &    364.377 &  68\\
I.A$_{12}^{i.c.}$(2) &     0.6687176921 &     0.3113447440 &    58.5101436950 &    368.722 &  68\\
I.A$_{13}^{i.c.}$(2) &     0.5219331515 &     0.1537833421 &    50.5549444042 &    390.548 &  72\\
I.A$_{14}^{i.c.}$(2) &     0.6484066991 &     0.5971552405 &    81.0318562446 &    409.726 &  76\\
I.A$_{15}^{i.c.}$(2) &     0.6807496308 &     0.5512579156 &    84.6308430084 &    432.167 &  80\\
I.A$_{16}^{i.c.}$(2) &     0.6508103868 &     0.6834247439 &   105.2312919998 &    471.737 &  88\\
I.A$_{17}^{i.c.}$(2) &     0.6869542921 &     0.6559982684 &   111.7375242360 &    494.496 &  92\\
I.A$_{18}^{i.c.}$(2) &     0.4665900313 &     0.2785470858 &    67.3569385398 &    520.654 &  96\\
I.A$_{19}^{i.c.}$(2) &     0.6534853531 &     0.1865830089 &    80.1996471047 &    542.402 &  100\\
I.A$_{20}^{i.c.}$(2) &     0.6640665882 &     0.6805220686 &   131.1248770198 &    579.098 &  108\\
I.A$_{21}^{i.c.}$(2) &     0.6597778792 &     0.4513736509 &   104.8697669510 &    606.485 &  112\\
I.A$_{22}^{i.c.}$(2) &     0.6471520650 &     0.4020332252 &   103.0148018894 &    628.563 &  116\\
I.A$_{23}^{i.c.}$(2) &     0.6626286520 &     0.2839544976 &   104.4563049755 &    672.449 &  124\\
I.A$_{24}^{i.c.}$(2) &     0.5253644523 &     0.2725303838 &    90.8016748047 &    672.470 &  124\\
I.A$_{25}^{i.c.}$(2) &     0.5466058934 &     0.5202407423 &   112.2826008711 &    691.899 &  128\\
I.A$_{26}^{i.c.}$(2) &     0.4524709149 &     0.1937913189 &    86.2950694409 &    694.359 &  128\\
I.A$_{27}^{i.c.}$(2) &     0.7385582230 &     0.4623589255 &   143.0327822206 &    736.360 &  136\\
I.A$_{28}^{i.c.}$(2) &     0.5178453979 &     0.4523792202 &   109.8866781782 &    736.361 &  136\\
I.A$_{29}^{i.c.}$(2) &     0.6895973623 &     0.6795598105 &   186.6182866360 &    793.825 &  148\\
I.A$_{30}^{i.c.}$(2) &     0.7417737283 &     0.5576355792 &   172.6862188871 &    799.409 &  148\\
I.A$_{31}^{i.c.}$(2) &     0.6787848022 &     0.3324577477 &   130.4345522210 &    802.428 &  148\\
I.A$_{32}^{i.c.}$(2) &     0.4942916225 &     0.3122096699 &   107.6378558600 &    802.502 &  148\\
I.A$_{33}^{i.c.}$(2) &     0.6616192451 &     0.2773985935 &   127.4910900115 &    824.310 &  152\\
I.A$_{34}^{i.c.}$(2) &     0.4972412503 &     0.2031457421 &   108.9018118241 &    846.169 &  156\\
I.A$_{35}^{i.c.}$(2) &     0.4457731294 &     0.1076713105 &   105.3479284426 &    867.997 &  160\\
I.A$_{36}^{i.c.}$(2) &     0.5807922587 &     0.4312476531 &   138.4192610066 &    888.319 &  164\\
I.A$_{37}^{i.c.}$(2) &     0.6450450089 &     0.3979743783 &   148.4035753878 &    910.372 &  168\\
I.A$_{38}^{i.c.}$(2) &     0.6059803235 &     0.5138426153 &   163.3555963264 &    951.591 &  176\\
I.A$_{39}^{i.c.}$(2) &     0.6668050678 &     0.5220530211 &   177.2812672926 &    951.589 &  176\\
I.A$_{40}^{i.c.}$(2) &     0.5306754259 &     0.3039555306 &   131.3126893593 &    954.331 &  176\\
I.A$_{41}^{i.c.}$(2) &     0.6720090611 &     0.3203793706 &   152.7891565944 &    954.299 &  176\\
I.A$_{42}^{i.c.}$(2) &     0.4750713117 &     0.2602297809 &   126.1180059947 &    976.273 &  180\\
I.A$_{43}^{i.c.}$(2) &     0.6349837177 &     0.4678756830 &   169.5497785789 &    996.045 &  184\\
I.A$_{44}^{i.c.}$(2) &     0.5469768158 &     0.2106173499 &   134.0431589510 &    997.999 &  184\\
I.A$_{45}^{i.c.}$(2) &     0.4363598168 &     0.2067947916 &   123.2142494342 &    998.179 &  184\\
I.A$_{46}^{i.c.}$(2) &     0.5105790195 &     0.4381247712 &   149.3966262626 &   1018.209 &  188\\
I.A$_{47}^{i.c.}$(2) &     0.4826350873 &     0.1356931812 &   127.6057515007 &   1019.813 &  188\\
I.A$_{48}^{i.c.}$(2) &     0.6581637706 &     0.4419205771 &   174.3387691062 &   1018.181 &  188\\
I.A$_{49}^{i.c.}$(2) &     0.6208957754 &     0.3798674379 &   166.4674004829 &   1062.279 &  196\\
I.A$_{50}^{i.c.}$(2) &     0.5283765904 &     0.5013997243 &   169.5876212839 &   1081.461 &  200\\

\hline
\end{tabular*}
{\rule{\temptablewidth}{1pt}}
\end{center}
\end{table*}

\begin{table*}
\tabcolsep 0pt \caption{Initial conditions and periods $T$ of the periodic three-body orbits for class I.A and I.B in the case of  $\bm{r}_1(0)=(-1,0)=-\bm{r}_2(0)$,  $\dot{\bm{r}}_1(0)=(v_1,v_2)=\dot{\bm{r}}_2(0)$ and $\bm{r}_3(0)=(0,0)$, $\dot{\bm{r}}_3(0)=(-2v_1/m_3, -2v_2/m_3)$ when $G=1$ and $m_1=m_2=1$ and $m_3=2$ by means of the search grid $4000\times 4000$ in the interval $T_0\in[0,200]$, where $T^*=T |E|^{3/2}$ is its scale-invariant period, $L_f$ is the length of the free group element.  } \label{table-S5} \vspace*{-12pt}
\begin{center}
\def\temptablewidth{1\textwidth}
{\rule{\temptablewidth}{1pt}}
\begin{tabular*}{\temptablewidth}{@{\extracolsep{\fill}}lccccc}
\hline
Class and number  & $v_1$ & $v_2$  & $T$  & $T^*$ & $L_f$\\
\hline
I.A$_{51}^{i.c.}$(2) &     0.5485296197 &     0.4651956166 &   174.5302500004 &   1125.911 &  208\\
I.A$_{52}^{i.c.}$(2) &     0.5048487414 &     0.2581606824 &   148.9271778598 &   1128.085 &  208\\
I.A$_{53}^{i.c.}$(2) &     0.6605145396 &     0.2693390090 &   173.5846701186 &   1128.030 &  208\\
I.A$_{54}^{i.c.}$(2) &     0.4653563186 &     0.2118208325 &   144.9873979687 &   1149.975 &  212\\
I.A$_{55}^{i.c.}$(2) &     0.5563868856 &     0.4227642462 &   176.9274017831 &   1170.145 &  216\\
I.A$_{56}^{i.c.}$(2) &     0.6360544648 &     0.3001819188 &   191.2539223702 &   1258.032 &  232\\
I.A$_{57}^{i.c.}$(2) &     0.6601780173 &     0.2666833376 &   196.6368084515 &   1279.890 &  236\\
I.A$_{58}^{i.c.}$(2) &     0.5619933924 &     0.1668615748 &   177.7791904073 &   1323.487 &  244\\
I.A$_{59}^{i.c.}$(2) &     0.4301622650 &     0.0900627168 &   161.2579898217 &   1345.457 &  248\\
I.A$_{60}^{i.c.}$(2) &     0.5707073620 &     0.3140095729 &   198.9345552114 &   1388.021 &  256\\
I.A$_{61}^{i.c.}$(2) &     0.4475731896 &     0.2162152621 &   181.3252835399 &   1453.790 &  268\\
I.A$_{62}^{i.c.}$(2) &     0.4807274686 &     0.1737864193 &   186.0049759663 &   1475.446 &  272\\
I.B$_{1}^{i.c.}$(2) &     0.4136863353 &     0.9410539017 &    12.5429961223 &     46.245 &  10\\
I.B$_{2}^{i.c.}$(2) &     0.3534308283 &     0.8508079166 &    12.2692512978 &     57.560 &  12\\
I.B$_{3}^{i.c.}$(2) &     0.0413741936 &     0.9956426300 &    15.5793823937 &     62.100 &  16\\
I.B$_{4}^{i.c.}$(2) &     0.7096842091 &     0.7687592147 &    27.3003058481 &     95.893 &  18\\
I.B$_{5}^{i.c.}$(2) &     0.6799645758 &     0.5887672136 &    24.2551964847 &    118.671 &  22\\
I.B$_{6}^{i.c.}$(2) &     0.6378164212 &     0.3886052513 &    22.6313155247 &    140.904 &  26\\
I.B$_{7}^{i.c.}$(2) &     0.6699852230 &     0.7085344248 &    48.5438772348 &    203.302 &  38\\
I.B$_{8}^{i.c.}$(2) &     0.6843384814 &     0.6317419448 &    49.1655025482 &    226.070 &  42\\
I.B$_{9}^{i.c.}$(2) &     0.6919910882 &     0.5460326131 &    49.1547189071 &    248.549 &  46\\
I.B$_{10}^{i.c.}$(2) &     0.6587938097 &     0.4454867422 &    46.5359882210 &    270.778 &  50\\
I.B$_{11}^{i.c.}$(2) &     0.6746780747 &     0.3259416999 &    47.1768704466 &    292.788 &  54\\
I.B$_{12}^{i.c.}$(2) &     0.4644102263 &     0.1242585136 &    38.7788080029 &    314.634 &  58\\
I.B$_{13}^{i.c.}$(2) &     0.6916935954 &     0.6979815204 &    75.4612881629 &    310.678 &  58\\
I.B$_{14}^{i.c.}$(2) &     0.6858800687 &     0.6461043683 &    74.1577191496 &    333.444 &  62\\
I.B$_{15}^{i.c.}$(2) &     0.7655652032 &     0.5318113366 &    82.4365268312 &    378.439 &  70\\
I.B$_{16}^{i.c.}$(2) &     0.6615614523 &     0.4629426722 &    70.1179389107 &    400.634 &  74\\
I.B$_{17}^{i.c.}$(2) &     0.6909413108 &     0.6906229000 &   100.2197830082 &    418.047 &  78\\
I.B$_{18}^{i.c.}$(2) &     0.6660110017 &     0.3013230386 &    69.9586100682 &    444.655 &  82\\
I.B$_{19}^{i.c.}$(2) &     0.5841885389 &     0.1723383387 &    64.0297299896 &    466.471 &  86\\
I.B$_{20}^{i.c.}$(2) &     0.7431942127 &     0.5700610999 &   106.8202805388 &    485.913 &  90\\
I.B$_{21}^{i.c.}$(2) &     0.5876675319 &     0.5574609749 &    85.3743868704 &    485.912 &  90\\
I.B$_{22}^{i.c.}$(2) &     0.5795678824 &     0.4658293684 &    84.8335185161 &    530.488 &  98\\
I.B$_{23}^{i.c.}$(2) &     0.6903638351 &     0.6860300463 &   124.9278103447 &    525.413 &  98\\
I.B$_{24}^{i.c.}$(2) &     0.6528942111 &     0.4180607376 &    92.2682840912 &    552.599 &  102\\
I.B$_{25}^{i.c.}$(2) &     0.7613651698 &     0.3029365683 &   106.3455407448 &    596.558 &  110\\
I.B$_{26}^{i.c.}$(2) &     0.4956617783 &     0.2749768423 &    78.7423557565 &    596.558 &  110\\
I.B$_{27}^{i.c.}$(2) &     0.5322323384 &     0.5465974633 &   101.0708650395 &    615.814 &  114\\
I.B$_{28}^{i.c.}$(2) &     0.6796765126 &     0.5536965825 &   120.7390738313 &    615.783 &  114\\
I.B$_{29}^{i.c.}$(2) &     0.6677734178 &     0.5160072240 &   118.3019983217 &    638.107 &  118\\
I.B$_{30}^{i.c.}$(2) &     0.4935620999 &     0.4634807883 &    97.3610217793 &    660.367 &  122\\
I.B$_{31}^{i.c.}$(2) &     0.5632072523 &     0.5622457909 &   124.3998006645 &    723.269 &  134\\
I.B$_{32}^{i.c.}$(2) &     0.7355327264 &     0.2899738941 &   127.7436324479 &    748.388 &  138\\
I.B$_{33}^{i.c.}$(2) &     0.6897103691 &     0.6805154515 &   174.2881413969 &    740.143 &  138\\
I.B$_{34}^{i.c.}$(2) &     0.5586512240 &     0.2714718822 &   103.9534809964 &    748.388 &  138\\
I.B$_{35}^{i.c.}$(2) &     0.5333261360 &     0.4979744794 &   120.6208513685 &    767.971 &  142\\
I.B$_{36}^{i.c.}$(2) &     0.6546992766 &     0.2061185669 &   114.7714470452 &    770.194 &  142\\
I.B$_{37}^{i.c.}$(2) &     0.6637383695 &     0.4797530605 &   140.7636176347 &    790.183 &  146\\
I.B$_{38}^{i.c.}$(2) &     0.6877572827 &     0.0842798120 &   118.9127597836 &    791.939 &  146\\
I.B$_{39}^{i.c.}$(2) &     0.6498528395 &     0.4085255224 &   137.8351993001 &    834.412 &  154\\
I.B$_{40}^{i.c.}$(2) &     0.7624410411 &     0.3440126143 &   160.8845261259 &    878.414 &  162\\
\hline
\end{tabular*}
{\rule{\temptablewidth}{1pt}}
\end{center}
\end{table*}

\begin{table*}
\tabcolsep 0pt \caption{Initial conditions and periods $T$ of the periodic three-body orbits for class I.B, I.C, II.A and II.B in the case of  $\bm{r}_1(0)=(-1,0)=-\bm{r}_2(0)$,  $\dot{\bm{r}}_1(0)=(v_1,v_2)=\dot{\bm{r}}_2(0)$ and $\bm{r}_3(0)=(0,0)$, $\dot{\bm{r}}_3(0)=(-2v_1/m_3, -2v_2/m_3)$ when $G=1$ and $m_1=m_2=1$ and $m_3=2$ by means of the search grid $4000\times 4000$ in the interval $T_0\in[0,200]$, where $T^*=T |E|^{3/2}$ is its scale-invariant period, $L_f$ is the length of the free group element.  } \label{table-S5} \vspace*{-12pt}
\begin{center}
\def\temptablewidth{1\textwidth}
{\rule{\temptablewidth}{1pt}}
\begin{tabular*}{\temptablewidth}{@{\extracolsep{\fill}}lccccc}
\hline
Class and number  & $v_1$ & $v_2$  & $T$  & $T^*$ & $L_f$\\
\hline
I.B$_{41}^{i.c.}$(2) &     0.6879826248 &     0.6653325096 &   199.6963836583 &    870.272 &  162\\
I.B$_{42}^{i.c.}$(2) &     0.6240256618 &     0.5323918480 &   156.2803805272 &    875.506 &  162\\
I.B$_{43}^{i.c.}$(2) &     0.4563106153 &     0.2611707573 &   114.8148902213 &    900.372 &  166\\
I.B$_{44}^{i.c.}$(2) &     0.6612633816 &     0.2749034316 &   139.0125537029 &    900.240 &  166\\
I.B$_{45}^{i.c.}$(2) &     0.5221782655 &     0.2071155435 &   121.1696096366 &    922.081 &  170\\
I.B$_{46}^{i.c.}$(2) &     0.5567455518 &     0.5648247735 &   164.5620595602 &    960.621 &  178\\
I.B$_{47}^{i.c.}$(2) &     0.6790474390 &     0.4222695611 &   166.7970949155 &    964.291 &  178\\
I.B$_{48}^{i.c.}$(2) &     0.6790840621 &     0.5778537084 &   193.5243647671 &    960.603 &  178\\
I.B$_{49}^{i.c.}$(2) &     0.6791350775 &     0.5558668618 &   193.0705870639 &    983.015 &  182\\
I.B$_{50}^{i.c.}$(2) &     0.5552591136 &     0.5217230932 &   164.7814492815 &   1005.379 &  186\\
I.B$_{51}^{i.c.}$(2) &     0.6758170233 &     0.5099997641 &   191.3878276034 &   1027.651 &  190\\
I.B$_{52}^{i.c.}$(2) &     0.6701256280 &     0.3155625312 &   164.1342676824 &   1030.233 &  190\\
I.B$_{53}^{i.c.}$(2) &     0.6642332036 &     0.4838180744 &   187.8413617236 &   1049.878 &  194\\
I.B$_{54}^{i.c.}$(2) &     0.6607235336 &     0.2709395969 &   162.0595886706 &   1052.100 &  194\\
I.B$_{55}^{i.c.}$(2) &     0.5730352235 &     0.2138677593 &   147.7800306334 &   1073.921 &  198\\
I.B$_{56}^{i.c.}$(2) &     0.5024476446 &     0.1455936846 &   139.4317930155 &   1095.727 &  202\\
I.B$_{57}^{i.c.}$(2) &     0.6603346018 &     0.2679312271 &   185.1104590104 &   1203.960 &  222\\
I.B$_{58}^{i.c.}$(2) &     0.4894090722 &     0.4358021895 &   176.1804659126 &   1224.076 &  226\\
I.B$_{59}^{i.c.}$(2) &     0.5813604720 &     0.4153081963 &   192.0257778397 &   1246.108 &  230\\
I.B$_{60}^{i.c.}$(2) &     0.4448123384 &     0.1589471380 &   152.8745723698 &   1247.733 &  230\\
I.B$_{61}^{i.c.}$(2) &     0.5415992006 &     0.1607640861 &   164.4006039689 &   1247.564 &  230\\
I.B$_{62}^{i.c.}$(2) &     0.4192735320 &     0.0717710277 &   150.8160159945 &   1269.555 &  234\\
I.B$_{63}^{i.c.}$(2) &     0.4417759692 &     0.1037595968 &   171.9555602120 &   1421.360 &  262\\
I.B$_{64}^{i.c.}$(2) &     0.4886058039 &     0.3067872636 &   195.0016721994 &   1464.084 &  270\\
I.B$_{65}^{i.c.}$(2) &     0.4950064193 &     0.1779001099 &   197.8654949145 &   1551.356 &  286\\
$I.C_{1}^{i.c.}$(2) &     0.4463851548 &     0.2849973719 &   113.7803515858 &    889.512 &  164\\
$I.C_{2}^{i.c.}$(2) &     0.6539824310 &     0.1949540976 &   183.4448758069 &   1236.666 &  228\\
II.A$_{1}^{i.c.}$(2) &     0.6563704629 &     0.7210981740 &    35.7193985834 &    149.610 &  28\\
II.A$_{2}^{i.c.}$(2) &     0.7043804200 &     0.2995216841 &    85.7364518416 &    520.587 &  96\\
II.A$_{3}^{i.c.}$(2) &     0.6277885723 &     0.5727334717 &   101.0259827620 &    539.638 &  100\\
II.A$_{4}^{i.c.}$(2) &     0.6705929917 &     0.5931586426 &   179.3690438002 &    884.414 &  164\\
II.B$_{1}^{i.c.}$(2) &     0.0488561235 &     0.8544148215 &    11.9723950175 &     63.308 &  16\\
II.B$_{2}^{i.c.}$(2) &     0.5050095375 &     0.8532423197 &    30.3164175587 &    122.281 &  28\\
II.B$_{3}^{i.c.}$(2) &     0.6912520905 &     0.8072079793 &    53.7137845333 &    180.218 &  34\\
II.B$_{4}^{i.c.}$(2) &     0.7165670952 &     0.7256175703 &   133.4837704301 &    502.525 &  94\\
II.B$_{5}^{i.c.}$(2) &     0.6928070373 &     0.6049731873 &   148.9269113715 &    700.779 &  130\\
II.B$_{6}^{i.c.}$(2) &     0.4965910911 &     0.4930949987 &   135.9238691725 &    897.850 &  166\\
II.B$_{7}^{i.c.}$(2) &     0.4302338078 &     0.2044723241 &   113.3177648518 &    922.286 &  170\\
\hline
\end{tabular*}
{\rule{\temptablewidth}{1pt}}
\end{center}
\end{table*}

\begin{table*}
\tabcolsep 0pt \caption{Initial conditions and periods $T$ of the periodic three-body orbits for class II.C in the case of  $\bm{r}_1(0)=(-1,0)=-\bm{r}_2(0)$,  $\dot{\bm{r}}_1(0)=(v_1,v_2)=\dot{\bm{r}}_2(0)$ and $\bm{r}_3(0)=(0,0)$, $\dot{\bm{r}}_3(0)=(-2v_1/m_3, -2v_2/m_3)$ when $G=1$ and $m_1=m_2=1$ and $m_3=2$ by means of the search grid $4000\times 4000$ in the interval $T_0\in[0,200]$, where $T^*=T |E|^{3/2}$ is its scale-invariant period, $L_f$ is the length of the free group element.  } \label{table-S5} \vspace*{-12pt}
\begin{center}
\def\temptablewidth{1\textwidth}
{\rule{\temptablewidth}{1pt}}
\begin{tabular*}{\temptablewidth}{@{\extracolsep{\fill}}lccccc}
\hline
Class and number  & $v_1$ & $v_2$  & $T$  & $T^*$ & $L_f$\\
\hline
II.C$_{1}^{i.c.}$(2) &     0.3419254129 &     0.6225042846 &     8.1277419699 &     53.018 &  10\\
II.C$_{2}^{i.c.}$(2) &     0.4183048895 &     0.8832477662 &    10.6835513193 &     44.526 &  10\\
II.C$_{3}^{i.c.}$(2) &     0.2315589634 &     0.5397849981 &     8.4364344582 &     62.741 &  12\\
II.C$_{4}^{i.c.}$(2) &     0.3482359951 &     0.7977005516 &    10.6521916661 &     54.931 &  12\\
II.C$_{5}^{i.c.}$(2) &     0.1747155595 &     0.9537742434 &    11.2448496818 &     47.676 &  12\\
II.C$_{6}^{i.c.}$(2) &     0.1876415041 &     0.4973714491 &     9.2174482315 &     71.945 &  14\\
II.C$_{7}^{i.c.}$(2) &     0.8585382523 &     0.3992014522 &    16.9140385366 &     75.336 &  14\\
II.C$_{8}^{i.c.}$(2) &     0.6504217297 &     0.7534481524 &    34.5441872395 &    138.069 &  26\\
II.C$_{9}^{i.c.}$(2) &     0.4066977900 &     0.2322976437 &    18.5607296551 &    151.910 &  28\\
II.C$_{10}^{i.c.}$(2) &     0.7592503197 &     0.5681249718 &    41.3559042212 &    183.635 &  34\\
II.C$_{11}^{i.c.}$(2) &     0.2066719875 &     0.9430283237 &    33.4271615500 &    143.057 &  36\\
II.C$_{12}^{i.c.}$(2) &     0.4146350243 &     0.3221010514 &    27.6479764521 &    216.939 &  40\\
II.C$_{13}^{i.c.}$(2) &     0.7609558501 &     0.2581080751 &    39.6352947801 &    227.807 &  42\\
II.C$_{14}^{i.c.}$(2) &     0.5254523354 &     0.5562034750 &    40.9263974560 &    248.593 &  46\\
II.C$_{15}^{i.c.}$(2) &     0.4618794775 &     0.3060780263 &    38.2278117270 &    292.838 &  54\\
II.C$_{16}^{i.c.}$(2) &     0.5661937162 &     0.2399247352 &    41.9294344254 &    303.721 &  56\\
II.C$_{17}^{i.c.}$(2) &     0.5603872773 &     0.4510742855 &    52.0486804950 &    335.710 &  62\\
II.C$_{18}^{i.c.}$(2) &     0.7141078611 &     0.7225193662 &    92.8206248892 &    352.914 &  66\\
II.C$_{19}^{i.c.}$(2) &     0.4649115785 &     0.3402622958 &    47.6265126218 &    357.844 &  66\\
II.C$_{20}^{i.c.}$(2) &     0.6167367951 &     0.5539133064 &    66.4559860020 &    367.232 &  68\\
II.C$_{21}^{i.c.}$(2) &     0.5048320727 &     0.2952984074 &    49.4689463532 &    368.745 &  68\\
II.C$_{22}^{i.c.}$(2) &     0.7436868264 &     0.4653634325 &    78.6445804901 &    400.652 &  74\\
II.C$_{23}^{i.c.}$(2) &     0.5441086738 &     0.5425314916 &    71.4678190796 &    432.178 &  80\\
II.C$_{24}^{i.c.}$(2) &     0.4885885152 &     0.3251789493 &    58.3003419715 &    433.754 &  80\\
II.C$_{25}^{i.c.}$(2) &     0.7413837805 &     0.3131678015 &    77.5137279648 &    444.662 &  82\\
II.C$_{26}^{i.c.}$(2) &     0.7045220682 &     0.7108410127 &   116.6767631628 &    460.295 &  86\\
II.C$_{27}^{i.c.}$(2) &     0.7275403415 &     0.4376751471 &    89.1194892943 &    476.631 &  88\\
II.C$_{28}^{i.c.}$(2) &     0.4858219061 &     0.3440221728 &    67.5630284675 &    498.750 &  92\\
II.C$_{29}^{i.c.}$(2) &     0.7303738305 &     0.5310663433 &   104.5878736683 &    508.262 &  94\\
II.C$_{30}^{i.c.}$(2) &     0.6947435987 &     0.4176207645 &    97.1791580997 &    552.600 &  102\\
II.C$_{31}^{i.c.}$(2) &     0.7275761133 &     0.1015030833 &    87.4537840006 &    553.278 &  102\\
II.C$_{32}^{i.c.}$(2) &     0.5070848684 &     0.4982410822 &    87.9447232560 &    573.200 &  106\\
II.C$_{33}^{i.c.}$(2) &     0.7735256256 &     0.3869085928 &   110.3782010705 &    574.666 &  106\\
II.C$_{34}^{i.c.}$(2) &     0.7472783026 &     0.5729156410 &   134.2794021508 &    604.591 &  112\\
II.C$_{35}^{i.c.}$(2) &     0.4578292895 &     0.2350514143 &    76.7938750498 &    607.518 &  112\\
II.C$_{36}^{i.c.}$(2) &     0.5277658520 &     0.4270090654 &    91.2353195797 &    617.546 &  114\\
II.C$_{37}^{i.c.}$(2) &     0.4898331715 &     0.4967647615 &    96.3577454631 &    638.141 &  118\\
II.C$_{38}^{i.c.}$(2) &     0.7273677980 &     0.3285959518 &   114.1715748005 &    661.513 &  122\\
II.C$_{39}^{i.c.}$(2) &     0.4892532461 &     0.4325728753 &    97.9924501867 &    682.497 &  126\\
II.C$_{40}^{i.c.}$(2) &     0.4195342908 &     0.2326473489 &    84.1841967872 &    683.533 &  126\\
II.C$_{41}^{i.c.}$(2) &     0.5695732932 &     0.4135235691 &   105.4982596512 &    693.508 &  128\\
II.C$_{42}^{i.c.}$(2) &     0.4280215163 &     0.1399822283 &    85.1290496907 &    705.284 &  130\\
II.C$_{43}^{i.c.}$(2) &     0.7172059976 &     0.4819919772 &   139.1173195582 &    725.261 &  134\\
II.C$_{44}^{i.c.}$(2) &     0.4718680015 &     0.2668205044 &    96.7216297615 &    748.464 &  138\\
II.C$_{45}^{i.c.}$(2) &     0.7232215799 &     0.4151840324 &   138.3471806807 &    758.459 &  140\\
II.C$_{46}^{i.c.}$(2) &     0.5404093775 &     0.5230173790 &   122.4016509480 &    756.840 &  140\\
II.C$_{47}^{i.c.}$(2) &     0.7525655073 &     0.2564337465 &   130.4554783876 &    759.328 &  140\\
II.C$_{48}^{i.c.}$(2) &     0.6936999256 &     0.5700462542 &   160.6270068068 &    788.200 &  146\\
II.C$_{49}^{i.c.}$(2) &     0.4843126014 &     0.0844399577 &    98.3846313825 &    791.988 &  146\\
II.C$_{50}^{i.c.}$(2) &     0.4621277678 &     0.2887580227 &   105.3750210612 &    813.495 &  150\\
\hline
\end{tabular*}
{\rule{\temptablewidth}{1pt}}
\end{center}
\end{table*}

\begin{table*}
\tabcolsep 0pt \caption{Initial conditions and periods $T$ of the periodic three-body orbits for class II.C and II.D in the case of  $\bm{r}_1(0)=(-1,0)=-\bm{r}_2(0)$,  $\dot{\bm{r}}_1(0)=(v_1,v_2)=\dot{\bm{r}}_2(0)$ and $\bm{r}_3(0)=(0,0)$, $\dot{\bm{r}}_3(0)=(-2v_1/m_3, -2v_2/m_3)$ when $G=1$ and $m_1=m_2=1$ and $m_3=2$ by means of the search grid $4000\times 4000$ in the interval $T_0\in[0,200]$, where $T^*=T |E|^{3/2}$ is its scale-invariant period, $L_f$ is the length of the free group element.  } \label{table-S5} \vspace*{-12pt}
\begin{center}
\def\temptablewidth{1\textwidth}
{\rule{\temptablewidth}{1pt}}
\begin{tabular*}{\temptablewidth}{@{\extracolsep{\fill}}lccccc}
\hline
Class and number  & $v_1$ & $v_2$  & $T$  & $T^*$ & $L_f$\\
\hline
II.C$_{51}^{i.c.}$(2) &     0.7409842733 &     0.2548024080 &   141.1189791990 &    835.242 &  154\\
II.C$_{52}^{i.c.}$(2) &     0.5527686761 &     0.4114559610 &   124.7340630824 &    834.417 &  154\\
II.C$_{53}^{i.c.}$(2) &     0.4411825868 &     0.2014775073 &   104.6229338883 &    846.269 &  156\\
II.C$_{54}^{i.c.}$(2) &     0.5903490824 &     0.5542274770 &   149.8355193390 &    853.145 &  158\\
II.C$_{55}^{i.c.}$(2) &     0.5095309361 &     0.1076273234 &   110.2367826286 &    867.906 &  160\\
II.C$_{56}^{i.c.}$(2) &     0.6476911727 &     0.6592310093 &   186.5647808498 &    870.284 &  162\\
II.C$_{57}^{i.c.}$(2) &     0.7542236731 &     0.2870913910 &   157.4602317313 &    900.278 &  166\\
II.C$_{58}^{i.c.}$(2) &     0.7225383025 &     0.5182878319 &   180.0585926431 &    897.807 &  166\\
II.C$_{59}^{i.c.}$(2) &     0.6829265182 &     0.3949227774 &   155.0007436922 &    910.373 &  168\\
II.C$_{60}^{i.c.}$(2) &     0.4944962109 &     0.3354960370 &   126.5930724317 &    932.507 &  172\\
II.C$_{61}^{i.c.}$(2) &     0.5731579573 &     0.5228404784 &   157.1449116997 &    940.441 &  174\\
II.C$_{62}^{i.c.}$(2) &     0.4230035790 &     0.1230593685 &   113.2202463977 &    944.017 &  174\\
II.C$_{63}^{i.c.}$(2) &     0.5348626093 &     0.1239575520 &   122.6812047920 &    943.827 &  174\\
II.C$_{64}^{i.c.}$(2) &     0.7365486620 &     0.5209890777 &   197.7892719451 &    962.741 &  178\\
II.C$_{65}^{i.c.}$(2) &     0.5405892439 &     0.4105686806 &   144.0844304603 &    975.325 &  180\\
II.C$_{66}^{i.c.}$(2) &     0.5365593027 &     0.2631585954 &   132.5304152818 &    976.192 &  180\\
II.C$_{67}^{i.c.}$(2) &     0.6963830870 &     0.5384939358 &   198.4069997112 &   1005.363 &  186\\
II.C$_{68}^{i.c.}$(2) &     0.5612174596 &     0.1365651140 &   135.9019435025 &   1019.753 &  188\\
II.C$_{69}^{i.c.}$(2) &     0.6047899899 &     0.5655352150 &   185.3068878768 &   1025.552 &  190\\
II.C$_{70}^{i.c.}$(2) &     0.4608970442 &     0.0782609721 &   125.8586099496 &   1030.711 &  190\\
II.C$_{71}^{i.c.}$(2) &     0.5090586559 &     0.4172474858 &   150.1955209051 &   1040.279 &  192\\
II.C$_{72}^{i.c.}$(2) &     0.5153130739 &     0.2807750264 &   139.9401650184 &   1041.217 &  192\\
II.C$_{73}^{i.c.}$(2) &     0.5711368382 &     0.4009174040 &   158.7702675567 &   1051.278 &  194\\
II.C$_{74}^{i.c.}$(2) &     0.7330058203 &     0.2797227895 &   178.0148431024 &   1052.111 &  194\\
II.C$_{75}^{i.c.}$(2) &     0.4930200888 &     0.2105623302 &   138.1032522068 &   1073.977 &  198\\
II.C$_{76}^{i.c.}$(2) &     0.7071493167 &     0.3380296379 &   185.1011957070 &   1095.217 &  202\\
II.C$_{77}^{i.c.}$(2) &     0.4843031700 &     0.3104804420 &   145.6710898949 &   1095.339 &  202\\
II.C$_{78}^{i.c.}$(2) &     0.7021624722 &     0.1513651325 &   169.6462689466 &   1095.681 &  202\\
II.C$_{79}^{i.c.}$(2) &     0.5309569319 &     0.4100676515 &   163.4279841939 &   1116.232 &  206\\
II.C$_{80}^{i.c.}$(2) &     0.5863327862 &     0.2631956974 &   160.1895523872 &   1128.034 &  208\\
II.C$_{81}^{i.c.}$(2) &     0.5042933121 &     0.4522661483 &   167.6967610047 &   1137.013 &  210\\
II.C$_{82}^{i.c.}$(2) &     0.5127148367 &     0.2132236730 &   150.2626044139 &   1149.888 &  212\\
II.C$_{83}^{i.c.}$(2) &     0.4203013342 &     0.1117122153 &   141.3424549265 &   1182.749 &  218\\
II.C$_{84}^{i.c.}$(2) &     0.5567049092 &     0.4018329071 &   177.6958840052 &   1192.184 &  220\\
II.C$_{85}^{i.c.}$(2) &     0.7135436004 &     0.1993230096 &   197.0626213511 &   1236.670 &  228\\
II.C$_{86}^{i.c.}$(2) &     0.6251526217 &     0.3253104492 &   188.2411712699 &   1236.133 &  228\\
II.C$_{87}^{i.c.}$(2) &     0.5237375027 &     0.4094955579 &   182.8253593805 &   1257.140 &  232\\
II.C$_{88}^{i.c.}$(2) &     0.4819316406 &     0.4834824592 &   192.6723508500 &   1298.529 &  240\\
II.C$_{89}^{i.c.}$(2) &     0.6074378575 &     0.1960385648 &   185.5428302100 &   1312.598 &  242\\
II.C$_{90}^{i.c.}$(2) &     0.4839930914 &     0.2707143755 &   175.6526252637 &   1345.022 &  248\\
II.C$_{91}^{i.c.}$(2) &     0.5837038254 &     0.3300225890 &   199.9369167063 &   1366.100 &  252\\
II.C$_{92}^{i.c.}$(2) &     0.4690144649 &     0.2161480199 &   174.4018734709 &   1377.782 &  254\\
II.C$_{93}^{i.c.}$(2) &     0.5003326814 &     0.1974567430 &   188.6007498773 &   1464.529 &  270\\
II.C$_{94}^{i.c.}$(2) &     0.5678878694 &     0.1486900134 &   199.7846561418 &   1486.225 &  274\\
II.C$_{95}^{i.c.}$(2) &     0.5017212512 &     0.1238005807 &   199.1773956809 &   1573.086 &  290\\
II.D$_{1}^{i.c.}$(2) &     0.7583850283 &     0.9342270211 &    20.3253125540 &     41.295 &  4\\
II.D$_{2}^{i.c.}$(2) &     0.3057224330 &     0.5215124257 &     8.8237067653 &     64.567 &  12\\
\hline
\end{tabular*}
{\rule{\temptablewidth}{1pt}}
\end{center}
\end{table*}

\begin{table*}
\tabcolsep 0pt \caption{Initial conditions and periods $T$ of the periodic three-body orbits in the case of  $\bm{r}_1(0)=(-1,0)=-\bm{r}_2(0)$,  $\dot{\bm{r}}_1(0)=(v_1,v_2)=\dot{\bm{r}}_2(0)$ and $\bm{r}_3(0)=(0,0)$, $\dot{\bm{r}}_3(0)=(-2v_1/m_3, -2v_2/m_3)$ when $G=1$ and $m_1=m_2=1$ and $m_3=4$ by means of the search grid $4000\times 4000$ in the interval $T_0\in[0,200]$, where $T^*=T |E|^{3/2}$ is its scale-invariant period, $L_f$ is the length of the free group element.  } \label{table-S5} \vspace*{-12pt}
\begin{center}
\def\temptablewidth{1\textwidth}
{\rule{\temptablewidth}{1pt}}
\begin{tabular*}{\temptablewidth}{@{\extracolsep{\fill}}lccccc}
\hline
Class and number  & $v_1$ & $v_2$  & $T$  & $T^*$ & $L_f$\\
\hline
I.A$_{1}^{i.c.}$(4) &     0.6886546484 &     1.5771573890 &    16.3239594553 &    133.417 &  12\\
I.A$_{2}^{i.c.}$(4) &     0.9911981217 &     0.7119472124 &    17.6507807837 &    276.852 &  24\\
I.A$_{3}^{i.c.}$(4) &     0.9714516067 &     0.9749602268 &    54.7446177671 &    736.894 &  64\\
I.A$_{4}^{i.c.}$(4) &     0.9854347005 &     0.8533367754 &    53.9920673975 &    783.838 &  68\\
I.A$_{5}^{i.c.}$(4) &     0.9956481414 &     0.5378987979 &    51.9746508441 &    877.069 &  76\\
I.A$_{6}^{i.c.}$(4) &     0.9081405883 &     0.2793470372 &    48.3394213297 &    923.388 &  80\\
I.A$_{7}^{i.c.}$(4) &     0.7820517397 &     0.9985331599 &    79.6878556097 &   1196.730 &  104\\
I.A$_{8}^{i.c.}$(4) &     0.8303276771 &     0.8619453491 &    80.6381542416 &   1290.761 &  112\\
I.A$_{9}^{i.c.}$(4) &     0.9875993951 &     0.7994230783 &    89.2570135846 &   1337.585 &  116\\
I.A$_{10}^{i.c.}$(4) &     0.9941229185 &     0.6131386486 &    87.2715581867 &   1430.813 &  124\\
I.A$_{11}^{i.c.}$(4) &     0.9962878480 &     0.4962235209 &    86.3035106391 &   1477.247 &  128\\
I.A$_{12}^{i.c.}$(4) &     0.8893383967 &     0.3498550324 &    80.0242474362 &   1523.553 &  132\\
I.A$_{13}^{i.c.}$(4) &     0.9737611115 &     0.9412093990 &   127.0151385872 &   1750.771 &  152\\
I.A$_{14}^{i.c.}$(4) &     0.8483553587 &     0.8712965762 &   114.1831544918 &   1797.704 &  156\\
I.A$_{15}^{i.c.}$(4) &     0.8494308475 &     0.7648099341 &   113.0730029261 &   1891.278 &  164\\
I.A$_{16}^{i.c.}$(4) &     0.9933657595 &     0.6428233219 &   122.5711891715 &   1984.535 &  172\\
I.A$_{17}^{i.c.}$(4) &     0.9951416390 &     0.5658643691 &   121.5966724212 &   2031.019 &  176\\
I.A$_{18}^{i.c.}$(4) &     0.9975784126 &     0.3688413653 &   119.6830441188 &   2123.741 &  184\\
I.A$_{19}^{i.c.}$(4) &     0.9982910210 &     0.2119134110 &   118.7483807712 &   2169.985 &  188\\
I.A$_{20}^{i.c.}$(4) &     0.9739258865 &     0.9354381836 &   163.1207379635 &   2257.704 &  196\\
I.A$_{21}^{i.c.}$(4) &     1.0006092513 &     0.9001922281 &   165.7348113442 &   2304.647 &  200\\
I.A$_{22}^{i.c.}$(4) &     0.9871838469 &     0.8086143731 &   160.8692729101 &   2398.304 &  208\\
I.A$_{23}^{i.c.}$(4) &     0.9892584028 &     0.7616755213 &   159.8521879007 &   2445.023 &  212\\
I.A$_{24}^{i.c.}$(4) &     0.9103595990 &     0.6516159482 &   149.0219605514 &   2538.245 &  220\\
I.A$_{25}^{i.c.}$(4) &     0.9929178705 &     0.6587699243 &   157.8716678998 &   2538.249 &  220\\
I.A$_{26}^{i.c.}$(4) &     0.9739648867 &     0.9317358197 &   199.2150728253 &   2764.636 &  240\\
I.A$_{27}^{i.c.}$(4) &     0.8889807236 &     0.8833502365 &   184.6126736271 &   2811.576 &  244\\
I.A$_{28}^{i.c.}$(4) &     0.9853196522 &     0.8652680791 &   198.4147566034 &   2858.458 &  248\\
I.A$_{29}^{i.c.}$(4) &     0.8671151375 &     0.8139325244 &   180.3491172929 &   2905.263 &  252\\
I.A$_{30}^{i.c.}$(4) &     0.8593214437 &     0.7428413082 &   178.2861322510 &   2998.700 &  260\\
I.A$_{31}^{i.c.}$(4) &     0.9926227094 &     0.6687318678 &   193.1725286382 &   3091.960 &  268\\
I.A$_{32}^{i.c.}$(4) &     0.9938888216 &     0.6227285557 &   192.1926675081 &   3138.490 &  272\\
I.A$_{33}^{i.c.}$(4) &     0.9949961979 &     0.5732651491 &   191.2190326488 &   3184.966 &  276\\
I.B$_{1}^{i.c.}$(4) &     0.9686212391 &     1.0035538399 &    36.6553191032 &    483.414 &  42\\
I.B$_{2}^{i.c.}$(4) &     0.9866754572 &     0.8199780135 &    35.8070111326 &    530.358 &  46\\
I.B$_{3}^{i.c.}$(4) &     0.9947340612 &     0.5860022689 &    34.8112412541 &    576.973 &  50\\
I.B$_{4}^{i.c.}$(4) &     0.9854028827 &     0.8697357271 &    72.2201261804 &   1037.310 &  90\\
I.B$_{5}^{i.c.}$(4) &     0.9890031845 &     0.7676837385 &    71.1011788123 &   1084.083 &  94\\
I.B$_{6}^{i.c.}$(4) &     0.9931162203 &     0.6518383667 &    70.1106814873 &   1130.696 &  98\\
I.B$_{7}^{i.c.}$(4) &     0.9960573144 &     0.5122336176 &    69.1389009393 &   1177.159 &  102\\
I.B$_{8}^{i.c.}$(4) &     0.9978825808 &     0.3190517769 &    68.1889247106 &   1223.481 &  106\\
I.B$_{9}^{i.c.}$(4) &     0.9735740630 &     0.9455048919 &   108.9564778767 &   1497.304 &  130\\
I.B$_{10}^{i.c.}$(4) &     0.8392318736 &     0.8671916976 &    97.2973539376 &   1544.233 &  134\\
I.B$_{11}^{i.c.}$(4) &     0.8458015116 &     0.7418783799 &    96.5606171482 &   1637.758 &  142\\
I.B$_{12}^{i.c.}$(4) &     0.9925099954 &     0.6724319595 &   105.4115215859 &   1684.407 &  146\\
I.B$_{13}^{i.c.}$(4) &     0.9964353852 &     0.4852677552 &   103.4683068013 &   1777.334 &  154\\
I.B$_{14}^{i.c.}$(4) &     0.9976499399 &     0.3583646437 &   102.5181725853 &   1823.655 &  158\\
I.B$_{15}^{i.c.}$(4) &     0.8319211104 &     0.9612129024 &   130.6811566596 &   1957.203 &  170\\
I.B$_{16}^{i.c.}$(4) &     0.9738675697 &     0.9379697056 &   145.0696448675 &   2004.238 &  174\\
I.B$_{17}^{i.c.}$(4) &     0.9856690147 &     0.8450868125 &   143.7791552562 &   2098.037 &  182\\
I.B$_{18}^{i.c.}$(4) &     0.9878345000 &     0.7942119413 &   142.7087243978 &   2144.810 &  186\\
I.B$_{19}^{i.c.}$(4) &     0.9901313441 &     0.7402972658 &   141.7024169129 &   2191.502 &  190\\
I.B$_{20}^{i.c.}$(4) &     0.8549437310 &     0.6783963636 &   128.6681308892 &   2238.064 &  194\\

\hline
\end{tabular*}
{\rule{\temptablewidth}{1pt}}
\end{center}
\end{table*}

\begin{table*}
\tabcolsep 0pt \caption{Initial conditions and periods $T$ of the periodic three-body orbits in the case of  $\bm{r}_1(0)=(-1,0)=-\bm{r}_2(0)$,  $\dot{\bm{r}}_1(0)=(v_1,v_2)=\dot{\bm{r}}_2(0)$ and $\bm{r}_3(0)=(0,0)$, $\dot{\bm{r}}_3(0)=(-2v_1/m_3, -2v_2/m_3)$ when $G=1$ and $m_1=m_2=1$ and $m_3=4$ by means of the search grid $4000\times 4000$ in the interval $T_0\in[0,200]$, where $T^*=T |E|^{3/2}$ is its scale-invariant period, $L_f$ is the length of the free group element.  } \label{table-S5} \vspace*{-12pt}
\begin{center}
\def\temptablewidth{1\textwidth}
{\rule{\temptablewidth}{1pt}}
\begin{tabular*}{\temptablewidth}{@{\extracolsep{\fill}}lccccc}
\hline
Class and number  & $v_1$ & $v_2$  & $T$  & $T^*$ & $L_f$\\
\hline

I.B$_{21}^{i.c.}$(4) &     0.8844781422 &     0.4726698354 &   128.6082000866 &   2377.481 &  206\\
I.B$_{22}^{i.c.}$(4) &     0.9966129597 &     0.4712277307 &   137.7981863183 &   2377.505 &  206\\
I.B$_{23}^{i.c.}$(4) &     0.9429283989 &     0.3750820300 &   132.1379728542 &   2423.826 &  210\\
I.B$_{24}^{i.c.}$(4) &     0.8923470368 &     0.9533088376 &   170.1093582167 &   2464.159 &  214\\
I.B$_{25}^{i.c.}$(4) &     0.8775052711 &     0.8805399726 &   166.4296589791 &   2558.109 &  222\\
I.B$_{26}^{i.c.}$(4) &     0.9920009809 &     0.6884914607 &   176.0140436201 &   2791.821 &  242\\
I.B$_{27}^{i.c.}$(4) &     0.9967158849 &     0.4626061527 &   172.1282612303 &   2977.676 &  258\\
I.B$_{28}^{i.c.}$(4) &     0.8634040830 &     0.5193955710 &   191.6136346322 &   3531.367 &  306\\
II.A$_{1}^{i.c.}$(4) &     0.8193973154 &     0.9283001929 &    80.5646820232 &   1243.829 &  108\\
II.A$_{2}^{i.c.}$(4) &     0.8723346349 &     0.9762996085 &   118.0177898281 &   1703.721 &  148\\
II.B$_{1}^{i.c.}$(4) &     0.9597587596 &     0.1504623356 &    33.0556151005 &    623.296 &  54\\
II.B$_{2}^{i.c.}$(4) &     0.8028948361 &     0.9363733297 &    63.8749643029 &    990.355 &  86\\
II.B$_{3}^{i.c.}$(4) &     0.8810167233 &     0.3087921811 &   159.1584664346 &   3070.198 &  266\\
II.C$_{1}^{i.c.}$(4) &     0.8711838791 &     0.4362455003 &    15.9419555705 &    300.076 &  26\\
II.C$_{2}^{i.c.}$(4) &     0.9037352027 &     0.7742282374 &    50.1414328025 &    807.223 &  70\\
II.C$_{3}^{i.c.}$(4) &     0.8360259826 &     0.7607468289 &    64.1746649704 &   1084.046 &  94\\
II.C$_{4}^{i.c.}$(4) &     0.8741665029 &     0.6579865343 &    81.0825827121 &   1407.540 &  122\\
II.C$_{5}^{i.c.}$(4) &     0.9222679496 &     0.2364217777 &    81.0640765674 &   1546.686 &  134\\
II.C$_{6}^{i.c.}$(4) &     0.7727255378 &     0.9813722168 &   109.9022712350 &   1680.144 &  146\\
II.C$_{7}^{i.c.}$(4) &     0.8493658092 &     0.8187152772 &   113.5938429128 &   1844.540 &  160\\
II.C$_{8}^{i.c.}$(4) &     0.8864141076 &     0.7338054599 &   115.1717551193 &   1914.642 &  166\\
II.C$_{9}^{i.c.}$(4) &     0.8572285903 &     0.5716988114 &   111.8713143932 &   2030.945 &  176\\
II.C$_{10}^{i.c.}$(4) &     0.9564378250 &     0.5203631283 &   117.8780060823 &   2054.229 &  178\\
II.C$_{11}^{i.c.}$(4) &     0.7920712457 &     0.9627630869 &   142.7572749740 &   2187.130 &  190\\
II.C$_{12}^{i.c.}$(4) &     0.8735053580 &     0.6463061859 &   129.5549674540 &   2261.370 &  196\\
II.C$_{13}^{i.c.}$(4) &     0.9164332753 &     0.2533703569 &   129.3612607156 &   2470.074 &  214\\
II.C$_{14}^{i.c.}$(4) &     0.7800111798 &     0.9764870056 &   173.2559877508 &   2646.981 &  230\\
II.C$_{15}^{i.c.}$(4) &     0.9103776228 &     0.5004553050 &   147.0538184568 &   2654.401 &  230\\
II.C$_{16}^{i.c.}$(4) &     0.8581304222 &     0.7667302298 &   162.3271446419 &   2698.506 &  234\\
II.C$_{17}^{i.c.}$(4) &     0.9287736223 &     0.8013022448 &   187.7523645240 &   2928.662 &  254\\
II.C$_{18}^{i.c.}$(4) &     0.9040085960 &     0.6611792882 &   181.5941857827 &   3091.954 &  268\\
II.C$_{19}^{i.c.}$(4) &     0.9415767476 &     0.5157985737 &   183.4012996390 &   3231.388 &  280\\

\hline
\end{tabular*}
{\rule{\temptablewidth}{1pt}}
\end{center}
\end{table*}

\begin{table*}
\tabcolsep 0pt \caption{Initial conditions and periods $T$ of the periodic three-body orbits in the case of  $\bm{r}_1(0)=(-1,0)=-\bm{r}_2(0)$,  $\dot{\bm{r}}_1(0)=(v_1,v_2)=\dot{\bm{r}}_2(0)$ and $\bm{r}_3(0)=(0,0)$, $\dot{\bm{r}}_3(0)=(-2v_1/m_3, -2v_2/m_3)$ when $G=1$ and $m_1=m_2=1$ and $m_3=5$ by means of the search grid $4000\times 4000$ in the interval $T_0\in[0,200]$, where $T^*=T |E|^{3/2}$ is its scale-invariant period, $L_f$ is the length of the free group element.  } \label{table-S5} \vspace*{-12pt}
\begin{center}
\def\temptablewidth{1\textwidth}
{\rule{\temptablewidth}{1pt}}
\begin{tabular*}{\temptablewidth}{@{\extracolsep{\fill}}lccccc}
\hline
Class and number  & $v_1$ & $v_2$  & $T$  & $T^*$ & $L_f$\\
\hline
I.A$_{1}^{i.c.}$(5) &     0.9216454928 &     0.8533067851 &    53.9388331345 &   1287.788 &  88\\
I.A$_{2}^{i.c.}$(5) &     0.8829781875 &     0.4605442830 &    87.3008518130 &   2401.081 &  164\\
I.A$_{3}^{i.c.}$(5) &     0.9894241760 &     0.7387990271 &   128.2645973796 &   3103.339 &  212\\
I.A$_{4}^{i.c.}$(5) &     0.9318317248 &     0.7011947975 &   160.3140070092 &   4040.338 &  276\\
I.A$_{5}^{i.c.}$(5) &     0.9521954098 &     0.6380537638 &   160.8351666149 &   4099.305 &  280\\
I.A$_{6}^{i.c.}$(5) &     0.9265055362 &     0.9102619607 &   199.1483651043 &   4623.496 &  316\\
I.A$_{7}^{i.c.}$(5) &     0.9414948707 &     0.7972769172 &   198.2855107755 &   4800.760 &  328\\
I.B$_{1}^{i.c.}$(5) &     0.9832102827 &     0.2610181545 &    35.4899632829 &    966.423 &  66\\
I.B$_{2}^{i.c.}$(5) &     0.9328981538 &     0.9308414732 &    72.7716713885 &   1667.838 &  114\\
I.B$_{3}^{i.c.}$(5) &     0.9272174116 &     0.7235415686 &    71.2641428314 &   1785.849 &  122\\
I.B$_{4}^{i.c.}$(5) &     0.9299236101 &     0.7420212724 &   107.1531608781 &   2664.097 &  182\\
I.B$_{5}^{i.c.}$(5) &     0.9364988380 &     0.4466549140 &   105.3306842497 &   2840.543 &  194\\
I.B$_{6}^{i.c.}$(5) &     0.9185948030 &     0.9105452579 &   144.4229235715 &   3365.161 &  230\\
I.B$_{7}^{i.c.}$(5) &     0.9451359068 &     0.8566746195 &   145.2514450303 &   3424.391 &  234\\
I.B$_{8}^{i.c.}$(5) &     0.9785669623 &     0.4482186236 &   142.8066562282 &   3777.715 &  258\\
I.B$_{9}^{i.c.}$(5) &     0.9275282956 &     0.9587654523 &   182.0138798215 &   4125.181 &  282\\
I.B$_{10}^{i.c.}$(5) &     0.9473861305 &     0.8410107115 &   181.4678831986 &   4302.643 &  294\\
I.B$_{11}^{i.c.}$(5) &     0.9750842086 &     0.5298658837 &   179.0967959379 &   4656.080 &  318\\
II.B$_{1}^{i.c.}$(5) &     0.9093722173 &     0.9517890604 &   108.3964675729 &   2486.840 &  170\\
II.C$_{1}^{i.c.}$(5) &     0.9557672644 &     0.7750957916 &    18.1014388854 &    439.118 &  30\\
II.C$_{2}^{i.c.}$(5) &     0.9440503938 &     0.4093724806 &    70.2927303619 &   1903.491 &  130\\
II.C$_{3}^{i.c.}$(5) &     0.9379933241 &     0.9019570416 &    90.9039033637 &   2106.988 &  144\\
II.C$_{4}^{i.c.}$(5) &     0.9796113137 &     0.4175042061 &    89.1520279744 &   2372.076 &  162\\
II.C$_{5}^{i.c.}$(5) &     0.9495004417 &     0.5575440062 &   106.5164092109 &   2781.860 &  190\\
II.C$_{6}^{i.c.}$(5) &     0.9541817480 &     0.3673032901 &   105.6505055608 &   2869.923 &  196\\
II.C$_{7}^{i.c.}$(5) &     0.9375025806 &     0.8054315397 &   144.0759759853 &   3483.402 &  238\\
II.C$_{8}^{i.c.}$(5) &     0.9589010451 &     0.9949666993 &   166.9876407055 &   3656.438 &  250\\
II.C$_{9}^{i.c.}$(5) &     0.9120508681 &     0.9598994535 &   162.8143855219 &   3715.515 &  254\\
II.C$_{10}^{i.c.}$(5) &     0.9740372231 &     0.9339142055 &   167.0634447329 &   3745.309 &  256\\
II.C$_{11}^{i.c.}$(5) &     0.9899851594 &     0.8685305652 &   167.1856631406 &   3834.035 &  262\\
II.C$_{12}^{i.c.}$(5) &     0.9940186923 &     0.9576808872 &   188.1862788803 &   4125.284 &  282\\
II.C$_{13}^{i.c.}$(5) &     0.9709140296 &     0.6030586716 &   161.6989567743 &   4128.753 &  282\\
II.C$_{14}^{i.c.}$(5) &     0.9481881607 &     0.8352390313 &   199.5740996229 &   4741.767 &  324\\
II.C$_{15}^{i.c.}$(5) &     0.9625713434 &     0.7097859647 &   198.5700198646 &   4918.722 &  336\\
II.C$_{16}^{i.c.}$(5) &     0.9736869452 &     0.5566185906 &   197.2313607831 &   5095.245 &  348\\
\hline
\end{tabular*}
{\rule{\temptablewidth}{1pt}}
\end{center}
\end{table*}

\begin{table*}
\tabcolsep 0pt \caption{Initial conditions and periods $T$ of the periodic three-body orbits in the case of  $\bm{r}_1(0)=(-1,0)=-\bm{r}_2(0)$,  $\dot{\bm{r}}_1(0)=(v_1,v_2)=\dot{\bm{r}}_2(0)$ and $\bm{r}_3(0)=(0,0)$, $\dot{\bm{r}}_3(0)=(-2v_1/m_3, -2v_2/m_3)$ when $G=1$ and $m_1=m_2=1$ and $m_3=8$ by means of the search grid $4000\times 4000$ in the interval $T_0\in[0,200]$, where $T^*=T |E|^{3/2}$ is its scale-invariant period, $L_f$ is the length of the free group element.  } \label{table-S5} \vspace*{-12pt}
\begin{center}
\def\temptablewidth{1\textwidth}
{\rule{\temptablewidth}{1pt}}
\begin{tabular*}{\temptablewidth}{@{\extracolsep{\fill}}lccccc}
\hline
Class and number  & $v_1$ & $v_2$  & $T$  & $T^*$ & $L_f$\\
\hline
I.A$_{1}^{i.c.}$(8) &     0.9423354093 &     0.7626700078 &    64.9545993260 &   3647.042 &  152\\
I.A$_{2}^{i.c.}$(8) &     0.9787845584 &     0.9658431710 &   110.1464503910 &   5854.357 &  244\\
I.A$_{3}^{i.c.}$(8) &     0.9362611866 &     0.9596217871 &   153.3805946371 &   8253.495 &  344\\
I.A$_{4}^{i.c.}$(8) &     0.9533667422 &     0.8663093755 &   196.1893235979 &  10749.369 &  448\\
I.B$_{1}^{i.c.}$(8) &     0.8766620953 &     0.9033499768 &    44.2205415861 &   2446.480 &  102\\
I.B$_{2}^{i.c.}$(8) &     0.9457785062 &     0.9086191760 &    87.3950286043 &   4750.723 &  198\\
I.B$_{3}^{i.c.}$(8) &     0.8915770541 &     0.8430352261 &    86.7011183501 &   4845.690 &  202\\
II.B$_{1}^{i.c.}$(8) &     0.9932653784 &     0.7033211462 &   173.7686930189 &   9742.249 &  406\\
II.C$_{1}^{i.c.}$(8) &     0.9097066120 &     0.8777115746 &    21.7197890174 &   1199.563 &  50\\
II.C$_{2}^{i.c.}$(8) &     0.9581798536 &     0.6904382924 &    43.1758329286 &   2447.457 &  102\\
II.C$_{3}^{i.c.}$(8) &     0.9847000702 &     0.8980796629 &    65.8092926553 &   3551.151 &  148\\
II.C$_{4}^{i.c.}$(8) &     0.9767252634 &     0.6098206184 &    64.6224044091 &   3695.308 &  154\\
II.C$_{5}^{i.c.}$(8) &     0.9326547771 &     0.8868649644 &    86.6725399311 &   4750.725 &  198\\
II.C$_{6}^{i.c.}$(8) &     0.9488792658 &     0.5640461916 &    86.1048324315 &   4990.663 &  208\\
II.C$_{7}^{i.c.}$(8) &     0.9932645266 &     0.5364010933 &   107.5645277200 &   6190.987 &  258\\
II.C$_{8}^{i.c.}$(8) &     0.9300079407 &     0.6918905419 &   129.5219662492 &   7389.860 &  308\\
II.C$_{9}^{i.c.}$(8) &     0.9215017059 &     0.7440903890 &   172.8653044238 &   9789.022 &  408\\
\hline
\end{tabular*}
{\rule{\temptablewidth}{1pt}}
\end{center}
\end{table*}

\begin{table*}
\tabcolsep 0pt \caption{Initial conditions and periods $T$ of the periodic three-body orbits for class I.A in the case of  $\bm{r}_1(0)=(-1,0)=-\bm{r}_2(0)$,  $\dot{\bm{r}}_1(0)=(v_1,v_2)=\dot{\bm{r}}_2(0)$ and $\bm{r}_3(0)=(0,0)$, $\dot{\bm{r}}_3(0)=(-2v_1/m_3, -2v_2/m_3)$ when $G=1$ and $m_1=m_2=1$ and $m_3=10$ by means of the search grid $4000\times 4000$ in the interval $T_0\in[0,200]$, where $T^*=T |E|^{3/2}$ is its scale-invariant period, $L_f$ is the length of the free group element.  } \label{table-S5} \vspace*{-12pt}
\begin{center}
\def\temptablewidth{1\textwidth}
{\rule{\temptablewidth}{1pt}}
\begin{tabular*}{\temptablewidth}{@{\extracolsep{\fill}}lccccc}
\hline
Class and number  & $v_1$ & $v_2$  & $T$  & $T^*$ & $L_f$\\
\hline
I.A$_{1}^{i.c.}$(10) &     0.8854191796 &     0.4245045883 &    24.1711064176 &   2056.279 &  68\\
I.A$_{2}^{i.c.}$(10) &     0.9689819274 &     0.2200633851 &    71.2682278153 &   6049.841 &  200\\
I.A$_{3}^{i.c.}$(10) &     0.9262313425 &     0.3318793235 &   119.5012751111 &  10162.520 &  336\\
I.B$_{1}^{i.c.}$(10) &     0.9399260818 &     0.3073705720 &    47.7043741335 &   4053.099 &  134\\
II.A$_{1}^{i.c.}$(10) &     0.9565884486 &     0.8490970249 &    24.2568013892 &   1935.907 &  64\\
II.B$_{1}^{i.c.}$(10) &     0.9986145269 &     0.5935168720 &    47.9360511650 &   3932.658 &  130\\
II.C$_{1}^{i.c.}$(10) &     0.9639216986 &     0.9744362737 &    48.1316203628 &   3751.177 &  124\\
II.C$_{2}^{i.c.}$(10) &     0.9254840675 &     0.9517296411 &    49.8685206945 &   3931.244 &  130\\
II.C$_{3}^{i.c.}$(10) &     0.9637413393 &     0.4820091574 &    95.6185104257 &   7985.803 &  264\\
\hline
\end{tabular*}
{\rule{\temptablewidth}{1pt}}
\end{center}
\end{table*}

\begin{sidewaystable}
\tabcolsep 0pt \caption{The free group elements for the periodic three-body orbits.} \label{table-S8} \vspace*{-12pt}
\begin{center}
\def\temptablewidth{1\textwidth}
{\rule{\temptablewidth}{1pt}}
\begin{tabular*}{\temptablewidth}{@{\extracolsep{\fill}}ll}
\hline
Class and number & \hspace{1.0cm}  free group element\\\hline
I.A$_{1}^{i.c.}$(0.5) & BabA\\
I.A$_{2}^{i.c.}$(0.5) & BaBabAbA\\
I.A$_{3}^{i.c.}$(0.5) & BaBaBabAbAbA\\
I.A$_{4}^{i.c.}$(0.5) & BaBAbaBabAbaBAbA\\
I.A$_{5}^{i.c.}$(0.5) & BaBAbAbaBabAbaBaBAbA\\
I.A$_{6}^{i.c.}$(0.5) & BabAbAbABabABaBaBabA\\
I.A$_{7}^{i.c.}$(0.5) & BAbAbAbAbabaBaBaBaBA\\
\hline
\end{tabular*}
{\rule{\temptablewidth}{1pt}}
\end{center}
\end{sidewaystable}

\begin{sidewaystable}
\tabcolsep 0pt \caption{The free group elements for the periodic three-body orbits.} \label{table-S8} \vspace*{-12pt}
\begin{center}
\def\temptablewidth{1\textwidth}
{\rule{\temptablewidth}{1pt}}
\begin{tabular*}{\temptablewidth}{@{\extracolsep{\fill}}ll}
\hline
Class and number & \hspace{1.0cm}  free group element\\\hline
I.A$_{41}^{i.c.}$(0.5) & BAbAbabABabaBabABAbAbabABaBABabABAbAbabaBaBABabABAbABabaBaBABabAbabABabaBaBA\\
I.A$_{42}^{i.c.}$(0.5) & BAbAbabABaBABaBAbabAbABAbaBabaBabABAbAbabaBaBABabAbabAbaBABaBabaBAbABAbABabaBaBA\\
I.A$_{43}^{i.c.}$(0.5) & BabAbaBAbaBabABabAbaBAbABabABaBAbaBAbABabABaBAbaBAbABabABaBAbaBabABabAbaBAbaBabA\\
I.A$_{44}^{i.c.}$(0.5) & BabABabABaBAbaBAbaBAbaBAbaBAbaBabABabABabABabABabAbaBAbaBAbaBAbaBAbaBAbABabABabA\\
I.A$_{45}^{i.c.}$(0.5) & BabABabABabaBAbaBAbaBAbaBAbaBABabABabABabABabABabABAbaBAbaBAbaBAbaBAbabABabABabA\\
I.A$_{46}^{i.c.}$(0.5) & BabABabaBabaBAbABAbaBAbabAbaBABaBABabABabABabABAbABAbaBabaBAbaBABaBAbabAbabABabA\\
I.A$_{47}^{i.c.}$(0.5) & BabAbaBAbaBAbaBabABabABabABaBAbaBAbaBAbABabABaBAbaBAbaBAbABabABabABabAbaBAbaBAbaBabA\\
\hline
\end{tabular*}
{\rule{\temptablewidth}{1pt}}
\end{center}
\end{sidewaystable}

\begin{sidewaystable}
\tabcolsep 0pt \caption{The free group elements for the periodic three-body orbits.} \label{table-S8} \vspace*{-12pt}
\begin{center}
\def\temptablewidth{1\textwidth}
{\rule{\temptablewidth}{1pt}}
\begin{tabular*}{\temptablewidth}{@{\extracolsep{\fill}}ll}
\hline
Class and number & \hspace{1.0cm}  free group element\\\hline
I.A$_{81}^{i.c.}$(0.5) & BabABabABabABabAbaBAbaBAbaBAbaBAbaBAbaBAbaBAbABabABabABabABabABabABabABabABaBAbaBAbaBAbaBAbaBAbaBAbaBAbaBa\\& bABabABabABabA\\
I.A$_{82}^{i.c.}$(0.5) & BabAbaBAbaBAbaBabABabABabAbaBAbaBAbABabABabABaBAbaBAbaBAbABabABaBAbaBAbaBAbABabABabABaBAbaBAbaBabABabABabA\\&baBAbaBAbaBabA\\
I.A$_{83}^{i.c.}$(0.5) & BabABabABabABabaBAbaBAbaBAbaBAbaBAbaBAbaBAbaBABabABabABabABabABabABabABabABAbaBAbaBAbaBAbaBAbaBAbaBAbaBAba\\&bABabABabABabA\\
I.A$_{84}^{i.c.}$(0.5) & BabAbaBAbaBAbABabABaBAbaBAbABabABabAbaBAbaBabABabAbaBAbaBAbABabABaBAbaBAbaBabABabAbaBAbaBabABabABaBAbaBAbA\\
\hline
\end{tabular*}
{\rule{\temptablewidth}{1pt}}
\end{center}
\end{sidewaystable}

\begin{sidewaystable}
\tabcolsep 0pt \caption{The free group elements for the periodic three-body orbits.} \label{table-S8} \vspace*{-12pt}
\begin{center}
\def\temptablewidth{1\textwidth}
{\rule{\temptablewidth}{1pt}}
\begin{tabular*}{\temptablewidth}{@{\extracolsep{\fill}}ll}
\hline
Class and number & \hspace{1.0cm}  free group element\\\hline
I.A$_{105}^{i.c.}$(0.5) & BAbAbabABaBABabAbabAbaBABaBAbabAbaBABaBabaBAbABAbaBabaBAbABAbABabaBabABAbAbabaBaBABabAbabABaBABaBAbabAbaB\\&ABaBAbabAbABAbaBabaBAbABAbaBabaBabABAbABabaBaBA\\
I.A$_{106}^{i.c.}$(0.5) & BAbAbabABabaBabABAbABAbaBabaBAbABAbaBAbaBAbabAbaBABaBAbabAbabABaBABabABAbAbabaBaBABabABAbABabaBabaBAbABAb\\&aBabaBAbaBAbaBABaBAbabAbaBABaBABabAbabABabaBaBA\\
I.A$_{107}^{i.c.}$(0.5) & BabABabABabABabABabAbaBAbaBAbaBAbaBAbaBAbaBAbaBAbaBAbaBAbABabABabABabABabABabABabABabABabABabABaBAbaBAbaB\\&AbaBAbaBAbaBAbaBAbaBAbaBAbaBabABabABabABabABabA\\
I.A$_{108}^{i.c.}$(0.5) & BabaBabABAbaBABaBAbabABabaBabABAbABABaBababAbabABaBABabABAbaBabaBAbabABaBABabABAbABabaBAbabAbaBABabABAbAB\\
\hline
\end{tabular*}
{\rule{\temptablewidth}{1pt}}
\end{center}
\end{sidewaystable}

\begin{sidewaystable}
\tabcolsep 0pt \caption{The free group elements for the periodic three-body orbits.} \label{table-S8} \vspace*{-12pt}
\begin{center}
\def\temptablewidth{1\textwidth}
{\rule{\temptablewidth}{1pt}}
\begin{tabular*}{\temptablewidth}{@{\extracolsep{\fill}}ll}
\hline
Class and number & \hspace{1.0cm}  free group element\\\hline
I.A$_{126}^{i.c.}$(0.5) & BabABabAbaBAbaBAbaBAbaBabABabABabABabAbaBAbaBAbaBAbABabABabABabABaBAbaBAbaBAbaBAbABabABabABabABaBAbaBAbaB\\&AbaBAbABabABabABabABaBAbaBAbaBAbaBabABabABabABabAbaBAbaBAbaBAbaBabABabA\\
I.A$_{127}^{i.c.}$(0.5) & BabABabaBAbaBAbaBAbaBABabABabABabABabaBAbaBAbaBAbaBABabABabABabABabaBAbaBAbaBAbaBABabABabABabABAbaBAbaBAb\\&aBAbabABabABabABabABAbaBAbaBAbaBAbabABabABabABabABAbaBAbaBAbaBAbabABabA\\
I.A$_{128}^{i.c.}$(0.5) & BAbAbabABabABabABabABabaBaBABAbAbabABaBABabABabABabaBabABAbAbabaBaBABabABabABabABabABAbAbabaBaBABabABabAB\\&abABabABAbAbabaBaBABabAbabABabABabABAbABabaBaBABAbAbabABabABabABabABabaBaBA\\
I.A$_{129}^{i.c.}$(0.5) & BabAbaBAbaBabABabABaBAbaBAbABabABaBAbaBAbaBabABaBAbaBAbaBabABabAbaBAbaBabABabABaBAbaBAbABabABaBAbaBAbABab\\
\hline
\end{tabular*}
{\rule{\temptablewidth}{1pt}}
\end{center}
\end{sidewaystable}

\begin{sidewaystable}
\tabcolsep 0pt \caption{The free group elements for the periodic three-body orbits.} \label{table-S8} \vspace*{-12pt}
\begin{center}
\def\temptablewidth{1\textwidth}
{\rule{\temptablewidth}{1pt}}
\begin{tabular*}{\temptablewidth}{@{\extracolsep{\fill}}ll}
\hline
Class and number & \hspace{1.0cm}  free group element\\\hline
I.B$_{1}^{i.c.}$(0.5) & BaBAbA\\
I.B$_{2}^{i.c.}$(0.5) & BaBaBAbAbA\\
I.B$_{3}^{i.c.}$(0.5) & BAbAbABABaBaBA\\
I.B$_{4}^{i.c.}$(0.5) & BabAbaBAbaBabA\\
I.B$_{5}^{i.c.}$(0.5) & BaBaBaBAbAbAbA\\
I.B$_{6}^{i.c.}$(0.5) & BabABAbaBAbaBABabA\\
I.B$_{7}^{i.c.}$(0.5) & BaBabaBaBAbAbabAbA\\
\hline
\end{tabular*}
{\rule{\temptablewidth}{1pt}}
\end{center}
\end{sidewaystable}

\begin{sidewaystable}
\tabcolsep 0pt \caption{The free group elements for the periodic three-body orbits.} \label{table-S8} \vspace*{-12pt}
\begin{center}
\def\temptablewidth{1\textwidth}
{\rule{\temptablewidth}{1pt}}
\begin{tabular*}{\temptablewidth}{@{\extracolsep{\fill}}ll}
\hline
Class and number & \hspace{1.0cm}  free group element\\\hline
I.B$_{41}^{i.c.}$(0.5) & BAbAbaBABaBabABAbAbABAbaBaBABabAbABABaBabABAbAbaBABaBaBABabAbABAbaBaBA\\
I.B$_{42}^{i.c.}$(0.5) & BabAbaBABaBaBABabAbaBABaBaBABabAbaBAbaBabABAbAbABAbaBabABAbAbABAbaBabA\\
I.B$_{43}^{i.c.}$(0.5) & BabABabABabAbabABaBAbabAbaBAbaBAbaBAbaBAbaBAbaBabaBAbABabaBabABabABabA\\
I.B$_{44}^{i.c.}$(0.5) & BAbAbabABaBABabAbabAbaBABaBAbabAbABABaBabaBAbABAbaBabaBabABAbABabaBaBA\\
I.B$_{45}^{i.c.}$(0.5) & BabABabaBAbaBAbaBABabABabABabaBAbaBAbaBAbabABabABabABAbaBAbaBAbabABabA\\
I.B$_{46}^{i.c.}$(0.5) & BabABaBAbaBAbaBAbaBabABabABabABaBAbaBAbaBAbABabABabABabAbaBAbaBAbaBAbABabA\\
I.B$_{47}^{i.c.}$(0.5) & BabABabABabABabABabaBAbaBAbaBAbaBAbaBAbaBAbaBAbaBAbaBAbabABabABabABabABabA\\
\hline
\end{tabular*}
{\rule{\temptablewidth}{1pt}}
\end{center}
\end{sidewaystable}

\begin{sidewaystable}
\tabcolsep 0pt \caption{The free group elements for the periodic three-body orbits.} \label{table-S8} \vspace*{-12pt}
\begin{center}
\def\temptablewidth{1\textwidth}
{\rule{\temptablewidth}{1pt}}
\begin{tabular*}{\temptablewidth}{@{\extracolsep{\fill}}ll}
\hline
Class and number & \hspace{1.0cm}  free group element\\\hline
I.B$_{81}^{i.c.}$(0.5) & BAbABabaBabaBAbABAbaBAbaBAbabABabABabABAbABabaBabaBAbABABaBAbabAbabABaBABabABabABabaBAbaBAbaBABaBAbabA\\&babABaBA\\
I.B$_{82}^{i.c.}$(0.5) & BabABabABabABabABabABabABabAbaBAbaBAbaBAbaBAbaBAbaBAbaBAbaBAbaBAbaBAbaBAbaBAbaBAbaBabABabABabABabABabA\\&BabABabA\\
I.B$_{83}^{i.c.}$(0.5) & BabAbabABaBAbabAbABAbaBabaBAbABabaBabABAbAbabABaBAbabAbaBAbaBabaBAbABabaBaBABabAbabABaBAbabAbaBABaBaba\\&BAbABabaBabA\\
I.B$_{84}^{i.c.}$(0.5) & BabAbaBAbaBAbABabABaBAbaBAbaBabABabABaBAbaBAbABabABabAbaBAbaBabABabABaBAbaBAbABabABabAbaBAbaBAbABabABa\\
\hline
\end{tabular*}
{\rule{\temptablewidth}{1pt}}
\end{center}
\end{sidewaystable}

\begin{sidewaystable}
\tabcolsep 0pt \caption{The free group elements for the periodic three-body orbits.} \label{table-S8} \vspace*{-12pt}
\begin{center}
\def\temptablewidth{1\textwidth}
{\rule{\temptablewidth}{1pt}}
\begin{tabular*}{\temptablewidth}{@{\extracolsep{\fill}}ll}
\hline
Class and number & \hspace{1.0cm}  free group element\\\hline
I.B$_{101}^{i.c.}$(0.5) & BabAbaBAbaBabABabAbaBAbaBAbABabABaBAbaBAbABabABabAbaBAbaBabABabAbaBAbaBabABabAbaBAbaBabABabABaBAbaBAb\\&ABabABaBAbaBAbaBabABabAbaBAbaBabA\\
I.B$_{102}^{i.c.}$(0.5) & BabAbaBAbaBabABabABaBAbaBAbABabABaBAbaBAbABabABaBAbaBAbaBabABabAbaBAbaBabABabAbaBAbaBAbABabABaBAbaBAb\\&ABabABaBAbaBAbABabABabAbaBAbaBabA\\
I.B$_{103}^{i.c.}$(0.5) & BabABabABAbABAbaBAbaBAbaBABaBAbabAbabABaBABabABabABabABAbABAbaBAbaBAbaBAbaBABaBABabABabABabABAbABabaB\\&abaBAbABAbaBAbaBAbaBABaBABabABabA\\
I.B$_{104}^{i.c.}$(0.5) & BabABabABabABabABabABabABabABabABaBAbaBAbaBAbaBAbaBAbaBAbaBAbaBAbaBAbaBAbaBAbaBAbaBAbaBAbaBAbaBAbaBAb\\
\hline
\end{tabular*}
{\rule{\temptablewidth}{1pt}}
\end{center}
\end{sidewaystable}

\begin{sidewaystable}
\tabcolsep 0pt \caption{The free group elements for the periodic three-body orbits.} \label{table-S8} \vspace*{-12pt}
\begin{center}
\def\temptablewidth{1\textwidth}
{\rule{\temptablewidth}{1pt}}
\begin{tabular*}{\temptablewidth}{@{\extracolsep{\fill}}ll}
\hline
Class and number & \hspace{1.0cm}  free group element\\\hline
I.B$_{121}^{i.c.}$(0.5) & BAbAbabABaBABabAbabABaBABabAbabABaBABabAbaBABaBAbabAbaBABaBAbabAbaBABaBAbabAbABABaBabaBAbABAbaBabaBAb\\&ABAbaBabaBAbABAbaBabABAbABabaBabABAbABabaBabABAbABabaBaBA\\
I.B$_{122}^{i.c.}$(0.5) & BabABabABabABabABabABabABabABabABabABabAbaBAbaBAbaBAbaBAbaBAbaBAbaBAbaBAbaBAbaBAbaBAbaBAbaBAbaBAbaBAb\\&aBAbaBAbaBAbaBAbaBabABabABabABabABabABabABabABabABabABabA\\
I.B$_{123}^{i.c.}$(0.5) & BabABabAbaBAbaBAbaBAbaBAbABabABabABabABaBAbaBAbaBAbaBAbABabABabABabABabAbaBAbaBAbaBAbaBabABabABabABab\\&ABaBAbaBAbaBAbaBAbABabABabABabABaBAbaBAbaBAbaBAbaBabABabA\\
I.B$_{124}^{i.c.}$(0.5) & BabABabaBAbaBAbaBABabABabABabaBAbaBAbaBABabABabABabaBAbaBAbaBABabABabABabaBAbaBAbaBAbabABabABabABAbaB\\
\hline
\end{tabular*}
{\rule{\temptablewidth}{1pt}}
\end{center}
\end{sidewaystable}

\begin{sidewaystable}
\tabcolsep 0pt \caption{The free group elements for the periodic three-body orbits.} \label{table-S8} \vspace*{-12pt}
\begin{center}
\def\temptablewidth{1\textwidth}
{\rule{\temptablewidth}{1pt}}
\begin{tabular*}{\temptablewidth}{@{\extracolsep{\fill}}ll}
\hline
Class and number & \hspace{1.0cm}  free group element\\\hline

I.B$_{141}^{i.c.}$(0.5) & BabABabABabABabABAbaBAbaBAbaBAbaBAbaBAbaBAbaBAbabABabABabABabABabABabABabABabABAbaBAbaBAbaBAbaBAbaBAb\\&aBAbaBAbaBABabABabABabABabABabABabABabABabaBAbaBAbaBAbaBAbaBAbaBAbaBAbaBABabABabABabABabA\\
I.B$_{142}^{i.c.}$(0.5) & BabABabaBabaBAbaBAbaBAbabAbaBABaBABabABabaBabABAbABAbaBabaBAbaBABaBABabAbabABabABabABabaBabaBAbaBAbaB\\&AbabAbabABabABabABabaBabABAbABAbaBAbabAbaBABaBABabAbabABabABAbABAbaBabaBAbaBAbaBAbabAbabABabA\\
I.B$_{143}^{i.c.}$(0.5) & BAbAbabABaBAbabAbABAbaBabABAbAbabABaBAbabAbABAbaBabABAbAbabABaBAbabAbABAbaBabABAbAbabABaBAbabAbABABaB\\&abaBAbABabaBaBABabAbaBABaBabaBAbABabaBaBABabAbaBABaBabaBAbABabaBaBABabAbaBABaBabaBAbABabaBaBA\\
I.B$_{144}^{i.c.}$(0.5) & BabABabaBabABAbABAbaBabaBAbABAbaBAbabAbaBABaBAbabAbabABaBABabAbabABabABAbABabaBabABAbABAbaBabaBAbaBAb\\
\hline
\end{tabular*}
{\rule{\temptablewidth}{1pt}}
\end{center}
\end{sidewaystable}

\begin{sidewaystable}
\tabcolsep 0pt \caption{The free group elements for the periodic three-body orbits.} \label{table-S8} \vspace*{-12pt}
\begin{center}
\def\temptablewidth{1\textwidth}
{\rule{\temptablewidth}{1pt}}
\begin{tabular*}{\temptablewidth}{@{\extracolsep{\fill}}ll}
\hline
Class and number & \hspace{1.0cm}  free group element\\\hline
II.C$_{1}^{i.c.}$(0.5) & BaBAbaBAbA\\
II.C$_{2}^{i.c.}$(0.5) & BAbAbaBaBA\\
II.C$_{3}^{i.c.}$(0.5) & BAbAbABaBaBA\\
II.C$_{4}^{i.c.}$(0.5) & BAbAbAbaBaBaBA\\
II.C$_{5}^{i.c.}$(0.5) & BaBaBAbaBAbAbA\\
II.C$_{6}^{i.c.}$(0.5) & BAbAbabABabaBaBA\\
II.C$_{7}^{i.c.}$(0.5) & BabAbaBabAbaBabA\\
\hline
\end{tabular*}
{\rule{\temptablewidth}{1pt}}
\end{center}
\end{sidewaystable}

\begin{sidewaystable}
\tabcolsep 0pt \caption{The free group elements for the periodic three-body orbits.} \label{table-S8} \vspace*{-12pt}
\begin{center}
\def\temptablewidth{1\textwidth}
{\rule{\temptablewidth}{1pt}}
\begin{tabular*}{\temptablewidth}{@{\extracolsep{\fill}}ll}
\hline
Class and number & \hspace{1.0cm}  free group element\\\hline
II.C$_{41}^{i.c.}$(0.5) & BabAbaBAbaBAbaBAbABabABabABaBAbaBAbaBAbaBabA\\
II.C$_{42}^{i.c.}$(0.5) & BAbAbabABaBABabABabABabABabABabABabABAbABabaBaBA\\
II.C$_{43}^{i.c.}$(0.5) & BabABabABaBABabAbaBABaBababAbABAbaBabABAbABabABabA\\
II.C$_{44}^{i.c.}$(0.5) & BabAbaBabAAbaBabAbaBAbaBabAbaBAbaBabAbaBBabAbaBabA\\
II.C$_{45}^{i.c.}$(0.5) & BAbAbAbaBABaBaBAbabAbAbABABaBaBabaBAbAbABAbaBaBaBA\\
II.C$_{46}^{i.c.}$(0.5) & BAbABabaBabABAbABAbaBAbaBAbaBAbaBABaBABabAbabABaBA\\
II.C$_{47}^{i.c.}$(0.5) & BabAbabABaBabaBAbABabaBabABabAbabABaBAbabAbABabaBabA\\
\hline
\end{tabular*}
{\rule{\temptablewidth}{1pt}}
\end{center}
\end{sidewaystable}

\begin{sidewaystable}
\tabcolsep 0pt \caption{The free group elements for the periodic three-body orbits.} \label{table-S8} \vspace*{-12pt}
\begin{center}
\def\temptablewidth{1\textwidth}
{\rule{\temptablewidth}{1pt}}
\begin{tabular*}{\temptablewidth}{@{\extracolsep{\fill}}ll}
\hline
Class and number & \hspace{1.0cm}  free group element\\\hline
II.C$_{81}^{i.c.}$(0.5) & BabABabABabABabABabABabAbabABaBAbabAbABABaBabaBAbABabaBabABabABabABabABabABabA\\
II.C$_{82}^{i.c.}$(0.5) & BabAbaBAbaBabABabAbaBAbaBabABaBAbaBAbABabABaBAbaBAbABabAbaBAbaBabABabAbaBAbaBabA\\
II.C$_{83}^{i.c.}$(0.5) & BabABabABaBAbaBAbaBAbaBAbaBAbABabABabABabABabABabABaBAbaBAbaBAbaBAbaBAbABabABabA\\
II.C$_{84}^{i.c.}$(0.5) & BAbAbAbaBABaBaBAbabAbAbaBABaBaBAbabAbAbaBAbaBaBabaBAbAbABAbaBaBabaBAbAbABAbaBaBaBA\\
II.C$_{85}^{i.c.}$(0.5) & BabABabABabABAbABabaBabaBAbABAbaBabaBAbABABaBAbabAbaBABaBAbabAbabABaBABabABabABabA\\
II.C$_{86}^{i.c.}$(0.5) & BAbAbabABabaBabABAbABAbaBabaBAbABAbaBabaBAbabAbaBABaBAbabAbaBABaBABabAbabABabaBaBA\\
II.C$_{87}^{i.c.}$(0.5) & BabaBabABAbaBAbaBAbabABabABabABAbABAbaBababAbaBABaBABabABabABabaBAbaBAbaBABabAbabA\\
\hline
\end{tabular*}
{\rule{\temptablewidth}{1pt}}
\end{center}
\end{sidewaystable}

\begin{sidewaystable}
\tabcolsep 0pt \caption{The free group elements for the periodic three-body orbits.} \label{table-S8} \vspace*{-12pt}
\begin{center}
\def\temptablewidth{1\textwidth}
{\rule{\temptablewidth}{1pt}}
\begin{tabular*}{\temptablewidth}{@{\extracolsep{\fill}}ll}
\hline
Class and number & \hspace{1.0cm}  free group element\\\hline
II.C$_{121}^{i.c.}$(0.5) & BAbABabaBabaBAbaBABaBABabABabABabaBAbaBAbaBABabAbabABabaBabABAbaBAbaBAbabABabABabABAbABAbaBAbabAbabA\\&BaBA\\
II.C$_{122}^{i.c.}$(0.5) & BAbABabaBabaBAbABAbaBAbabAbabABabABabABabABabaBAbaBAbaBAbaBAbabABabABabABabABabaBabaBAbaBABaBAbabAba\\&bABaBA\\
II.C$_{123}^{i.c.}$(0.5) & BAbAbabABabABabABAbABabaBabaBAbABAbaBabaBAbABAbaBAbaBAbaBAbaBABaBAbabAbaBABaBAbabAbabABaBABabABabABa\\&baBaBA\\
II.C$_{124}^{i.c.}$(0.5) & BabABabABabABAbABabaBabABAbABabaBabABAbABabABabABabABabABabABabABabABaBABabAbabABaBABabAbabABaBABabA\\
\hline
\end{tabular*}
{\rule{\temptablewidth}{1pt}}
\end{center}
\end{sidewaystable}

\begin{sidewaystable}
\tabcolsep 0pt \caption{The free group elements for the periodic three-body orbits.} \label{table-S8} \vspace*{-12pt}
\begin{center}
\def\temptablewidth{1\textwidth}
{\rule{\temptablewidth}{1pt}}
\begin{tabular*}{\temptablewidth}{@{\extracolsep{\fill}}ll}
\hline
Class and number & \hspace{1.0cm}  free group element\\\hline
II.C$_{141}^{i.c.}$(0.5) & BAbABabaBabaBAbaBAbaBABabAbabABabABAbABAbaBAbaBAbabABabABabABabABabaBAbaBAbaBABaBABabABabaBabABAbaBA\\&baBAbabAbabABaBA\\
II.C$_{142}^{i.c.}$(0.5) & BabABAbaBAbaBAbabABabABAbABAbaBAbabABabABabABAbaBAbaBAbabABabaBAbaBAbaBABabABabABabaBAbaBABaBABabABa\\&baBAbaBAbaBABabA\\
II.C$_{143}^{i.c.}$(0.5) & BAbAbabABabABabABabABabABabABAbABabaBabABAbAbabaBabABabABabABabABabABabAbabaBaBABabAbabABaBABabABabA\\&BabABabABabABabaBaBA\\
II.C$_{144}^{i.c.}$(0.5) & BabABabABabABabABabABabABabABabABAbaBAbaBAbaBAbaBAbaBAbaBAbaBAbaBAbaBAbaBAbaBAbaBAbaBAbaBABabABabABa\\
\hline
\end{tabular*}
{\rule{\temptablewidth}{1pt}}
\end{center}
\end{sidewaystable}

\begin{sidewaystable}
\tabcolsep 0pt \caption{The free group elements for the periodic three-body orbits.} \label{table-S8} \vspace*{-12pt}
\begin{center}
\def\temptablewidth{1\textwidth}
{\rule{\temptablewidth}{1pt}}
\begin{tabular*}{\temptablewidth}{@{\extracolsep{\fill}}ll}
\hline
Class and number & \hspace{1.0cm}  free group element\\\hline
II.C$_{161}^{i.c.}$(0.5) & BabAbabABaBABabAbabABaBABabAbabABaBABabAbabABaBABabAbabABaBABabABabABabABAbABabaBabABAbABabaBabABAbA\\&BabaBabABAbABabaBabABAbABabaBabA\\
II.C$_{162}^{i.c.}$(0.5) & BAbABabaBabABAbABabaBabABAbABabaBabABAbABabaBabABAbABabaBabABabABabABabABabAbabABaBABabAbabABaBABabA\\&babABaBABabAbabABaBABabAbabABaBA\\
II.C$_{163}^{i.c.}$(0.5) & BabABabABabAbabABaBABabAbaBABaBAbaBAbaBAbaBAbaBAbaBAbaBABaBAbabAbABABaBabaBAbABAbaBAbaBAbaBAbaBAbaBA\\&baBAbABAbaBabABAbABabaBabABabABabA\\
II.C$_{164}^{i.c.}$(0.5) & BabABabaBabaBAbaBABaBABabABabaBabABAbaBAbaBAbabAbabABabABAbABAbaBababAbaBABaBABabABabaBabaBAbaBAbaBA\\
\hline
\end{tabular*}
{\rule{\temptablewidth}{1pt}}
\end{center}
\end{sidewaystable}

\begin{sidewaystable}
\tabcolsep 0pt \caption{The free group elements for the periodic three-body orbits.} \label{table-S8} \vspace*{-12pt}
\begin{center}
\def\temptablewidth{1\textwidth}
{\rule{\temptablewidth}{1pt}}
\begin{tabular*}{\temptablewidth}{@{\extracolsep{\fill}}ll}
\hline
Class and number & \hspace{1.0cm}  free group element\\\hline
II.C$_{181}^{i.c.}$(0.5) & BabABabaBabaBAbaBAbaBAbabABabABabABAbABAbaBAbabAbaBABaBABabABabaBabaBAbABABaBAbabAbabABabABAbABAbaBa\\&baBAbaBABaBABabABabABabaBAbaBAbaBAbabAbabABabA\\
II.C$_{182}^{i.c.}$(0.5) & BAbAbabaBabaBAbABAbaBABabABabABabaBabaBAbaBABaBAbabAbabABaBABabABAbaBabaBAbabAbaBABabABAbABabaBabaBA\\&bABAbaBAbabAbabABabABabABAbaBABaBAbabAbabaBaBA\\
II.C$_{183}^{i.c.}$(0.5) & BabAbaBAbaBabABabAbaBAbaBAbABabABaBAbaBAbABabABaBAbaBAbABabABabAbaBAbaBabABabAbaBAbaBabABabABaBAbaBA\\&bABabABaBAbaBAbABabABaBAbaBAbaBabABabAbaBAbaBabA\\
II.C$_{184}^{i.c.}$(0.5) & BabaBabABAbABAbaBAbabAbaBABaBABabAbabABabABabABAbaBAbaBAbabAbaBABaBABabAbabaBabABAbABAbaBabaBAbaBAba\\
\hline
\end{tabular*}
{\rule{\temptablewidth}{1pt}}
\end{center}
\end{sidewaystable}

\begin{sidewaystable}
\tabcolsep 0pt \caption{The free group elements for the periodic three-body orbits.} \label{table-S8} \vspace*{-12pt}
\begin{center}
\def\temptablewidth{1\textwidth}
{\rule{\temptablewidth}{1pt}}
\begin{tabular*}{\temptablewidth}{@{\extracolsep{\fill}}ll}
\hline
Class and number & \hspace{1.0cm}  free group element\\\hline
II.C$_{201}^{i.c.}$(0.5) & BabaBabABAbaBABaBAbabABabaBabABAbaBABaBABabABabaBabaBAbaBABaBABabABabaBabaBAbABABaBAbabAbabABabABAbA\\&BAbaBAbabAbabABabABAbABAbaBABabAbabABabaBAbABAbaBABabAbabA\\
II.C$_{202}^{i.c.}$(0.5) & BabAbaBAbaBAbaBabABabABabAbaBAbaBAbABabABabABaBAbaBAbaBAbABabABabAbaBAbaBAbABabABabABaBAbaBAbaBabABa\\&bABaBAbaBAbaBAbABabABabABaBAbaBAbaBabABabABabAbaBAbaBAbaBabA\\
II.C$_{203}^{i.c.}$(0.5) & BabABabABabAbaBAbaBAbaBAbaBAbaBabABabABabABabABabAbaBAbaBAbaBAbaBAbaBAbABabABabABabABabABaBAbaBAbaBA\\&baBAbaBAbaBabABabABabABabABabAbaBAbaBAbaBAbaBAbaBabABabABabA\\
II.C$_{204}^{i.c.}$(0.5) & BabABabABabABabABabABaBAbaBAbaBAbaBAbaBAbaBAbaBAbaBAbaBAbABAbaBabABabABabABabABabABabABabABabABabAba\\
\hline
\end{tabular*}
{\rule{\temptablewidth}{1pt}}
\end{center}
\end{sidewaystable}

\begin{sidewaystable}
\tabcolsep 0pt \caption{The free group elements for the periodic three-body orbits.} \label{table-S8} \vspace*{-12pt}
\begin{center}
\def\temptablewidth{1\textwidth}
{\rule{\temptablewidth}{1pt}}
\begin{tabular*}{\temptablewidth}{@{\extracolsep{\fill}}ll}
\hline
Class and number & \hspace{1.0cm}  free group element\\\hline
II.C$_{221}^{i.c.}$(0.5) & BabABabABabABabABabABabAbaBAbaBAbaBAbaBAbaBAbaBAbaBAbaBAbaBAbaBAbABabABabABabABabABabABabABabABabABa\\&bABabABaBAbaBAbaBAbaBAbaBAbaBAbaBAbaBAbaBAbaBAbaBabABabABabABabABabABabA\\
II.C$_{222}^{i.c.}$(0.5) & BabAbabABaBABabAbabABaBAbabAbaBABaBAbabAbABAbaBabaBAbABAbaBabABAbABabaBabABAbABabaBaBABAbAbabABaBABa\\&bAbabABaBABabAbaBABaBAbabAbaBABaBabaBAbABAbaBabaBAbABabaBabABAbABabaBabA\\
II.C$_{223}^{i.c.}$(0.5) & BabAbaBAbaBAbaBabABabABabABaBAbaBAbaBAbABabABabABaBAbaBAbaBabABabABabAbaBAbaBAbaBabABabABabAbaBAbaBA\\&baBabABabABabAbaBAbaBAbABabABabABaBAbaBAbaBAbABabABabABabAbaBAbaBAbaBabA\\
II.C$_{224}^{i.c.}$(0.5) & BabABabAbaBAbaBAbaBAbABabABabABabABaBAbaBAbaBAbaBAbABabABabABabAbaBAbaBAbaBAbaBAbABabABabABaBAbaBAba\\
\hline
\end{tabular*}
{\rule{\temptablewidth}{1pt}}
\end{center}
\end{sidewaystable}

\clearpage

\begin{sidewaystable}
\tabcolsep 0pt \caption{The free group elements for the periodic three-body orbits.} \label{table-S8} \vspace*{-12pt}
\begin{center}
\def\temptablewidth{1\textwidth}
{\rule{\temptablewidth}{1pt}}
\begin{tabular*}{\temptablewidth}{@{\extracolsep{\fill}}ll}
\hline
Class and number & \hspace{1.0cm}  free group element\\\hline
II.C$_{241}^{i.c.}$(0.5) & BabABabABabABabABAbaBAbaBAbaBAbaBAbaBAbaBAbaBAbabABabABabABabABabABabABabABabABabaBAbaBAbaBAbaBAbaBA\\&baBAbaBAbabABabABabABabABabABabABabABabABabaBAbaBAbaBAbaBAbaBAbaBAbaBAbaBABabABabABabABabA\\
II.C$_{242}^{i.c.}$(0.5) & BabABabABabABabABabaBabABAbABabaBabABAbABabaBabABAbABabABabABabABabABabABabABabABabABAbABabaBaBABAbA\\&babABaBABabABabABabABabABabABabABabABabABaBABabAbabABaBABabAbabABaBABabAbabABabABabABabABabA\\
II.C$_{243}^{i.c.}$(0.5) & BabABabaBAbaBAbaBABabABabABabaBAbaBAbaBABabABabABabaBAbaBAbaBABabABabABabaBAbaBAbaBABabABabABabaBAba\\&bABabABabABAbaBAbaBAbabABabABabABAbaBAbaBAbabABabABabABAbaBAbaBAbabABabABabABAbaBAbaBAbabABabA\\
II.C$_{244}^{i.c.}$(0.5) & BabABabaBAbaBAbaBABabABabABabABAbaBAbaBAbaBABabABabABabABAbaBAbaBAbabABabABabABAbaBAbaBAbaBABabABabA\\
\hline
\end{tabular*}
{\rule{\temptablewidth}{1pt}}
\end{center}
\end{sidewaystable}

\begin{sidewaystable}
\tabcolsep 0pt \caption{The free group elements for the periodic three-body orbits.} \label{table-S8} \vspace*{-12pt}
\begin{center}
\def\temptablewidth{1\textwidth}
{\rule{\temptablewidth}{1pt}}
\begin{tabular*}{\temptablewidth}{@{\extracolsep{\fill}}ll}
\hline
Class and number & \hspace{1.0cm}  free group element\\\hline
I.A$_{1}^{i.c.}$(0.75) & BabA\\
I.A$_{2}^{i.c.}$(0.75) & BaBabAbA\\
I.A$_{3}^{i.c.}$(0.75) & BabaBABabABAbabA\\
I.A$_{4}^{i.c.}$(0.75) & BaBAbaBabAbaBAbA\\
I.A$_{5}^{i.c.}$(0.75) & BabAbABabABaBabA\\
I.A$_{6}^{i.c.}$(0.75) & BAbAbAbabaBaBaBA\\
I.A$_{7}^{i.c.}$(0.75) & BabAbaBAbABabABaBAbaBabA\\
\hline
\end{tabular*}
{\rule{\temptablewidth}{1pt}}
\end{center}
\end{sidewaystable}

\begin{sidewaystable}
\tabcolsep 0pt \caption{The free group elements for the periodic three-body orbits.} \label{table-S8} \vspace*{-12pt}
\begin{center}
\def\temptablewidth{1\textwidth}
{\rule{\temptablewidth}{1pt}}
\begin{tabular*}{\temptablewidth}{@{\extracolsep{\fill}}ll}
\hline
Class and number & \hspace{1.0cm}  free group element\\\hline
I.A$_{51}^{i.c.}$(0.75) & BabaBABabABAbaBABabaBAbabABAbaBABabaBAbabABabaBABabABAbabABabaBAbabABAbaBABabaBAbabABAbaBABabABAbabA\\
I.A$_{52}^{i.c.}$(0.75) & BabABAbaBAbaBABabABabABAbaBAbabABabABabaBAbaBAbabABabABabaBAbaBAbabABabABabaBAbaBABabABabABAbaBAbaBABabA\\
I.A$_{53}^{i.c.}$(0.75) & BabaBAbabABabaBAbabABabaBAbaBABabABAbaBABabABAbaBABabABAbaBABabABAbaBABabABAbaBAbabABabaBAbabABabaBAbabA\\
I.A$_{54}^{i.c.}$(0.75) & BAbaBABabaBAbabaBababABabaBABabABABaBABAbaBABabaBAbabaBAbabABAbaBABAbABABabABAbabABababAbabaBAbabABAbaBA\\
I.A$_{55}^{i.c.}$(0.75) & BabaBABabaBAbabABAbabABabaBABabABAbabABAbaBABabaBABabABAbabABAbaBABabaBABabABAbabABabaBABabaBAbabABAbabA\\
I.A$_{56}^{i.c.}$(0.75) & BAbabABAbabABAbabABAbabABAbabABAbabABAbabABAbabABAbabaBABabaBABabaBABabaBABabaBABabaBABabaBABabaBABabaBA\\
I.A$_{57}^{i.c.}$(0.75) & BabaBABabABAbaBABabABAbaBABabaBAbabABabaBAbabABabaBABabABAbabABabaBAbabABabaBAbabABAbaBABabABAbaBABabABA\\
\hline
\end{tabular*}
{\rule{\temptablewidth}{1pt}}
\end{center}
\end{sidewaystable}

\begin{sidewaystable}
\tabcolsep 0pt \caption{The free group elements for the periodic three-body orbits.} \label{table-S8} \vspace*{-12pt}
\begin{center}
\def\temptablewidth{1\textwidth}
{\rule{\temptablewidth}{1pt}}
\begin{tabular*}{\temptablewidth}{@{\extracolsep{\fill}}ll}
\hline
Class and number & \hspace{1.0cm}  free group element\\\hline
I.A$_{76}^{i.c.}$(0.75) & BabaBABabaBABabaBABabaBAbabABAbabABAbabABAbaBABabaBABabaBABabaBABabABAbabABAbabABAbabABAbaBABabaBABabaB\\&ABabaBAbabABAbabABAbabABAbabA\\
I.A$_{77}^{i.c.}$(0.75) & BabaBABabaBABabaBABabaBABabaBABabaBABabaBABabaBABabaBABabaBABabaBABabABAbabABAbabABAbabABAbabABAbabABAb\\&abABAbabABAbabABAbabABAbabABAbabA\\
I.A$_{78}^{i.c.}$(0.75) & BabaBAbabABAbaBABabABAbabABabaBAbabABAbaBABabABAbabABabaBAbabABAbaBABabABAbaBABabaBAbabABabaBABabABAbaB\\&ABabaBAbabABabaBABabABAbaBABabaBAbabA\\
I.A$_{79}^{i.c.}$(0.75) & BAbabABAbabABAbabABAbabABAbabABAbabABAbabABAbabABAbabABAbabABAbabABAbabaBABabaBABabaBABabaBABabaBABabaB\\
\hline
\end{tabular*}
{\rule{\temptablewidth}{1pt}}
\end{center}
\end{sidewaystable}

\begin{sidewaystable}
\tabcolsep 0pt \caption{The free group elements for the periodic three-body orbits.} \label{table-S8} \vspace*{-12pt}
\begin{center}
\def\temptablewidth{1\textwidth}
{\rule{\temptablewidth}{1pt}}
\begin{tabular*}{\temptablewidth}{@{\extracolsep{\fill}}ll}
\hline
Class and number & \hspace{1.0cm}  free group element\\\hline
I.A$_{98}^{i.c.}$(0.75) & BAbabABAbabABAbabABAbabABAbabABABabaBABabaBABabaBABabaBABabaBABababABAbabABAbabABAbabABAbabABAbabaBABab\\&aBABabaBABabaBABabaBABababABAbabABAbabABAbabABAbabABAbabABABabaBABabaBABabaBABabaBABabaBA\\
I.A$_{99}^{i.c.}$(0.75) & BAbabABAbabABAbabABAbabABAbabaBAbabABABabaBABabaBABabaBABabaBABababABAbaBABAbabABAbabABAbabABAbabABAbab\\&aBABabaBABabaBABabaBABabaBABAbaBABababABAbabABAbabABAbabABAbabABABabaBAbabaBABabaBABabaBABabaBABa\\&baBA\\
I.A$_{100}^{i.c.}$(0.75) & BAbaBABababABabaBABAbaBABababABabaBABAbaBABababAbabaBABAbABABabaBAbabaBABabABABabaBAbabaBABabABABabaBA\\&babaBAbabABABabABAbabaBAbabABABabABAbabaBAbabABABaBABAbabaBababABAbaBABAbabABababABAbaBABAbabABab\\
\hline
\end{tabular*}
{\rule{\temptablewidth}{1pt}}
\end{center}
\end{sidewaystable}

\begin{sidewaystable}
\tabcolsep 0pt \caption{The free group elements for the periodic three-body orbits.} \label{table-S8} \vspace*{-12pt}
\begin{center}
\def\temptablewidth{1\textwidth}
{\rule{\temptablewidth}{1pt}}
\begin{tabular*}{\temptablewidth}{@{\extracolsep{\fill}}ll}
\hline
Class and number & \hspace{1.0cm}  free group element\\\hline
I.B$_{31}^{i.c.}$(0.75) & BabaBAbaBABabABabaBAbaBABabABabaBAbabABabABAbaBAbabABabABAbaBAbabA\\
I.B$_{32}^{i.c.}$(0.75) & BabABabaBAbaBAbaBABabABabABabaBAbaBAbaBAbabABabABabABAbaBAbaBAbabABabA\\
I.B$_{33}^{i.c.}$(0.75) & BabABAbABAbaBAbaBABabABabABAbABAbaBAbaBABaBABabABabABAbaBAbaBABaBABabA\\
I.B$_{34}^{i.c.}$(0.75) & BabaBABabaBABabaBABabaBABabaBABabaBAbabABAbabABAbabABAbabABAbabABAbabA\\
I.B$_{35}^{i.c.}$(0.75) & BabABAbaBAbabABabABAbaBAbabABabABAbaBAbaBABabABabaBAbaBABabABabaBAbaBABabA\\
I.B$_{36}^{i.c.}$(0.75) & BabaBABabABAbabABabaBAbabABAbaBABabaBAbabABAbaBABabaBAbabABabaBABabABAbabA\\
I.B$_{37}^{i.c.}$(0.75) & BAbabABAbabABAbabABAbabABAbabABAbabABABabaBABabaBABabaBABabaBABabaBABabaBA\\
\hline
\end{tabular*}
{\rule{\temptablewidth}{1pt}}
\end{center}
\end{sidewaystable}

\begin{sidewaystable}
\tabcolsep 0pt \caption{The free group elements for the periodic three-body orbits.} \label{table-S8} \vspace*{-12pt}
\begin{center}
\def\temptablewidth{1\textwidth}
{\rule{\temptablewidth}{1pt}}
\begin{tabular*}{\temptablewidth}{@{\extracolsep{\fill}}ll}
\hline
Class and number & \hspace{1.0cm}  free group element\\\hline
I.B$_{71}^{i.c.}$(0.75) & BabABAbABAbaBAbaBABabABabABAbABAbaBAbaBABabABabABAbABAbaBAbaBABaBABabABabABAbaBAbaBABaBABabABabABAbaBAb\\&aBABaBABabA\\
I.B$_{72}^{i.c.}$(0.75) & BabaBABabaBABabaBABabABAbabABAbabABAbaBABabaBABabaBABabaBAbabABAbabABAbabABAbaBABabaBABabaBABabABAbabAB\\&AbabABAbabA\\
I.B$_{73}^{i.c.}$(0.75) & BAbaBABababABabaBABAbaBABababAbabaBABabABABabaBAbabaBABabABABabABAbabaBAbabABABabABAbabaBababABAbaBABAb\\&abABababABAbaBA\\
I.B$_{74}^{i.c.}$(0.75) & BabaBABabaBABabaBABabaBABabaBABabaBABabaBABabaBABabaBABabaBAbabABAbabABAbabABAbabABAbabABAbabABAbabABAb\\
\hline
\end{tabular*}
{\rule{\temptablewidth}{1pt}}
\end{center}
\end{sidewaystable}

\begin{sidewaystable}
\tabcolsep 0pt \caption{The free group elements for the periodic three-body orbits.} \label{table-S8} \vspace*{-12pt}
\begin{center}
\def\temptablewidth{1\textwidth}
{\rule{\temptablewidth}{1pt}}
\begin{tabular*}{\temptablewidth}{@{\extracolsep{\fill}}ll}
\hline
Class and number & \hspace{1.0cm}  free group element\\\hline
I.B$_{96}^{i.c.}$(0.75) & BabABAbABAbaBAbaBABabABabABAbABAbaBAbaBABabABabABAbABAbaBAbaBABabABabABAbABAbaBAbaBABaBABabABabABAbaBAb\\&aBABaBABabABabABAbaBAbaBABaBABabABabABAbaBAbaBABaBABabA\\
I.B$_{97}^{i.c.}$(0.75) & BabaBABabaBABabaBAbabABAbabABAbaBABabaBABabaBABabABAbabABAbabABabaBABabaBABabaBAbabABAbabABAbabABabaBAB\\&abaBABabABAbabABAbabABAbaBABabaBABabaBAbabABAbabABAbabA\\
I.B$_{98}^{i.c.}$(0.75) & BabaBABabABAbaBABabaBAbabABAbaBABabaBAbabABabaBABabABAbabABabaBABabABAbaBABabaBAbabABAbaBABabABAbabABab\\&aBABabABAbabABabaBAbabABAbaBABabaBAbabABAbaBABabABAbabA\\
I.B$_{99}^{i.c.}$(0.75) & BAbaBABabaBAbabaBAbabABABaBABAbaBABababAbabaBABabABABaBABAbabABababABabaBABabABABabABAbabABababABabaBAB\\
\hline
\end{tabular*}
{\rule{\temptablewidth}{1pt}}
\end{center}
\end{sidewaystable}

\begin{sidewaystable}
\tabcolsep 0pt \caption{The free group elements for the periodic three-body orbits.} \label{table-S8} \vspace*{-12pt}
\begin{center}
\def\temptablewidth{1\textwidth}
{\rule{\temptablewidth}{1pt}}
\begin{tabular*}{\temptablewidth}{@{\extracolsep{\fill}}ll}
\hline
Class and number & \hspace{1.0cm}  free group element\\\hline
I.C$_{1}^{i.c.}$(0.75) & BAbabABAbabABAbabABAbabABAbabABAbabABAbabABABabaBABabaBABabaBABabaBABabaBABabaBABabaBABAbabABAbabABA\\&babABAbabABAbabABAbabABAbabABABabaBABabaBABabaBABabaBABabaBABabaBABabaBA\\
II.A$_{1}^{i.c.}$(0.75) & BabABaBAbaBabABabABabAbaBAbABabA\\
II.A$_{2}^{i.c.}$(0.75) & BabaBAbabABAbaBABabABAbabABabaBAbabABabaBABabABAbaBABabABAbabABabaBAbabABAbaBABabABAbaBABabaBAbabAB\\&abaBABabABAbaBABabABAbabABabaBAbabABabaBABabABAbaBABabaBAbabA\\
II.B$_{1}^{i.c.}$(0.75) & BaBAbaBabABaBAbABabAbaBAbA\\
II.B$_{2}^{i.c.}$(0.75) & BabABabAbaBAbaBAbaBAbaBabABabA\\
\hline
\end{tabular*}
{\rule{\temptablewidth}{1pt}}
\end{center}
\end{sidewaystable}

\begin{sidewaystable}
\tabcolsep 0pt \caption{The free group elements for the periodic three-body orbits.} \label{table-S8} \vspace*{-12pt}
\begin{center}
\def\temptablewidth{1\textwidth}
{\rule{\temptablewidth}{1pt}}
\begin{tabular*}{\temptablewidth}{@{\extracolsep{\fill}}ll}
\hline
Class and number & \hspace{1.0cm}  free group element\\\hline
II.C$_{31}^{i.c.}$(0.75) & BabABAbaBAbaBABaBABabABAbABAbaBAbaBABabA\\
II.C$_{32}^{i.c.}$(0.75) & BabaBABAbaBABababAbabaBababABAbaBABAbabA\\
II.C$_{33}^{i.c.}$(0.75) & BabAbaBAbABabAbaBAbABaBAbaBabABaBAbaBabA\\
II.C$_{34}^{i.c.}$(0.75) & BabAbaBAbABaBAbABaBabAbABaBAbABaBAbaBabA\\
II.C$_{35}^{i.c.}$(0.75) & BaBAbaBabABaBAbaBAbABaBAbaBAbABabAbaBAbA\\
II.C$_{36}^{i.c.}$(0.75) & BAbaBABababABabaBABAbaBABAbabABababABAbaBA\\
II.C$_{37}^{i.c.}$(0.75) & BabABabaBAbaBAbabAbabABabaBabaBAbaBAbabABabA\\
\hline
\end{tabular*}
{\rule{\temptablewidth}{1pt}}
\end{center}
\end{sidewaystable}

\begin{sidewaystable}
\tabcolsep 0pt \caption{The free group elements for the periodic three-body orbits.} \label{table-S8} \vspace*{-12pt}
\begin{center}
\def\temptablewidth{1\textwidth}
{\rule{\temptablewidth}{1pt}}
\begin{tabular*}{\temptablewidth}{@{\extracolsep{\fill}}ll}
\hline
Class and number & \hspace{1.0cm}  free group element\\\hline

II.C$_{81}^{i.c.}$(0.75) & BabaBAbabABabaBAbabABabaBAbabABabaBABabABAbabABabaBAbabABabaBAbabABabaBAbabA\\
II.C$_{82}^{i.c.}$(0.75) & BabABAbABAbaBABaBABabABabABAbaBAbaBABabABAbaBAbaBABabABabABAbABAbaBABaBABabA\\
II.C$_{83}^{i.c.}$(0.75) & BabAbaBabABaBAbaBabABaBAbABabABaBAbaBabAbaBAbABabABaBAbABabAbaBAbABabAbaBabA\\
II.C$_{84}^{i.c.}$(0.75) & BAbABABabaBAbabABababABabaBABabABAbaBABAbaBABabABAbabABababABabaBAbabABABaBA\\
II.C$_{85}^{i.c.}$(0.75) & BAbaBABabaBAbabABababABabaBABabABAbaBABAbaBABabABAbabABababABabaBAbabABAbaBA\\
II.C$_{86}^{i.c.}$(0.75) & BababABAbaBABAbabABababABAbaBABAbabABababABabaBABAbaBABababABabaBABAbaBABababA\\
II.C$_{87}^{i.c.}$(0.75) & BabABabaBabaBAbaBAbabAbabABabABabaBAbaBAbaBAbabABabABabaBabaBAbaBAbabAbabABabA\\
\hline
\end{tabular*}
{\rule{\temptablewidth}{1pt}}
\end{center}
\end{sidewaystable}

\begin{sidewaystable}
\tabcolsep 0pt \caption{The free group elements for the periodic three-body orbits.} \label{table-S8} \vspace*{-12pt}
\begin{center}
\def\temptablewidth{1\textwidth}
{\rule{\temptablewidth}{1pt}}
\begin{tabular*}{\temptablewidth}{@{\extracolsep{\fill}}ll}
\hline
Class and number & \hspace{1.0cm}  free group element\\\hline
II.C$_{121}^{i.c.}$(0.75) & BabABabaBAbaBAbabAbabABabABabaBabaBAbaBAbabABabABabaBAbaBAbaBAbabABabABabaBAbaBAbabAbabABabABabaB\\&abaBAbaBAbabABabA\\
II.C$_{122}^{i.c.}$(0.75) & BAbaBABabaBAbabaBababABabaBABabABAbaBABAbaBABabaBAbabABababABabaBAbabABAbaBABAbaBABabABAbabABabab\\&AbabaBAbabABAbaBA\\
II.C$_{123}^{i.c.}$(0.75) & BabaBABAbaBABabaBABabaBABabaBABabABABabABAbabABAbabABAbaBABAbaBABabaBABabaBABabABABabABAbabABAbab\\&ABAbabABAbaBABAbabA\\
II.C$_{124}^{i.c.}$(0.75) & BababABAbaBABAbabABAbabABAbabABAbabABAbabABAbabABABabABAbabaBABabABABabaBABabaBABabaBABabaBABabaB\\
\hline
\end{tabular*}
{\rule{\temptablewidth}{1pt}}
\end{center}
\end{sidewaystable}

\begin{sidewaystable}
\tabcolsep 0pt \caption{The free group elements for the periodic three-body orbits.} \label{table-S8} \vspace*{-12pt}
\begin{center}
\def\temptablewidth{1\textwidth}
{\rule{\temptablewidth}{1pt}}
\begin{tabular*}{\temptablewidth}{@{\extracolsep{\fill}}ll}
\hline
Class and number & \hspace{1.0cm}  free group element\\\hline
II.C$_{141}^{i.c.}$(0.75) & BabaBABabaBABabaBAbabABAbabABAbaBABabaBABabaBAbabABAbabABAbaBABabaBABabaBAbabABAbabABAbaBABabaBAB\\&abaBAbabABAbabABAbaBABabaBABabaBAbabABAbabABAbabA\\
II.C$_{142}^{i.c.}$(0.75) & BabaBABabABAbabABabaBAbabABAbaBABabaBAbabABAbaBABabABAbabABabaBABabABAbabABabaBABabABAbabABabaBAB\\&abABAbaBABabaBAbabABAbaBABabaBAbabABabaBABabABAbabA\\
II.C$_{143}^{i.c.}$(0.75) & BabaBABabABAbaBABabABAbabABabaBAbabABabaBABabABAbaBABabaBAbabABabaBABabABAbaBABabABAbabABabaBAbab\\&ABAbaBABabABAbabABabaBAbabABabaBABabABAbaBABabABAbabA\\
II.C$_{144}^{i.c.}$(0.75) & BababABAbaBABAbabABababABAbabABAbabABAbabABAbabABAbabaBAbabaBABabABABabaBAbabaBAbabABABabABAbabaB\\
\hline
\end{tabular*}
{\rule{\temptablewidth}{1pt}}
\end{center}
\end{sidewaystable}

\begin{sidewaystable}
\tabcolsep 0pt \caption{The free group elements for the periodic three-body orbits.} \label{table-S8} \vspace*{-12pt}
\begin{center}
\def\temptablewidth{1\textwidth}
{\rule{\temptablewidth}{1pt}}
\begin{tabular*}{\temptablewidth}{@{\extracolsep{\fill}}ll}
\hline
Class and number & \hspace{1.0cm}  free group element\\\hline
II.C$_{156}^{i.c.}$(0.75) & BabaBABabaBABabABAbabABAbabABabaBABabaBABabABAbabABAbabABabaBABabaBABabABAbabABAbabABAbaBABabaBAB\\&abaBABabABAbabABAbabABabaBABabaBABabABAbabABAbabABabaBABabaBABabABAbabABAbabA\\
II.C$_{157}^{i.c.}$(0.75) & BAbaBABababABabaBABabABABabaBAbabaBAbabABABaBABAbabABababABabaBABAbaBABababAbabaBABabABABabABAbab\\&aBababABAbaBABAbabABababABabaBABAbABABabaBAbabaBAbabABABabABAbabABababABAbaBA\\
II.C$_{158}^{i.c.}$(0.75) & BabaBABabABAbaBABabaBAbabABabaBABabABAbabABabaBAbabABAbaBABabABAbabABabaBABabABAbabABabaBAbabABab\\&aBABabABAbabABabaBABabABAbaBABabaBAbabABabaBABabABAbabABabaBAbabABAbaBABabABAbabA\\
II.C$_{159}^{i.c.}$(0.75) & BAbabABAbabABAbabABAbabABAbabaBABabABABabaBABabaBABabaBABAbabABababABAbabABAbabABAbabABABabABABab\\
\hline
\end{tabular*}
{\rule{\temptablewidth}{1pt}}
\end{center}
\end{sidewaystable}

\begin{sidewaystable}
\tabcolsep 0pt \caption{The free group elements for the periodic three-body orbits.} \label{table-S8} \vspace*{-12pt}
\begin{center}
\def\temptablewidth{1\textwidth}
{\rule{\temptablewidth}{1pt}}
\begin{tabular*}{\temptablewidth}{@{\extracolsep{\fill}}ll}
\hline
Class and number & \hspace{1.0cm}  free group element\\\hline
I.A$_{1}^{i.c.}$(2) & BAbabaBA\\
I.A$_{2}^{i.c.}$(2) & BABabababABA\\
I.A$_{3}^{i.c.}$(2) & BABababaBABABabababABABAbababABA\\
I.A$_{4}^{i.c.}$(2) & BABabababaBABABABabababABABABAbabababABA\\
I.A$_{5}^{i.c.}$(2) & BABAbabababABABABAbabababaBABABABabababaBABA\\
I.A$_{6}^{i.c.}$(2) & BAbababABABAbababABABAbabaBABABababaBABABababaBA\\
I.A$_{7}^{i.c.}$(2) & BABababaBABABababaBABABabababABABAbababABABAbababABA\\
\hline
\end{tabular*}
{\rule{\temptablewidth}{1pt}}
\end{center}
\end{sidewaystable}

\begin{sidewaystable}
\tabcolsep 0pt \caption{The free group elements for the periodic three-body orbits.} \label{table-S8} \vspace*{-12pt}
\begin{center}
\def\temptablewidth{1\textwidth}
{\rule{\temptablewidth}{1pt}}
\begin{tabular*}{\temptablewidth}{@{\extracolsep{\fill}}ll}
\hline
Class and number & \hspace{1.0cm}  free group element\\\hline
I.A$_{31}^{i.c.}$(2) & BABabababaBABABABabababaBABABAbabababABABABAbababaBABABABabababaBABABABabababABABABAbabababABABABAbababa\\&BABABABabababaBABABAbabababABABABAbabababABA\\
I.A$_{32}^{i.c.}$(2) & BABAbababaBABABABabababaBABABAbabababABABABAbababaBABABABabababaBABABAbabababaBABABAbabababABABABAbababa\\&BABABABabababaBABABAbabababABABABAbababaBABA\\
I.A$_{33}^{i.c.}$(2) & BABabababaBABABABabababaBABABABabababaBABABABabababaBABABABabababaBABABABabababABABABAbabababABABABAbaba\\&babABABABAbabababABABABAbabababABABABAbabababABA\\
I.A$_{34}^{i.c.}$(2) & BABAbabababABABABAbabababABABABAbabababABABABAbabababABABABAbabababABABABAbabababaBABABABabababaBABABABa\\
\hline
\end{tabular*}
{\rule{\temptablewidth}{1pt}}
\end{center}
\end{sidewaystable}

\begin{sidewaystable}

\tabcolsep 0pt \caption{The free group elements for the periodic three-body orbits.} \label{table-S8} \vspace*{-12pt}
\begin{center}
\def\temptablewidth{1\textwidth}
{\rule{\temptablewidth}{1pt}}
\begin{tabular*}{\temptablewidth}{@{\extracolsep{\fill}}ll}
\hline
Class and number & \hspace{1.0cm}  free group element\\\hline
I.A$_{51}^{i.c.}$(2) & BABabababABABABabababABABABAbababaBABABAbababaBABABAbababaBABABAbababaBABABAbabababABABABabababABABABaba\\&babABABABabababABABABabababaBABABAbababaBABABAbababaBABABAbababaBABABAbababaBABABABabababABABABabababABA\\
I.A$_{52}^{i.c.}$(2) & BABAbabababABABABabababaBABABABabababaBABABABabababaBABABABabababaBABABABabababaBABABABabababABABABAbaba\\&babaBABABABabababABABABAbabababABABABAbabababABABABAbabababABABABAbabababABABABAbabababABABABabababaBABA\\
I.A$_{53}^{i.c.}$(2) & BABabababaBABABABabababaBABABABabababaBABABABabababaBABABABabababaBABABABabababaBABABABabababaBABABABaba\\&babABABABAbabababABABABAbabababABABABAbabababABABABAbabababABABABAbabababABABABAbabababABABABAbabababABA\\
I.A$_{54}^{i.c.}$(2) & BABAbabababABABABAbabababABABABababababABABABAbabababABABABAbabababABABABABabababABABABAbabababABABABAba\\
\hline
\end{tabular*}
{\rule{\temptablewidth}{1pt}}
\end{center}
\end{sidewaystable}

\begin{sidewaystable}
\tabcolsep 0pt \caption{The free group elements for the periodic three-body orbits.} \label{table-S8} \vspace*{-12pt}
\begin{center}
\def\temptablewidth{1\textwidth}
{\rule{\temptablewidth}{1pt}}
\begin{tabular*}{\temptablewidth}{@{\extracolsep{\fill}}ll}
\hline
Class and number & \hspace{1.0cm}  free group element\\\hline
I.B$_{11}^{i.c.}$(2) & BABabababaBABABABabababaBABABAbabababABABABAbabababABA\\
I.B$_{12}^{i.c.}$(2) & BABAbabababABABABAbabababABABABABabababaBABABABabababaBABA\\
I.B$_{13}^{i.c.}$(2) & BAbababABABAbababABABAbababABABababaBABABababaBABABababaBA\\
I.B$_{14}^{i.c.}$(2) & BABababaBABABababaBABABababaBABABAbababABABAbababABABAbababABA\\
I.B$_{15}^{i.c.}$(2) & BABabababABABABababaBABABAbababaBABABAbababaBABABAbababABABABabababABA\\
I.B$_{16}^{i.c.}$(2) & BABabababABABABabababaBABABAbababaBABABAbababaBABABAbabababABABABabababABA\\
I.B$_{17}^{i.c.}$(2) & BAbababABABAbababABABAbababABABAbababABABababaBABABababaBABABababaBABABababaBA\\
\hline
\end{tabular*}
{\rule{\temptablewidth}{1pt}}
\end{center}
\end{sidewaystable}

\clearpage

\begin{sidewaystable}
\tabcolsep 0pt \caption{The free group elements for the periodic three-body orbits.} \label{table-S8} \vspace*{-12pt}
\begin{center}
\def\temptablewidth{1\textwidth}
{\rule{\temptablewidth}{1pt}}
\begin{tabular*}{\temptablewidth}{@{\extracolsep{\fill}}ll}
\hline
Class and number & \hspace{1.0cm}  free group element\\\hline
I.B$_{41}^{i.c.}$(2) & BABababaBABABababaBABABababaBABABababaBABABababaBABABababaBABABababaBABABababaBABABAbababABABAbababABABA\\&bababABABAbababABABAbababABABAbababABABAbababABABAbababABA\\
I.B$_{42}^{i.c.}$(2) & BABabababABABAbababaBABABAbababaBABABAbababABABABabababABABABabababABABAbababaBABABAbababaBABABabababABA\\&BABabababABABABababaBABABAbababaBABABAbababaBABABabababABA\\
I.B$_{43}^{i.c.}$(2) & BABAbabababABABABabababaBABABABabababaBABABABabababaBABABABabababaBABABAbabababABABABABabababaBABABAbaba\\&babABABABAbabababABABABAbabababABABABAbabababABABABabababaBABA\\
I.B$_{44}^{i.c.}$(2) & BABabababaBABABABabababaBABABABabababaBABABABabababaBABABABabababaBABABABabababaBABABAbabababABABABAbaba\\
\hline
\end{tabular*}
{\rule{\temptablewidth}{1pt}}
\end{center}
\end{sidewaystable}

\begin{sidewaystable}
\tabcolsep 0pt \caption{The free group elements for the periodic three-body orbits.} \label{table-S8} \vspace*{-12pt}
\begin{center}
\def\temptablewidth{1\textwidth}
{\rule{\temptablewidth}{1pt}}
\begin{tabular*}{\temptablewidth}{@{\extracolsep{\fill}}ll}
\hline
Class and number & \hspace{1.0cm}  free group element\\\hline
I.B$_{61}^{i.c.}$(2) & BABAbabababABABABAbabababABABABAbabababaBABABABabababaBABABABabababaBABABABababababABABABAbabababABABABA\\&babababABABABABabababaBABABABabababaBABABABababababABABABAbabababABABABAbabababABABABAbabababaBABA\\&BABabababaBABABABabababaBABA\\
I.B$_{62}^{i.c.}$(2) & BABAbabababABABABAbabababaBABABABabababaBABABABAbabababABABABAbabababABABABABabababaBABABABababababABABA\\&BAbabababABABABABabababaBABABABababababABABABAbabababABABABABabababaBABABABabababaBABABABAbabababA\\&BABABAbabababaBABABABabababaBABA\\
I.B$_{63}^{i.c.}$(2) & BABAbabababABABABAbabababABABABABabababaBABABABababababABABABAbabababABABABAbabababaBABABABabababaBABABA\\
\hline
\end{tabular*}
{\rule{\temptablewidth}{1pt}}
\end{center}
\end{sidewaystable}

\begin{sidewaystable}
\tabcolsep 0pt \caption{The free group elements for the periodic three-body orbits.} \label{table-S8} \vspace*{-12pt}
\begin{center}
\def\temptablewidth{1\textwidth}
{\rule{\temptablewidth}{1pt}}
\begin{tabular*}{\temptablewidth}{@{\extracolsep{\fill}}ll}
\hline
Class and number & \hspace{1.0cm}  free group element\\\hline
II.C$_{11}^{i.c.}$(2) & BabAAbaBBabABaaBAbaBAbbABabAAbaBBabA\\
II.C$_{12}^{i.c.}$(2) & BABAbababaBABABABabababABABABAbababaBABA\\
II.C$_{13}^{i.c.}$(2) & BABabababaBABABABababababABABABAbabababABA\\
II.C$_{14}^{i.c.}$(2) & BABababaBABABAbababaBABABAbababaBABABAbababABA\\
II.C$_{15}^{i.c.}$(2) & BABAbababaBABABABabababaBABABAbabababABABABAbababaBABA\\
II.C$_{16}^{i.c.}$(2) & BABAbabababABABABAbabababABABABabababaBABABABabababaBABA\\
II.C$_{17}^{i.c.}$(2) & BABabababABABABabababaBABABAbababaBABABAbabababABABABabababABA\\
\hline
\end{tabular*}
{\rule{\temptablewidth}{1pt}}
\end{center}
\end{sidewaystable}

\begin{sidewaystable}
\tabcolsep 0pt \caption{The free group elements for the periodic three-body orbits.} \label{table-S8} \vspace*{-12pt}
\begin{center}
\def\temptablewidth{1\textwidth}
{\rule{\temptablewidth}{1pt}}
\begin{tabular*}{\temptablewidth}{@{\extracolsep{\fill}}ll}
\hline
Class and number & \hspace{1.0cm}  free group element\\\hline
II.C$_{41}^{i.c.}$(2) & BABabababABABABAbababaBABABABabababABABABabababaBABABAbabababABABABabababaBABABAbabababABABABabababABAB\\&ABAbababaBABABABabababABA\\
II.C$_{42}^{i.c.}$(2) & BABAbabababABABABAbabababABABABABabababaBABABABababababABABABababababABABABababababABABABAbabababABABAB\\&ABabababaBABABABabababaBABA\\
II.C$_{43}^{i.c.}$(2) & BABabababABABABabababABABABabababABABABAbababaBABABAbababaBABABAbababaBABABAbababaBABABAbababaBABABABab\\&ababABABABabababABABABabababABA\\
II.C$_{44}^{i.c.}$(2) & BABAbabababABABABabababaBABABABabababaBABABABabababaBABABABabababaBABABAbabababABABABAbabababABABABAbab\\
\hline
\end{tabular*}
{\rule{\temptablewidth}{1pt}}
\end{center}
\end{sidewaystable}

\begin{sidewaystable}
\tabcolsep 0pt \caption{The free group elements for the periodic three-body orbits.} \label{table-S8} \vspace*{-12pt}
\begin{center}
\def\temptablewidth{1\textwidth}
{\rule{\temptablewidth}{1pt}}
\begin{tabular*}{\temptablewidth}{@{\extracolsep{\fill}}ll}
\hline
Class and number & \hspace{1.0cm}  free group element\\\hline
II.C$_{61}^{i.c.}$(2) & BABabababABABABababaBABABAbababaBABABAbababaBABABabababABABABabababABABABababaBABABAbababaBABABAbababAB\\&ABABabababABABABabababABABAbababaBABABAbababaBABABAbababABABABabababABA\\
II.C$_{62}^{i.c.}$(2) & BABAbabababABABABAbabababaBABABABabababaBABABABababababABABABAbabababABABABABabababABABABABabababABABAB\\&ABabababaBABABABababababABABABAbabababABABABAbabababaBABABABabababaBABA\\
II.C$_{63}^{i.c.}$(2) & BABAbabababABABABAbabababABABABAbabababaBABABABabababaBABABABababababABABABAbabababABABABABabababaBABAB\\&ABababababABABABAbabababABABABAbabababaBABABABabababaBABABABabababaBABA\\
II.C$_{64}^{i.c.}$(2) & BABabababABABABabababABABABabababABABABabababABABABababaBABABAbababaBABABAbababaBABABAbababaBABABAbabab\\
\hline
\end{tabular*}
{\rule{\temptablewidth}{1pt}}
\end{center}
\end{sidewaystable}

\begin{sidewaystable}
\tabcolsep 0pt \caption{The free group elements for the periodic three-body orbits.} \label{table-S8} \vspace*{-12pt}
\begin{center}
\def\temptablewidth{1\textwidth}
{\rule{\temptablewidth}{1pt}}
\begin{tabular*}{\temptablewidth}{@{\extracolsep{\fill}}ll}
\hline
Class and number & \hspace{1.0cm}  free group element\\\hline
II.C$_{81}^{i.c.}$(2) & BABabababABABABabababABABABAbababaBABABAbababaBABABABabababABABABabababABABABabababABABABabababaBABABAb\\&ababaBABABAbabababABABABabababABABABabababABABABabababABABABAbababaBABABAbababaBA\\&BABABabababABABABabababABA\\
II.C$_{82}^{i.c.}$(2) & BABAbabababABABABAbabababABABABAbabababABABABAbabababABABABAbabababABABABAbabababABABABABabababaBABABAB\\&abababABABABAbabababABABABABabababaBABABABabababaBABABABabababaBABABABabababaBABA\\&BABabababaBABABABabababaBABA\\
II.C$_{83}^{i.c.}$(2) & BABAbabababABABABAbabababaBABABABabababaBABABABababababABABABAbabababABABABABabababaBABABABababababABAB\\
\hline
\end{tabular*}
{\rule{\temptablewidth}{1pt}}
\end{center}
\end{sidewaystable}

\begin{sidewaystable}
\tabcolsep 0pt \caption{The free group elements for the periodic three-body orbits.} \label{table-S8} \vspace*{-12pt}
\begin{center}
\def\temptablewidth{1\textwidth}
{\rule{\temptablewidth}{1pt}}
\begin{tabular*}{\temptablewidth}{@{\extracolsep{\fill}}ll}
\hline
Class and number & \hspace{1.0cm}  free group element\\\hline
I.A$_{1}^{i.c.}$(4) & BABabababABA\\
I.A$_{2}^{i.c.}$(4) & BABABAbabababababaBABABA\\
I.A$_{3}^{i.c.}$(4) & BABABabababababaBABABABABABabababababABABABABABAbabababababABABA\\
I.A$_{4}^{i.c.}$(4) & BABABAbabababababABABABABABAbabababababaBABABABABABabababababaBABABA\\
I.A$_{5}^{i.c.}$(4) & BABABAbababababababABABABABABABAbabababababaBABABABABABABababababababaBABABA\\
I.A$_{6}^{i.c.}$(4) & BABABABababababababaBABABABABABABabababababababABABABABABABAbababababababABABABA\\
I.A$_{7}^{i.c.}$(4) & BABABabababababABABABABABAbabababababABABABABABabababababABABABABABabababababaBABABABABABabababababABABA\\
\hline
\end{tabular*}
{\rule{\temptablewidth}{1pt}}
\end{center}
\end{sidewaystable}

\begin{sidewaystable}
\tabcolsep 0pt \caption{The free group elements for the periodic three-body orbits.} \label{table-S8} \vspace*{-12pt}
\begin{center}
\def\temptablewidth{1\textwidth}
{\rule{\temptablewidth}{1pt}}
\begin{tabular*}{\temptablewidth}{@{\extracolsep{\fill}}ll}
\hline
Class and number & \hspace{1.0cm}  free group element\\\hline
I.A$_{27}^{i.c.}$(4) & BABABAbabababababABABABABABAbabababababABABABABABAbabababababABABABABABAbabababababABABABABABAbabababab\\&abABABABABABAbabababababaBABABABABABabababababaBABABABABABabababababaBABABABABABabababababaBABABABABABaba\\&babababaBABABABABABabababababaBABABA\\
I.A$_{28}^{i.c.}$(4) & BABABAbabababababABABABABABAbabababababABABABABABABabababababaBABABABABABababababababABABABABABAbababab\\&ababABABABABABAbabababababaBABABABABABabababababaBABABABABABababababababABABABABABAbabababababABABABABABA\\&BabababababaBABABABABABabababababaBABABA\\
I.A$_{29}^{i.c.}$(4) & BABABAbabababababABABABABABABabababababaBABABABABABAbabababababABABABABABAbabababababaBABABABABABababab\\
\hline
\end{tabular*}
{\rule{\temptablewidth}{1pt}}
\end{center}
\end{sidewaystable}

\begin{sidewaystable}
\tabcolsep 0pt \caption{The free group elements for the periodic three-body orbits.} \label{table-S8} \vspace*{-12pt}
\begin{center}
\def\temptablewidth{1\textwidth}
{\rule{\temptablewidth}{1pt}}
\begin{tabular*}{\temptablewidth}{@{\extracolsep{\fill}}ll}
\hline
Class and number & \hspace{1.0cm}  free group element\\\hline
I.B$_{16}^{i.c.}$(4) & BABABabababababaBABABABABABabababababaBABABABABABabababababaBABABABABABabababababaBABABABABAbababababab\\&ABABABABABAbabababababABABABABABAbabababababABABABABABAbabababababABABA\\
I.B$_{17}^{i.c.}$(4) & BABABAbabababababABABABABABAbabababababaBABABABABABababababababABABABABABAbabababababABABABABABABababab\\&ababaBABABABABABababababababABABABABABAbabababababaBABABABABABabababababaBABABA\\
I.B$_{18}^{i.c.}$(4) & BABABAbabababababABABABABABABababababababABABABABABAbabababababaBABABABABABAbabababababABABABABABABabab\\&abababaBABABABABABAbabababababaBABABABABABababababababABABABABABABabababababaBABABA\\
I.B$_{19}^{i.c.}$(4) & BABABAbabababababaBABABABABABAbabababababaBABABABABABababababababABABABABABABababababababABABABABABABab\\
\hline
\end{tabular*}%
{\rule{\temptablewidth}{1pt}}
\end{center}
\end{sidewaystable}

\end{document}